\documentclass[a4paper, 11pt]{report}
\pdfoutput=1   
\usepackage{chngcntr}   
\counterwithout{footnote}{chapter}

\usepackage{vmargin}
\setmarginsrb{3cm}{2cm}{2cm}{1.3cm}{0pt}{0mm}{12pt}{.9cm}   
\usepackage[pdfpagelabels]{hyperref}
\usepackage{parskip}
\usepackage{multirow}
\usepackage{longtable}
\usepackage{listings}
\usepackage{float}
\usepackage{morefloats}
\usepackage{graphicx}
\usepackage{amsmath}
\usepackage{amsfonts}
\usepackage{amssymb}
\usepackage{amsthm}
\usepackage{bm}
\usepackage{cite}
\usepackage{appendix}
\usepackage{chngcntr}
\usepackage{etoolbox}

\usepackage{array}
\newcolumntype{L}[1]{>{\raggedright\let\newline\\\arraybackslash\hspace{0pt}}p{#1}}
\newcolumntype{C}[1]{>{\centering\let\newline\\\arraybackslash\hspace{0pt}}p{#1}}
\newcolumntype{R}[1]{>{\raggedleft\let\newline\\\arraybackslash\hspace{0pt}}p{#1}}

\usepackage[dvipsnames]{xcolor}

\newcommand{\rEq}[1]{Eq.~\eqref{#1}}   

\newcommand{\rSec}[1]{Section~\ref{#1}}

\newcommand{\rApp}[1]{Appendix~\ref{#1}}

\newcommand{\p}[1]{\textsc{#1}}   

\newcommand{\sla}[1]{\ifmmode%
  \setbox0=\hbox{$#1$}%
  \setbox1=\hbox to\wd0{\hss$/$\hss}\else%
  \setbox0=\hbox{#1}%
  \setbox1=\hbox to\wd0{\hss/\hss}\fi%
  #1\hskip-\wd0\box1 }

\newcommand{\GeV}{\,\mathrm{GeV}}
\def\mathunderline#1#2{\color{#1}\underline{{\color{black}#2}}\color{red}}

\newcommand{\cO}{{\mathcal{O}}}
\newcommand{\cL}{{\mathcal{L}}}

\newcommand{\nn}{\nonumber \\}  
\newcommand{\nnl}{\nonumber \\}  
\newcommand{\hc}{\mathrm{h.c.}}
\newcommand{\eps}{\varepsilon}
\newcommand{\bvec}{\begin{pmatrix}}
\newcommand{\evec}{\end{pmatrix}}

\usepackage{graphicx}
\usepackage{colortbl}
\usepackage{amssymb}
\usepackage{epstopdf}
\usepackage{caption}
\usepackage{subcaption}
\newcommand{\code}[1]{\mbox{\texttt{#1}}}
\newcommand{\OO}{\ensuremath{\mathcal{O}}}
\newcommand{\pdp}{\ensuremath{\phi^\dagger\phi}}
\renewcommand{\phi}{\ensuremath{\varphi}}

\newcommand{\sss}{\scriptscriptstyle}
\newcommand{\bpm}{\begin{pmatrix}}      
\newcommand{\epm}{\end{pmatrix}}

\newcommand{\Op}[1]{\OO_{\sss #1}}
\newcommand{\Opp}[2]{\OO_{\sss #1}^{\sss #2}}

\newcommand{\Cp}[1]{C_{\sss #1}}
\newcommand{\Cpp}[2]{C_{\sss #1}^{\sss #2}}

\definecolor{lightgray}{rgb}{0.83, 0.83, 0.83}

 \def\lra#1{\overset{\text{\scriptsize$\leftrightarrow$}}{#1}}

\newcommand{\HRule}{\rule{\linewidth}{0.5mm}}
\renewcommand*{\thefootnote}{\alph{footnote}}

\begin{document}
\hypersetup{pageanchor=false}
\thispagestyle{empty}

	\begin{flushright}
	  {\large
            \textbf{\href{https://cds.cern.ch/record/2801789}{LHCHWG-2022-001}}} \\[0.5cm]	
		{\large 	\textrm{May 16, 2022}} \\[2.0cm]
	\end{flushright}

	\begin{center}

	\textsc{\Large 	\href{https://twiki.cern.ch/twiki/bin/view/LHCPhysics/LHCHWG}{LHC Higgs Working Group}\footnote{\href{https://twiki.cern.ch/twiki/bin/view/LHCPhysics/LHCHWG}{\sl https://twiki.cern.ch/twiki/bin/view/LHCPhysics/LHCHWG}}} \\[0.5cm]
	\textsc{\Large 	Public Note} \\[1.5cm]
	
	\HRule \\[0.9cm]
	\textbf{\Large \href{https://twiki.cern.ch/twiki/bin/view/LHCPhysics/HiggsOffshellTaskForce}{Off-shell Higgs Interpretations Task Force}\footnote{\href{https://twiki.cern.ch/twiki/bin/view/LHCPhysics/HiggsOffshellTaskForce}{\sl https://twiki.cern.ch/twiki/bin/view/LHCPhysics/HiggsOffshellTaskForce}}} \\[.8cm]
	\textbf{\Large Models and Effective Field Theories Subgroup Report} \\[0.8cm]
	\HRule \\[1.5cm]

	\textrm{\large Aleksandr Azatov $^{1,2,}$\footnote{\href{mailto:aazatov@sissa.it}{aazatov@sissa.it}}, Jorge de Blas $^{3,}$\footnote{\href{mailto:deblasm@ugr.es}{deblasm@ugr.es}}, Adam Falkowski$^{4,}$\footnote{\href{mailto:adam.falkowski@ijclab.in2p3.fr}{adam.falkowski@ijclab.in2p3.fr}}, Andrei V. Gritsan$^{5,}$\footnote{\href{mailto:gritsan@jhu.edu}{gritsan@jhu.edu}}, Christophe Grojean$^{6,7,}$\footnote{\href{mailto:christophe.grojean@desy.de}{christophe.grojean@desy.de}},  Lucas Kang$^{5,}$\footnote{\href{mailto:lkang12@jhu.edu}{lkang12@jhu.edu}}, Nikolas Kauer$^{8,}$\footnote{\href{mailto:n.kauer@rhul.ac.uk}{n.kauer@rhul.ac.uk}} (ed.), Ennio Salvioni$^{9,10,}$\footnote{\href{mailto:ennio.salvioni@unipd.it}{ennio.salvioni@unipd.it}},  {Ulascan Sarica$^{11,}$\footnote{\href{mailto:ulascan.sarica@cern.ch}{ulascan.sarica@cern.ch}},} Marion Thomas$^{12,}$\footnote{\href{mailto:marion.thomas@manchester.ac.uk}{marion.thomas@manchester.ac.uk}} and Eleni Vryonidou$^{12,}$\footnote{\href{mailto:eleni.vryonidou@manchester.ac.uk}{eleni.vryonidou@manchester.ac.uk}}} \\[0.3cm]	
	\textit{$^{1}$ SISSA International School for Advanced Studies, Via Bonomea 265, 34136 Trieste, Italy} \\
	\textit{$^{2}$ INFN -- Sezione di Trieste, Via Bonomea 265, 34136 Trieste, Italy} \\
	\textit{$^{3}$ CAFPE and Departamento de F\'isica Te\'orica y del Cosmos, Universidad de Granada, Campus de Fuentenueva, 18071 Granada, Spain}  \\
	\textit{$^{4}$ Universit\'{e} Paris-Saclay, CNRS/IN2P3, IJCLab, 91405 Orsay, France} \\
	\textit{$^{5}$ Dept.\ of Physics and Astronomy, Johns Hopkins University, Baltimore, MD 21218, USA} \\
	\textit{$^{6}$ Deutsches Elektronen-Synchrotron DESY, Notkestr.\ 85, 22607 Hamburg, Germany} \\
	\textit{$^{7}$ Institut f{\"u}r Physik, Humboldt-Universit{\"a}t zu Berlin, Newstonstr.\ 15, 12489 Berlin, Germany} \\
	\textit{$^{8}$ Dept.\ of Physics, Royal Holloway, University of London, Egham Hill, Egham TW20 0EX, UK}\!\!\!\!\!  \\
	\textit{$^{9}$ Dipartimento di Fisica e Astronomia, Universit\`{a} di Padova, Via Marzolo 8, 35131 Padua, Italy}\!\!\!\!\! \\
	\textit{$^{10}$ INFN -- Sezione di Padova, Via Marzolo 8, 35131 Padua, Italy}  \\
	\textit{$^{11}$ Dept.\ of Physics, University of California at Santa Barbara, Santa Barbara, CA 93106, USA} \\
	\textit{$^{12}$ Dept.\ of Physics and Astronomy, University of Manchester, Manchester M13 9PL, UK}  \\

	\end{center}

        \newpage
        \thispagestyle{empty}



\mbox{}\vspace*{3em}
\begin{center}
	\textbf{Abstract}
\end{center}
This report presents the results of the Models and Effective Field Theories Subgroup of the Off-Shell Interpretations Task Force in the LHC Higgs Working Group.  The main goal of the subgroup was to discuss and advance the potential impact of off-shell Higgs measurements on searches for BSM physics carried out in the EFT framework or as benchmark model studies.
In the first contribution, the off-shell potential to resolve flat directions in parameter space for on-shell measurements is studied.  Furthermore, the sensitivity of off-shell measurements to SMEFT dimension-6 operators for the $gg\to ZZ$ process is discussed, and studies of explicit models that are testable in off-shell production are reviewed.
In the second contribution, the SMEFT effects in the off-shell gluon fusion and electroweak processes are discussed.  Subsequently, the computation of integrated and differential effects using SMEFT@NLO and MG5\_aMC@NLO, or JHUGen and MCFM,  is demonstrated.  On that basis, a study of the prospects of obtaining additional SMEFT constraints -- beyond those from existing global fits -- by utilising the off-shell process is presented.
For clarification, a revised introduction, definition and discussion of the Higgs basis parametrisation of the SMEFT is given in the third contribution.
In short notes on the SMEFT, the Higgs basis with an additional constraint is discussed and relations between the Higgs and Warsaw bases are presented.
Lastly, an overview of EFT calculations and tools is given.

\newpage

\hypersetup{pageanchor=false}
\tableofcontents
\thispagestyle{empty}
\newpage
\hypersetup{pageanchor=true}

\hypersetup{pageanchor=true}
\renewcommand*{\thefootnote}{\arabic{footnote}}
\setcounter{footnote}{0}
\setcounter{page}{1}


\chapter{Introduction}

After the discovery of the 125\,GeV Higgs boson \cite{ATLAS:2012yve,CMS:2012qbp}, the search for physics beyond the Standard Model (SM) continues. Two complementary frameworks to study New Physics exist. On the one hand, if New Physics is heavier than the energy scales directly probed in the experiments, its effects can be well captured by an Effective Field Theory (EFT), where the SM Lagrangian is extended by terms consisting of Wilson coefficients multiplied by higher-dimensional operators constructed from the SM fields, which respect the SM gauge symmetries. On the other hand, one may consider explicit extensions of the SM with additional dynamical degrees of freedom and related model parameters, which may feature additional gauge or other symmetries.  By comparing predictions of such EFTs/explicit models with experimental measurements, bounds on Wilson coefficients and model parameters, respectively, can be obtained. The two approaches are complementary.  EFTs facilitate a generic parameterisation of New Physics that originates at higher scales, allowing us to remain rather agnostic about the exact structure of the SM extension in the ultraviolet, but have a limited range of validity.  On the other hand, light degrees of freedom can only be described by explicit models.  It is therefore essential to analyse experimental data using both descriptions.

Experimental data collected at the Large Hadron Collider (LHC) for Higgs production in gluon fusion or vector boson fusion (VBF) with subsequent decay to a pair of $Z$ or $W$ bosons contains an $\mathcal{O}(10\%)$ signal contribution from the off-shell high-mass region above $2M_V$\cite{Glover:1988rg,Kauer:2012hd,Caola:2013yja,Campbell:2013una}.\footnote{Note that off-shell effects can also induce a shift of the Higgs mass peak \cite{Dicus:1987fk,Dixon:2003yb,Martin:2012xc,Dixon:2013haa}.} The situation is similar at a high-energy $e^+ e^-$ collider.  The off-shell high-mass Higgs signal has distinctly different properties than the dominant ``on-peak'' signal contribution, where the Higgs is nearly on-shell.  It thus has the potential to provide complementary information in searches for Beyond-SM (BSM) physics.  First, the off-shell signal above $2M_V$ has a plateau-like distribution, which extends to even higher energies in VBF than in gluon fusion.  Therefore, the typical invariant mass of the off-shell Higgs is much higher than in the on-shell case.  Consequently, all high-energy effects are more pronounced and the sensitivity to certain Wilson coefficients is enhanced.  Secondly, unlike the on-peak signal, the off-shell signal is affected by substantial signal-background interference.  Thirdly, unlike the on-peak signal, the off-shell signal is essentially independent of the total Higgs width.  This has enabled $\mathcal{O}(1)$ bounds on the Higgs width from LHC data relative to the SM prediction~\cite{CMS:2014quz,ATLAS:2015cuo,ATLAS:2018jym,CMS:2022ley}.  It came as a surprise, given that the direct Higgs width measurement at the LHC \cite{CMS:2017dib} is severely limited by the mass resolution, which is approximately 600 times larger than the SM prediction.\footnote{The Higgs width bound from combining on- and off-shell data relies on assuming no energy dependence of the relevant Higgs couplings.}

The comprehensive task that suggests itself is to determine the subset of EFT operators that are most sensitive to off-shell data and to examine in detail the prospects of off-shell Higgs data providing better or complementary constraints than those obtained in global fits.  The ability of the off-shell data to resolve degeneracies in parameter space, which was exploited to obtain the mentioned Higgs width bounds, may more generally be useful to resolve flat directions for on-shell measurements in the parameter space of explicit BSM realisations.  Regarding the Higgs width constraint analysis, it would be interesting to understand in more detail how the scaling violation for the total Higgs width can arise and what constraints can be achieved at colliders.

When assessing the potential impact of off-shell Higgs measurements on searches for BSM physics carried out in the EFT framework or as benchmark model studies, the following issues deserve consideration.

The EFT studies discussed in this report make use of the SMEFT, consistently with the current experimental status. The ``minimal'' list of SMEFT operators and accordingly couplings that deserve priority in the initial stage of analyses needs to be determined.\footnote{The discussion and results in Sections~\ref{sec:smefteffggZZ} and \ref{sec:vryothomresults} of this report address this issue.}

Different bases of the SMEFT have been developed.  At leading order (LO), some off-shell studies prefer the Higgs basis over the Warsaw basis.\footnote{Beyond LO it becomes difficult to argue for a particular basis, except for practical considerations, such as tools availability.}  It is hence important to clarify the motivation and details of this basis, and its relation to the Warsaw basis.\footnote{Chapter~\ref{ch:higgsbasis} and Section~\ref{sec:rel-higgs-warsaw} in this report address this issue.}

Initially, it is suggestive to analyse the impact of SMEFT operators on the off-shell process individually and to ignore correlations with other processes.  Subsequently, however, the interplay with other channels due to shared couplings, e.g.\ in top production, and dependencies between operators cannot be ignored.  
The following questions arise: How should bounds, especially more competitive ones, on relevant Wilson coefficients obtained in other channels be taken into account?  How can operators be disentangled?  Can independent subsets of operators be found?

From a practical point of view, the number of degrees of freedom in fits is limited.  A key question is therefore: With what quantities should the available degrees of freedom be associated?  When going beyond the ``minimal'' list mentioned above, what approach should be taken in expanding the fit variables?  
From an experimental point of view, a full EFT analysis of Higgs data is extremely challenging, even without consideration of non-Higgs-sector operators. 
A priori, it is suggestive to focus on the operators affecting the $HVV$, $Hgg$ and $Ht\bar{t}$ couplings that appear in off-shell Higgs production in gluon fusion; and to deprioritise consideration of non-Higgs EFT couplings, which primarily affect backgrounds, whenever they are better constrainted by non-Higgs data.
While other $HVV$ and $Hf\bar{f}$ operators may also contribute, they are expected to be less important.

The off-shell enhancement is not only present in gluon-fusion Higgs production. As mentioned, Higgs production in VBF features a similar enhancement.  
We note that the VBF channel is less model dependent, since it is not loop induced, but yields less data overall.  At high energies, however, its signal is similar to or larger than the gluon-fusion signal.  The additional sensitivity provided by the VBF/$VH$ channel data should also be analysed.

As mentioned above, a two-pronged approach is essential.  The EFT-based studies need to be complemented by studies of explicit SM extensions.  Which poses the questions: What types of BSM benchmark models should be analysed? Which new particles can be probed by off-shell Higgs measurements?  More specifically: Are common models sufficient, for example the MSSM, 2HDM and SM+scalar?  What toy models should be studied?  Should composite inspired models be studied?  Can toy models be nominated that cover the two extreme scenarios: no exotic/undetected width, but large detectable effects in off-shell measurements; a sizeable exotic/undetected width, but no detectable deviation in off-shell measurements? It would be useful to identify which regions of the parameter spaces of common models are more likely to be revealed in exotic Higgs decays or in off-shell data. For the BSM benchmark models as well as for the SMEFT, a more detailed analysis of the additional sensitivity provided by the off-shell process when combined with constraints from top or electroweak (EW) measurements for instance appears well justified.

At a later stage, the following issues deserve consideration.  

The additional operators in the SMEFT do not only affect the signal, but also the background and the Higgs width.\footnote{For the interfering background to the off-shell process, this is illustrated in Figure~\ref{fig:FeynmanDiagrams}.}
Therefore, SMEFT effects should be taken into account in background amplitudes and, for consistency, also in the Higgs width, in order to clarify if they can be significant. If that is the case, then statistical data analysis methods will be required that permit to include BSM effects in the background when determining bounds on Wilson coefficients.  This is of course not just an issue for the EFT modelling of New Physics, but similarly affects model parameters in studies of BSM benchmark models.

At this more advanced stage, next-to-leading order (NLO) QCD and EW corrections should be taken into account if they are available.  In particular for QCD, calculations and tools are available, for instance to take into account NLO QCD corrections in the SMEFT.\footnote{An overview of calculations and tools is given in Chapter~\ref{ch:tools}.}  BSM@NLO in general, especially for dimension-8 operators, appears to be beyond current capabilities.  But, neglected BSM higher order effects can be estimated to assess the validity of the EFT expansion.  Such studies or studies that include available QCD or EW corrections and find large effects may motivate future efforts towards general BSM@NLO, in particular if and when significant deviations from the SM are observed.  From a technical point of view, BSM@NLO is more feasible for VBF/$VH$ processes, which proceed at tree level and are hence more amenable for proof-of-concept studies.  For loop-induced processes like gluon-fusion Higgs production, the technical threshold is significantly higher.

An interesting proposal that may warrant more attention in the future is to study a specific BSM benchmark model which is extended with higher-dimensional operators -- in particular if a SM deviation is observed.  The latter would also provide guidance as to what underlying model should be chosen.  Treating the Higgs width as a free parameter in a SMEFT fit would imply the assumption of additional BSM physics to account for the possible Higgs width deviation.  It would be instructive if these additional degrees of freedom could be modelled explicitly, for instance through additional fields in the underlying model.

The report is structured as follows: 
In Chapter~\ref{ch:ofsforBSM}, the potential impact of off-shell Higgs measurements on BSM physics is re-examined, in particular the prospects to resolve flat directions in parameter space for on-shell measurements.  Furthermore, the sensitivity of off-shell measurements to SMEFT dimension-6 operators for the $gg\to ZZ$ process and studies of explicit models that are testable in off-shell production are reviewed.   
In Chapter~\ref{ch:ofsSMEFT}, SMEFT effects in the off-shell process of Higgs production and decay to a weak-boson pair are discussed, and the computation of integrated and differential effects using SMEFT@NLO and MG5\_aMC@NLO, or JHUGen and MCFM, is demonstrated.  Furthermore, a study of the prospects of obtaining additional SMEFT constraints beyond those from global fits by utilising the off-shell process is presented.
In Chapter~\ref{ch:higgsbasis}, a revised introduction, definition and discussion of the Higgs basis parametrisation of the SMEFT is given, including an appendix on notation and conventions.
Chapter~\ref{ch:shortnotes} contains short notes on the SMEFT.  The first is on the Higgs basis with an additional constraint.  The second is on the relation between the Higgs and Warsaw bases.
In Chapter~\ref{ch:tools}, an overview of EFT calculations and tools is given.
The report closes with a summary and final remarks in Chapter~\ref{ch:conclusions}.


\chapter[What can off-shell Higgs measurements tell us about BSM physics?]{What can off-shell Higgs measurements tell us about BSM physics?\footnote{contributed by A.\ Azatov, J.\ de Blas, C.\ Grojean, E.\ Salvioni}\label{ch:ofsforBSM}}

In this chapter we briefly re-examine the potential impact of off-shell Higgs measurements on Beyond the Standard Model (BSM) physics.
\section{Going beyond a universal flat direction}
\begin{figure}[t]
    \centering
    \includegraphics[width=0.4\textwidth]{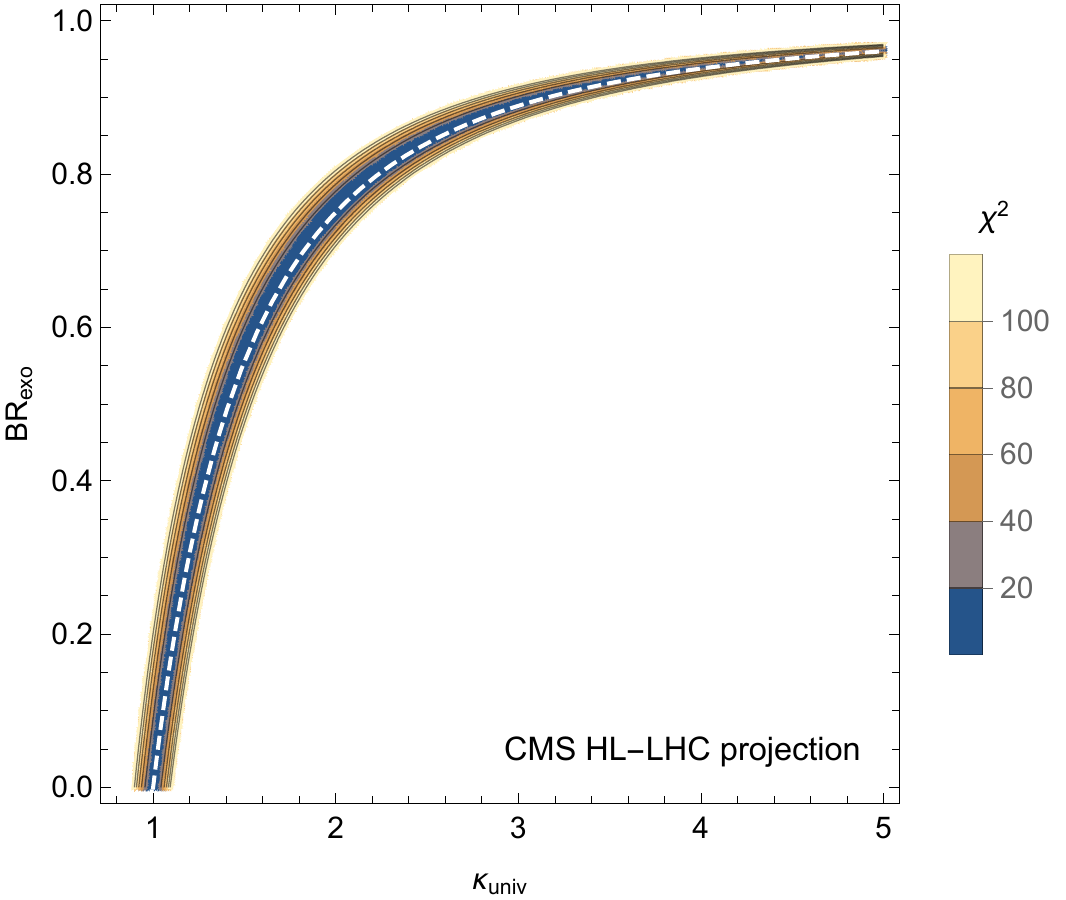}
    \caption{The flat direction that affects on-shell Higgs measurements if a universal coupling rescaling $\kappa_{\rm univ}$ and the presence of an untagged branching ratio $\mathrm{BR}_{\rm exo}$ are assumed. The white dashed line corresponds to the exact relation $\mathrm{BR}_{\rm exo} = (\kappa_{\rm univ}^2 - 1) / \kappa_{\rm univ}^2$ for $\kappa_{\rm univ} > 1$. The colored regions correspond to $\chi^2$ contours for the projection to the HL-LHC of CMS on-shell Higgs measurements, assuming data will agree with the SM~\cite{deBlas:2019rxi}.}
    \label{fig:flat_dir}
\end{figure}

To begin we review the original proposal by Caola and Melnikov~\cite{Caola:2013yja}, who pointed out that the off-shell channel can lift a flat direction plaguing LHC on-shell Higgs measurements: if the Higgs couplings are universally rescaled, $g_{hii} = \kappa_{\rm univ} g_{hii}^{\rm SM}$, and the Higgs width is modified according to $\Gamma_h = \kappa_{\rm univ}^4 \Gamma_h^{\rm SM}$, on-shell rates remain identical to the Standard Model (SM). Genuinely new contributions to the Higgs width can be categorized in {\it invisible} and {\it untagged}. Since the former is already constrained to $\mathrm{BR}_{\rm inv} < 0.11$ at $95\%$ CL by direct measurement~\cite{ATLAS:2020cjb,CMS:2022qva} (up to $139$ fb$^{-1}$ at $13$~TeV), in this note we focus on the presence of an untagged partial width, in which case the flat direction is along
\begin{equation}
\mathrm{BR}_{\rm exo} = \frac{\kappa_{\rm univ}^2 - 1}{ \kappa_{\rm univ}^2 }\,,
\end{equation}
as can be seen in Fig.~\ref{fig:flat_dir}. Notice that, importantly, the flat direction is present for $\kappa_{\rm univ} > 1$. Caola and Melnikov observed that the off-shell rate is $d\sigma_{gg\to h^\ast \to ZZ} / d s \propto g_{hgg}^2 g_{hZZ}^2/s^2 = \kappa_{\rm univ}^4 (g_{hgg}^{\rm SM})^2 (g^{\rm SM}_{hZZ})^2/s^2$, where $\sqrt{s} = m_{4\ell}$. Hence, an upper bound on the cross section in the large$\,$-$\,m_{4\ell}$ region translates into an upper bound on $\kappa_{\rm univ}$ (and therefore into an upper bound on the Higgs width, under the above set of assumptions).

How can such a universal flat direction be realized in a concrete BSM setup? As a simple illustration, we consider a scalar extension of the SM containing the following interactions,
\begin{equation} \label{eq:BSM_toy}
\mathcal{L}_{\rm BSM} \ni \frac{c_H}{2f^2} (\partial_\mu |H|^2 )^2 - \lambda_{H\varphi} |H|^2 \varphi^2 \,,
\end{equation}
where $H$ is the SM Higgs doublet, and $\varphi$ is a real scalar that decays dominantly to hadrons, for example a color-singlet or -octet decaying to $gg$ (a singlet would decay through higher-dimensional operators, such as $\varphi\, G_{\mu\nu} G^{\mu\nu}$). The dimension-6 operator in Eq.~\eqref{eq:BSM_toy} gives a universal rescaling $g_{hii}/g_{hii}^{\rm SM} = 1 - c_H v^2/f^2$. For $m_\varphi < m_h/2$, the untagged Higgs width also acquires a contribution from $h\to \varphi \varphi$ mediated by the dimension-4 operator, with $\Gamma \sim \lambda_{H\varphi}^2 v^2 / (8\pi m_h)$ for $m_\varphi \ll m_h/2$. Thus, two conditions need to be met to obtain a universal flat direction: first, $c_H$ must be {\it negative}, which is only realized in somewhat exotic (albeit possible) theories, e.g. models with electroweak triplet scalars or non-compact coset spaces; second, there must be an accidental ``conspiracy'' relating the a priori-independent quantities $c_H/f^2$ and $\lambda_{H\varphi}$ in the appropriate fashion. 

\vspace{0.2cm}
\noindent {\bf  Models with a universal flat direction?}
\vspace{0.1cm}

To make these issues more explicit, we inspect how $c_H < 0$ can arise from integrating out triplet scalars at tree level~\cite{Low:2009di}. Considering both a real triplet $\phi_r^a$ with hypercharge $Y = 0$ and a complex triplet $\Phi_c^a$ with $Y = 1$, the relevant pieces of the UV Lagrangian are
\begin{equation}
\mathcal{L}_{\rm UV} = \frac{1}{2} \partial_\mu \phi_r^a  \partial^\mu \phi_r^a - \frac{1}{2} M_r^2 \phi_r^a \phi_r^a + \beta_r f \phi_r^a H^\dagger \frac{\sigma^a}{2} H + \partial_\mu \Phi_c^{a \ast} \partial^\mu \Phi_c^a - M_c^2 \Phi_c^{a\ast} \Phi_c^a + \beta_c f \big( \Phi_c^{a \ast} H^T \epsilon \frac{\sigma^a}{2} H + \mathrm{h.c.} \big),
\end{equation}
where $\epsilon \equiv i \sigma^2$. Integrating out the heavy scalars we obtain in the SILH basis for the EFT~\cite{Giudice:2007fh}
\begin{equation}
\frac{c_H}{f^2} = - \frac{\beta_r^2 f^2}{2 M_r^4} -  \frac{\beta_c^2 f^2}{2 M_c^4}\,, \qquad \frac{c_T}{f^2} = \frac{\beta_r^2 f^2}{4 M_r^4} -  \frac{\beta_c^2 f^2}{2 M_c^4}\,,
\end{equation}
where $c_H$ is manifestly negative and $c_T$ is the coefficient of $( H^\dagger \hspace{-1.5mm} \stackrel{\leftrightarrow}{D_\mu} \hspace{-1.3mm} H)^2 / (2 f^2)$. In scenarios with custodial symmetry, such as the Georgi-Machacek (GM) model~\cite{Georgi:1985nv,Chanowitz:1985ug}, the real and complex triplets satisfy the relations $ \beta_r^2 f^2 = 2 \beta_c^2 f^2$ and $M_r^2 = M_c^2$, resulting in $c_T = 0$ and $c_H /f^2 = - 3 \beta_r^2 f^2 / (4 M_r^4)$. However, under the assumption that the mass of the triplets is sufficiently large that they can be integrated out, the GM model does not contain a possible candidate for the light scalar $\varphi$; the latter can of course be added to the model as an additional singlet, but an ad-hoc suitable relation between $\beta_r f, M_r$, and $\lambda_{H\varphi}$ would need to be imposed in order to sit along the flat direction discussed above. 

A second possibility to obtain $c_H < 0$ is via a non-compact coset. An example that has been discussed in the literature is $SO(4,1)/SO(4)$~\cite{Alonso:2016btr}, giving rise to $H$ as a Goldstone doublet within a consistent effective theory. The price to pay is that the effective theory cannot be UV-completed by an ordinary QFT, since the latter cannot have the non-compact $SO(4,1)$ as a linearly realized global symmetry group; see Ref.~\cite{Liu:2016idz} for related discussions. As in the GM model, to realize a universal flat direction a genuinely new contribution to the Higgs width is required. This could be obtained by extending the coset to include additional Goldstones, one of which may be identified with the light $\varphi$ (in this case, the role of the last operator in Eq.~\eqref{eq:BSM_toy} could also be played by $c_{H\varphi} \partial_\mu |H|^2 \partial^\mu \varphi^2/f^2$), but the necessary parametric relation would again need to be accidental.

The above examples make it clear that a BSM theory needs to satisfy specific conditions in order for a universal flat direction to be realized, as already emphasized in Ref.~\cite{Azatov:2014jga}. For this reason, we next ask whether the fit to on-shell Higgs data allows for (approximately) flat directions even when the assumption of coupling universality is relaxed.

\vspace{0.2cm}
\noindent {\bf Relaxing coupling universality}
\vspace{0.1cm}

As a first step in the exploration of the impact of off-shell measurements on more general BSM scenarios, we depart from coupling universality by allowing the rescaling of the $hb\bar{b}$ interaction to be different from the others, thus focusing on the $( \tilde{\kappa}_{\rm univ}, \kappa_b , \mathrm{BR}_{\rm exo})$ parameter space. The rationale for choosing this parametrization is that, since in the SM the total Higgs width is dominated by $\kappa_b$, the on-shell global fit has an approximate flat direction in the $(\kappa_b, \mathrm{BR}_{\rm exo})$ plane: for a given $\kappa_b < 1$ there exists a value of $\mathrm{BR}_{\rm exo}$ that maintains the total width SM-like. This flat direction is lifted by observables that test directly the $h\to b\bar{b}$ decay, whose sensitivity will be somewhat limited even at the HL-LHC.\footnote{Gluon fusion production, $h\to \gamma\gamma$ and $h\to Z\gamma$ are also sensitive to $\kappa_b$, but only very weakly.} The best channel is expected to be $Zh, h\to b\bar{b}$ ($Wh, h\to b\bar{b}$), where CMS projects an uncertainty of $6.5\%$ ($9.4\%$)\cite{Cepeda:2019klc}. This is followed by $t\bar{t}h, h \to b\bar{b}$, with a projected uncertainty weaker by approximately a factor $2$. In this context off-shell Higgs production can be regarded as providing complementary information to on-shell $h\to b\bar{b}$ channels. 

\begin{figure}[t!]
	\centering
\raisebox{0\height}{\includegraphics[width=0.46\textwidth]{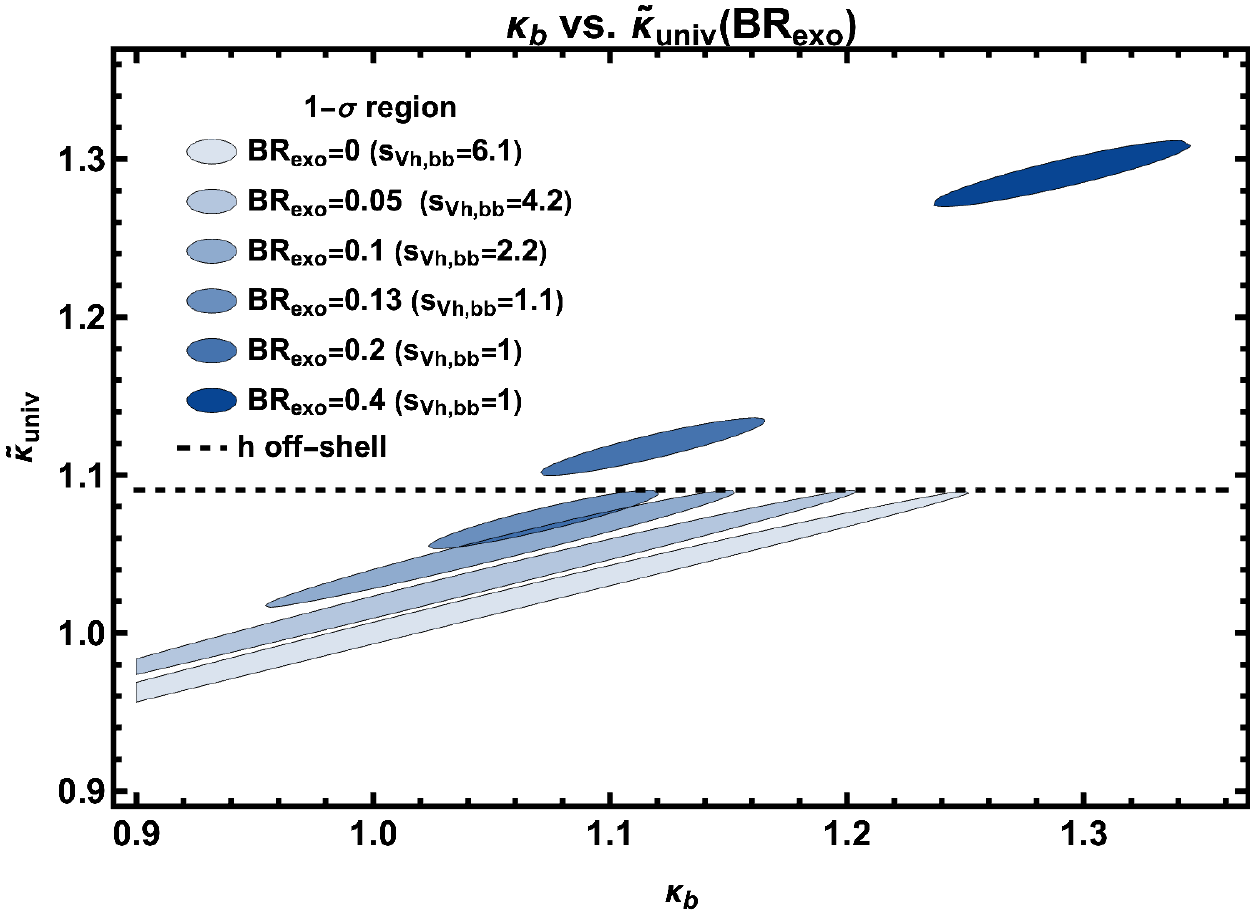}} 
\hspace{2mm}
\raisebox{+0.003\height}{\includegraphics[width=0.466\textwidth]{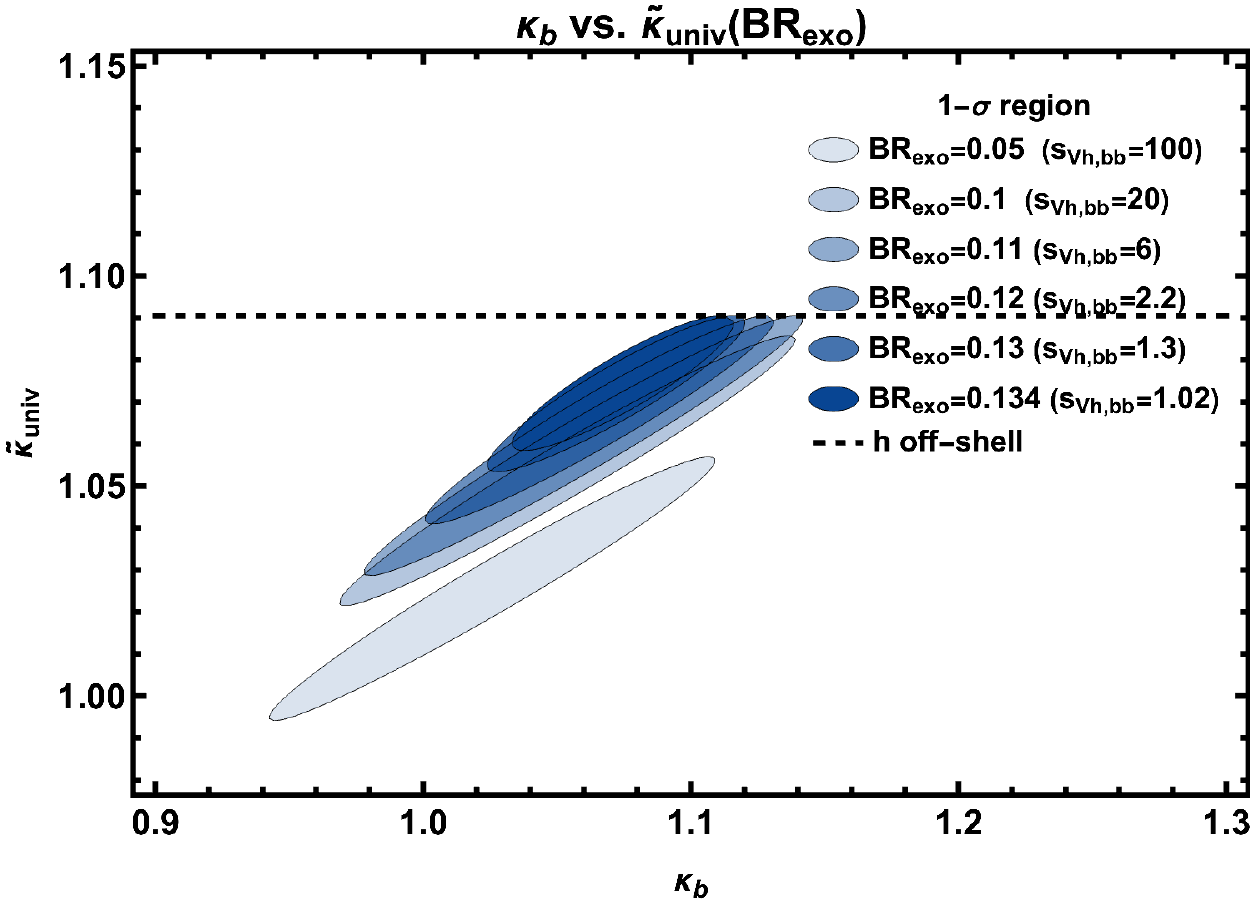}} 
	\caption{HL-LHC comparison of the off-shell sensitivity to the on-shell fit, for the $( \tilde{\kappa}_{\rm univ}, \kappa_b , \mathrm{BR}_{\rm exo})$ parameter space. {\it Left:} in the on-shell fit only $Vh, h\to b\bar{b}$ is considered among the direct probes of the $hb\bar{b}$ coupling, with expected uncertainty relaxed by a factor $s_{Vh, b\bar{b}}\,$. {\it Right:} the on-shell fit includes all $h\to b\bar{b}$ modes, and we relax the uncertainty of $Vh, h\to b\bar{b}$ only. }
	\label{fig:3dim_space}
\end{figure}

We illustrate this in the left panel of Fig.~\ref{fig:3dim_space}, where each shaded ellipse covers the region of the $(\kappa_b , \tilde{\kappa}_{\rm univ})$ plane allowed at $1\sigma$ by the on-shell CMS HL-LHC fit~\cite{Cepeda:2019klc,deBlas:2019rxi}, assuming the indicated value of $\mathrm{BR}_{\rm exo}$ and relaxing the uncertainty on the leading $Vh, h\to b\bar{b}$ measurement by the indicated multiplicative factor $s_{Vh, b\bar{b}}\,$. Subleading direct probes of the $hb\bar{b}$ coupling, in particular $t\bar{t} h, h\to b\bar{b}$, are excluded from the fit. In addition, the dashed line shows the upper limit of the allowed range of $\tilde{\kappa}_{\rm univ}$ as found from the off-shell contribution to $gg\to 4\ell$ at the HL-LHC~\cite{Azatov:2016xik}. Using a binned fit and neglecting systematic uncertainties we find $0.57 < \tilde{\kappa}_{\rm univ} < 1.09$ at $1\sigma$. 

The implications of our results can be evaluated by considering a few benchmarks: for a relatively large $\mathrm{BR}_{\rm exo} = 0.2$, off-shell is guaranteed to have stronger sensitivity than $Vh, h\to b\bar{b}$ even for nominal uncertainties on the latter; for intermediate $\mathrm{BR}_{\rm exo} = 0.1$, off-shell would provide genuinely new information for $s_{Vh, b\bar{b}} \approx 2$ (i.e., if the uncertainty turns out to be a factor $2$ weaker than currently expected); for small $\mathrm{BR}_{\rm exo} = 0.05$, the performance in $Vh, h\to b\bar{b}$ would need to be much ($\approx 4$ times) worse than currently projected for the off-shell constraint to be competitive. The right panel of Fig.~\ref{fig:3dim_space} shows similar results, but including all $h\to b\bar{b}$ channels in the on-shell fit. We see that if $\mathrm{BR}_{\rm exo} \lesssim 0.1$ off-shell cannot provide useful information even if $s_{Vh, b\bar{b}}\,$ is very large, due to the $t\bar{t}h, h \to b\bar{b}$ sensitivity which takes over in that limit.

\vspace{0.2cm}
\noindent {\bf Impact of observing deviations from the SM in on-shell $h\to b\bar{b}$ channels}
\vspace{0.1cm}

\begin{figure}[t!]
	\centering
\raisebox{0\height}{\includegraphics[width=0.45\textwidth]{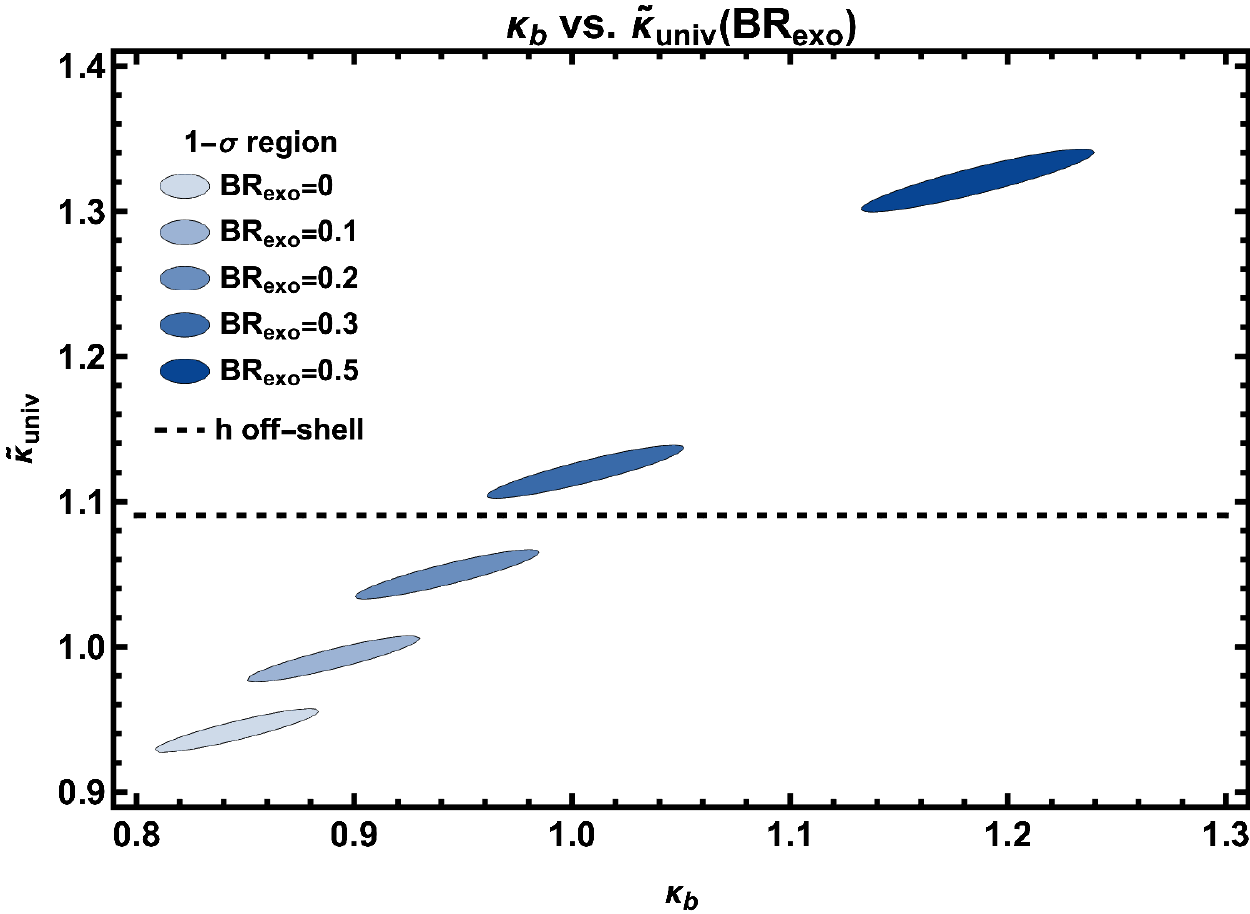}} 
\hspace{2mm}
\raisebox{-0.00\height}{\includegraphics[width=0.456\textwidth]{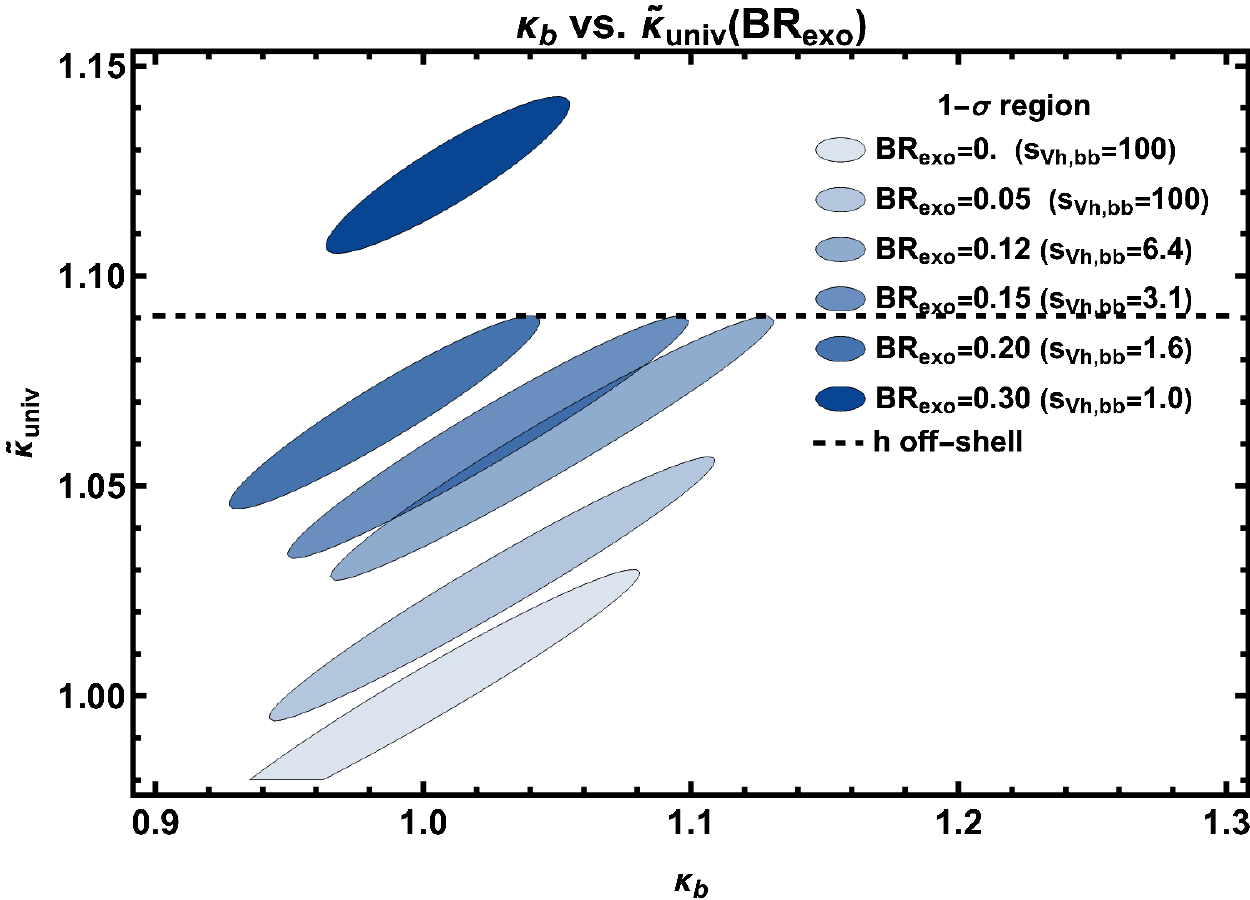}} 
	\caption{HL-LHC comparison of the off-shell sensitivity to the on-shell fit, assuming the $Vh, h\to b\bar{b}$ to deviate from the SM as $\mu_{Vh, h\to b\bar{b}} = 0.75$. {\it Left:} nominal uncertainties are assumed for $h\to b\bar{b}$ channels. {\it Right:} we relax the uncertainty of $Vh, h\to b\bar{b}$ by the multiplicative factor $s_{Vh, b\bar{b}}\,$. }
	\label{fig:3dim_space_deviationSM}
\end{figure}

As a further step in our analysis we consider a scenario where the $Vh, h\to b\bar{b}$ rate is observed to be lower than the SM prediction at the HL-LHC. We choose $\mu_{Vh, h\to b\bar{b}} = 0.75$, corresponding to the lower edge of the current $1\sigma$ uncertainty band. Figure~\ref{fig:3dim_space_deviationSM} shows the corresponding contours in the $(\kappa_b, \tilde{\kappa}_{\rm univ})$ plane. In the left panel we assume nominal uncertainties for the $h\to b\bar{b}$ channels, in which case off-shell provides a relevant constraint as long as $\mathrm{BR}_{\rm exo} > 0.23$. In the right panel the $s_{Vh, b\bar{b}}\,$ factor is varied as well, showing in particular that for $\mathrm{BR}_{\rm exo} < 0.15$ off-shell would be competitive only in the presence of a strong relaxation (by a factor $> 3$) of the $Vh, h\to b\bar{b}$ uncertainty. 

For the sake of illustration we give the analytical relation between the three parameters under consideration, $( \tilde{\kappa}_{\rm univ}, \kappa_b , \mathrm{BR}_{\rm exo})$, and three experimental rates that determine them, normalized to the SM predictions:
\begin{itemize}
\item $\mu_{\rm on}$, the on-shell rate for $gg\to h \to ZZ^\ast$ (or equivalently, any other on-shell process that does not involve $h\to b\bar{b}$); 
\item $\mu_{Vh, b\bar{b}}\,$, the on-shell rate for $q\bar{q}\to V (h\to b\bar{b})$; 
\item $\mu_{\rm off}$, the rate for the $gg\to 4\ell$ process in the kinematic region $m_{4\ell} \in [250, 1500]\;\mathrm{GeV}$, which includes the off-shell Higgs contribution. 
\end{itemize}
We find the relations
{\small
\begin{equation}
\mathrm{BR}_{\rm exo} = 1 - \mu_{\rm on} \frac{\tilde{\kappa}_{\rm univ}^2 ( 1 - \mathrm{BR}^{b\bar{b}}_{\rm SM} ) + \kappa_b^2 \mathrm{BR}^{b\bar{b}}_{\rm SM}}{\tilde{\kappa}_{\rm univ}^4}\,,\ \ \kappa_b^2 = \frac{\mu_{Vh, b\bar{b}}}{\mu_{\rm on}} \,\tilde{\kappa}_{\rm univ}^2\,,\ \ \tilde{\kappa}_{\rm univ}^2 = \frac{ \pm \sqrt{b^2 + 4  ( \mu_{\rm off} - a) c} \,- b}{2c}\,,
\end{equation}
}%
where $\mathrm{BR}_{\rm SM}^{b\bar{b}} = 0.58$ and the last relation is derived by inverting $\mu_{\rm off} = a + b \tilde{\kappa}_{\rm univ}^2 + c \tilde{\kappa}_{\rm univ}^4\,$, where $\{a, b, c\} =\{1.07, - 0.20, 0.13\}$.\footnote{The corresponding HL-LHC sensitivity derived using a single inclusive bin is $0.50 < \tilde{\kappa}_{\rm univ} < 1.13$ at $1\sigma$, moderately weaker than the result obtained dividing the $m_{4\ell} \in [250, 1500]\;\mathrm{GeV}$ region into $5$ bins, $0.57 < \tilde{\kappa}_{\rm univ} < 1.09$ as already mentioned.} Both solutions exist for $\mu_{\rm off} < a = 1.07$, one having $\tilde{\kappa}_{\rm univ} > 1$ and one $\tilde{\kappa}_{\rm univ} < 1$, whereas for $\mu_{\rm off} > a$ only the ``$+$'' solution with $\tilde{\kappa}_{\rm univ} > 1$ remains.

\vspace{0.2cm}
\noindent {\bf  Summary}
\vspace{0.1cm}

In this note we have reconsidered the power of off-shell Higgs measurements to probe new physics. We started from the original proposal of off-shell observables as capable of lifting the universal flat direction plaguing on-shell Higgs rates. We have spelled out the conditions that a BSM theory must satisfy for such a universal flat direction to arise, finding that two specific ingredients must be simultaneously present: an enhancement of the Higgs couplings with respect to the SM, which requires models with extended scalar multiplets or non-compact cosets, and an accidental relation between the coupling modification and the BSM decay width of the Higgs. Motivated by these arguments we have made a first step away from coupling universality, considering a $( \tilde{\kappa}_{\rm univ}, \kappa_b , \mathrm{BR}_{\rm exo})$ parametrization. In this scenario on-shell rates display an approximate flat direction, which is broken both by on-shell observables involving $h\to b\bar{b}$, and by off-shell measurements. We have compared the two, finding that off-shell has the leading resolving power in the presence of a relatively large $\mathrm{BR}_{\rm exo}\gtrsim 0.2\,$; for smaller values, off-shell will be competitive if the uncertainties in the $Vh, h\to b\bar{b}$ channel turn out to be larger than currently projected. Finally, given the importance of the interplay with $h\to b\bar{b}$ observables, we have explored how the above conclusions would be affected if the $Vh, h\to b\bar{b}$ rate were observed to deviate from the SM prediction.

\section{SMEFT effects in $gg\to ZZ$ \label{sec:smefteffggZZ}}

We turn to discuss the sensitivity of off-shell measurements to SMEFT dimension-6 operators, considering the $gg\to ZZ$ process. In the Higgs basis (see Chapter~\ref{ch:higgsbasis}, whose conventions we follow) we count {\color{blue}$9$ $CP$-even} and {\color{red}$5$ $CP$-odd} coefficients,
{\footnotesize
\begin{align}
\label{eq:lagrangian}
\Delta \mathcal{L} \,=\, \frac{h}{v} \Big( {\color{blue}c_{gg}} \frac{g_s^2}{4}  G_{\mu\nu}^a G^{\mu\nu \, a} -  m_t {\mathunderline{red}{\mathunderline{blue}{ [\delta y_u]_{33} }}}   \bar{t}_L t_R \;+ &\; \mathrm{h.c.} + {\color{blue}\delta c_z} \frac{g_Z^2 v^2}{4} Z_\mu Z^\mu + {\color{blue}c_{zz}} \frac{g_Z^2}{4} Z_{\mu\nu}Z^{\mu\nu} + {\color{blue} c_{z \Box}} g_L^2 Z_\mu \partial_\nu Z^{\mu\nu} \nonumber \\ +\, {\color{red} \tilde{c}_{gg} } \frac{g_s^2}{4} G_{\mu\nu}^a \widetilde{G}_{\mu\nu}^a 
\,+\, {\color{red} \tilde{c}_{zz} } \frac{g_Z^2}{4} Z_{\mu\nu} \widetilde{Z}_{\mu\nu} \Big)& - g_Z {\color{blue} (\delta g_L^{Zu})_{33}} Z_\mu \bar{t}_L \gamma^\mu t_L - g_Z {\color{blue} (\delta g_R^{Zu})_{33} } Z_\mu \bar{t}_R \gamma^\mu t_R  \\  \,-\, \frac{m_t}{4v^2} \Big( 1 + \frac{h}{v} \Big) \Big( g_s \bar{t}_R & \sigma^{\mu\nu} T^a \mathunderline{red}{ \mathunderline{blue}{  [d_{Gu}]_{33} }} {\color{black} t_L G_{\mu\nu}^{a} + g_Z \bar{t}_R \sigma^{\mu\nu} T^a} \mathunderline{red}{\mathunderline{blue}{ [d_{Zu}]_{33} }} { \color{black} t_L Z_{\mu\nu} \Big) + \mathrm{h.c.}, }\nonumber
\end{align}
}
where $g_Z \equiv \sqrt{g_L^2 + g_Y^2}$ and complex coefficients are underlined in both colors. Notice that couplings involving the photon and the $W$ were omitted; these would be relevant for $gg\to WW$.

We focus here on the $CP$-even couplings. Several of these can be constrained from other measurements, in particular:
\begin{itemize}
\item One linear combination of $c_{gg}$ and $\mathrm{Re}\,[\delta y_u]_{33}$ determines the leading correction to Higgs production in gluon fusion at the LHC,
\begin{equation}
\frac{\sigma_{gg\to h}}{\sigma_{gg\to h}^{\rm SM}} \simeq \Big( 1  + 12 \pi^2 c_{gg} + \mathrm{Re}\,[\delta y_u]_{33} \Big)^2 .
\end{equation}
The remaining blind direction is lifted mainly by $t\bar{t}h$ production, which is sensitive to $\mathrm{Re}\,[\delta y_u]_{33}$ at tree-level.
\item $\delta c_z$ is directly probed by on-shell $h\to ZZ^\ast$ decays, as well as $h\to WW^\ast$ if $\delta m_w = 0$ is assumed (recall that $\delta c_w = \delta c_z + 4 \delta m_w$), a sensible approximation in the context of Higgs analyses, as discussed in Chapter~\ref{ch:higgsbasis}. 
\item The corrections to the $Z\bar{t}_L t_L$, $W\bar{t}_L b_L$ and $Z\bar{b}_L b_L$ interactions are not all independent at dimension $6$,
\begin{equation}
\delta g_L^{Wq} = V_{\rm CKM}^\dagger \delta g_L^{Zu} V_{\rm CKM} - \delta g_L^{Zd} \quad \to \quad (\delta g_L^{Wq})_{33} \simeq (\delta g_L^{Zu})_{33}  - (\delta g_L^{Zd})_{33}
\end{equation}
where the last relation considers only BSM modifications of the third generation quark gauge couplings.\footnote{Equivalently, in the Warsaw basis the $Z\bar{t}_L t_L$, $W\bar{t}_L b_L$ and $Z\bar{b}_L b_L$ corrections arise from only two operators, $Q_{\phi q}^{(1)33}$ and $Q_{\phi q}^{(3)33}\,$.} The coefficient $(\delta g_L^{Zd})_{33}$ has been measured with $O(10^{-3})$ accuracy at LEP, whereas $(\delta g_L^{Wq})_{33}$ is known to better than $10\%$ from single-top measurements at the LHC:  $(\delta g_L^{Wq})_{33} = v^2 \frac{C_{\phi q}^{(3)33}}{\Lambda^2}$ with $C_{\phi q}^{(3)33} /\Lambda^2$ constrained to be $\in [-2.66, 0.34]\,\mathrm{TeV}^{-2}$ at $68\%$ CL~\cite{Brivio:2019ius}, giving $-0.16 < (\delta g_L^{Wq})_{33} < 0.02$. This leaves relatively little room for new physics in $(\delta g_L^{Zu})_{33}$. On the contrary, $(\delta g_R^{Zu})_{33}$ is probed in $t\bar{t}Z$ production with limited accuracy~\cite{Brivio:2019ius}.

\item The gluon dipole coefficient $\mathrm{Re}\,[d_{Gu}]_{33}$ is rather strongly constrained by $t\bar{t}$ production~\cite{Brivio:2019ius}. The coefficient $[d_{Gu}]_{33}$ is in one-to-one correspondence with the Warsaw basis operator $Q_{uG}^{33}\,$, and $\mathrm{Re}\,[d_{Gu}]_{33} = - 2 \sqrt{2} \frac{v^3}{g_s m_t } \frac{C_{tG}}{\Lambda^2}$ where $C_{tG} /\Lambda^2$ was found to be $\in [0.30, 0.74]\,\mathrm{TeV}^{-2}$ at $68\%$ CL~\cite{Brivio:2019ius}.
\end{itemize}
The above discussion suggests it would be sensible to prioritize a certain subset of operators, where the potential sensitivity is most relevant, when presenting experimental results in off-shell Higgs production channels. First steps toward this goal are taken in Chapter~\ref{ch:ofsSMEFT}.

\vspace{-2mm}
\section{Brief review of models testable in off-shell production}
\vspace{-2mm}
To showcase the insight on ultraviolet physics that can be gained from off-shell measurements, here we review briefly the existing analyses of $gg\to ZZ$ sensitivity to {\it explicit models}. We note that many more off-shell studies have been written, but they are not mentioned here if their discussion was framed in EFT.
\begin{itemize}
\item Reference~\cite{Englert:2014ffa} considered very light and strongly-mixed stops in supersymmetry, whose impact comes primarily from Higgs-mediated $gg\to h^\ast \to ZZ$ diagrams. In addition, the role of a new scalar $\phi$ mixed with the 125~GeV Higgs was discussed, showing that its effect is limited to the resonant region $m(ZZ) \sim m_\phi\,$, whereas for $\sqrt{\hat{s}} \gg m_\phi$ the $h$-mediated and $\phi\,$-mediated amplitudes sum to give a SM-like result, as a consequence of perturbative unitarity. Similar considerations were offered in~\cite{Logan:2014ppa} for scalars belonging to larger electroweak representations, as appear e.g. in the GM model.

\item In~\cite{Azatov:2016xik} the impact of heavy vector-like quarks on $gg\to ZZ$ was analyzed, highlighting the important role of corrections to the $Zt\bar{t}$ couplings and therefore to the continuum amplitude (box diagrams).\footnote{See also~\cite{Cao:2020npb} for an interesting discussion of the role of polarization in probing the $Zt\bar{t}$ couplings.} First, a simplified model with one $SU(2)_L$-$\,$singlet, $Y = 2/3$ vector-like quark was discussed. Then, a realistic Composite Higgs setup with several top partner multiplets was introduced and the impact of $gg\to ZZ$ measurements on the parameter space was assessed. Vector-like fermions charged under color but carrying lepton-like electroweak quantum numbers were discussed in~\cite{Englert:2014aca}.

\item References~\cite{Goncalves:2017iub,Englert:2020gcp} considered extending the SM by a (complex or real) $Z_2$-symmetric scalar $S$, coupled to the Higgs as, e.g., $\mathcal{L} \supset - \lambda_S |H|^2 |S|^2$. The new scalar contributes to $gg\to ZZ$ via two-loop diagrams. For scalar mass $m_S > m_h/2$ this is a ``nightmare scenario'' for BSM physics, notoriously difficult to probe~\cite{Craig:2014lda,Ruhdorfer:2019utl} yet well motivated, for example by obtaining a first order EW phase transition or by neutral naturalness models with SM-singlet scalar top partners. Very recently, Ref.~\cite{Haisch:2022rkm} showed that application of a matrix-element based kinematic discriminant can strongly increase the sensitivity at the HL-LHC.

\item In~\cite{Goncalves:2018pkt} additional new physics that can appreciably impact off-shell measurements was studied, including two flat extra dimensions modifying the running of $y_t$ and a form factor for the $t\bar{t}h$ interaction aiming to approximate effects of Higgs and top compositeness.\footnote{In~\cite{Goncalves:2018pkt} the Quantum Critical Higgs model~\cite{Bellazzini:2015cgj} was also found to have an important impact on $gg\to ZZ$, but this was later shown~\cite{ShayeganShirazi:2019shw} to originate from a mistake in the treatment of the $hZZ$ form factor. We thank John Terning for a communication on this issue.} The same form factor was considered in a recent analysis of BSM sensitivity in $gg\to ZZ\to \ell^+ \ell^- \nu \bar{\nu}$~\cite{Goncalves:2020vyn}.
\end{itemize}


\chapter[Off-shell Higgs production in the SMEFT]{Off-shell Higgs production in the SMEFT\footnote{contributed by M.\ Thomas and E.\ Vryonidou (Section~\ref{sec:smeftatnlo}), A.V.\ Gritsan, L.\ Kang, and U.\ Sarica (Section~\ref{sec:jhugen})}
\label{ch:ofsSMEFT}}

\section{Studies with the SMEFTatNLO framework}
\label{sec:smeftatnlo}

Off-shell Higgs production is an interesting process in SMEFT, as it is modified by both Higgs and top interactions. As an example, the impact of this process in breaking the degeneracy between the top Yukawa and Higgs-gluon interactions has been discussed in the literature \cite{Azatov:2014jga,Azatov:2016xik}. The interesting interplay of top and Higgs interactions in this process extends also to the EW couplings of the top, for which this process can provide complementary information to the typical probes of top-pair and single-top and $Z$ associated production.

\subsection{Relevant Operators}

\begin{figure}
 \center 
 \includegraphics[scale=0.8]{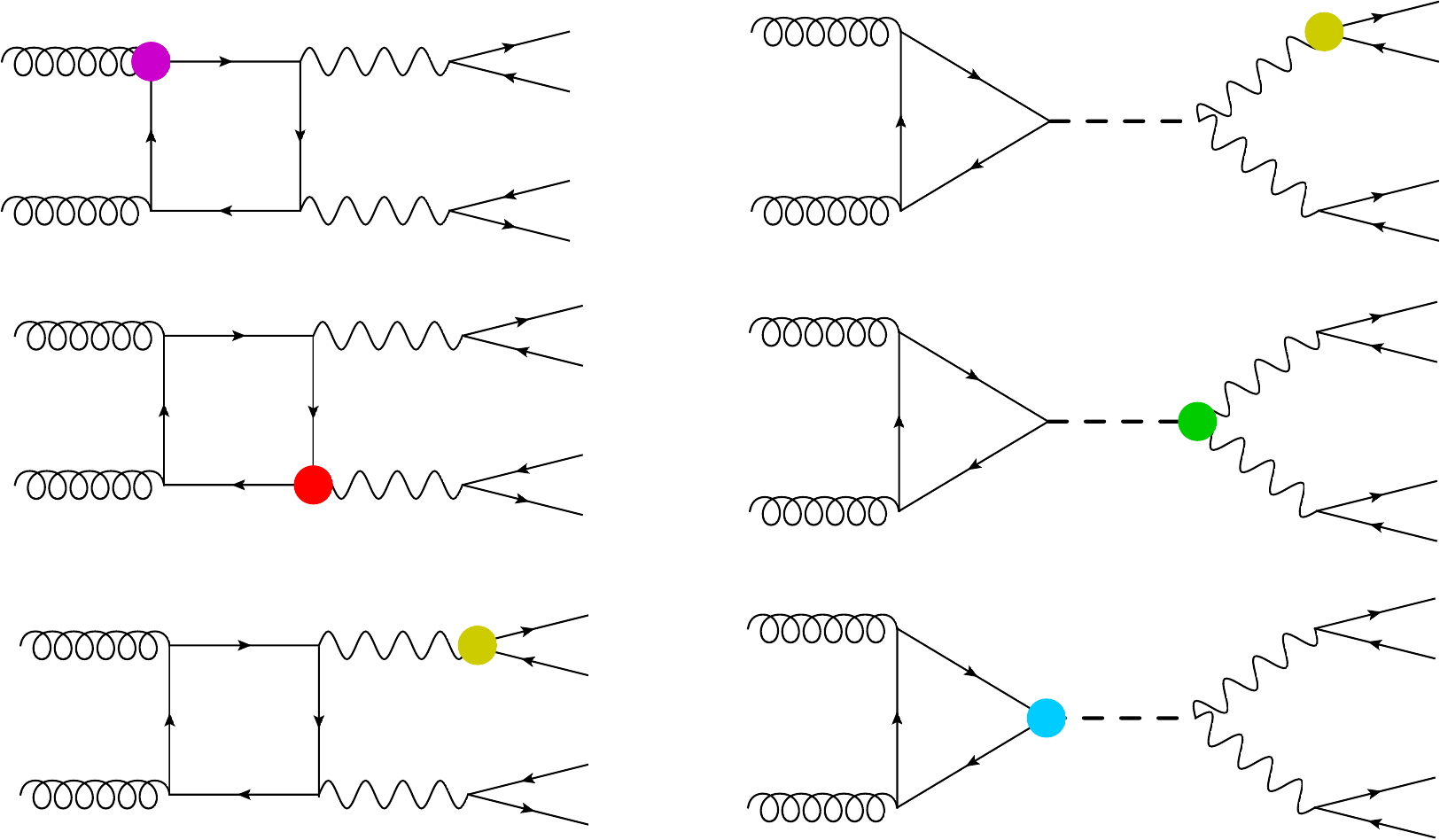}
 \caption{\label{fig:FeynmanDiagrams}
Example Feynman diagrams for off-shell Higgs production and corresponding background with EFT insertions.}
\end{figure}

Typical Feynman diagrams for the production of 4 leptons in gluon fusion, i.e. off-shell Higgs production and the gluon fusion background are shown in Fig.~\ref{fig:FeynmanDiagrams}. Possible EFT insertions are denoted by coloured blobs. Whilst not shown in the diagrams, operators will also modify the Higgs width and gauge boson widths. 

As a starting point, the table below presents the list of $CP$-even gauge and 2-fermion SMEFT operators from the Warsaw basis~\cite{Grzadkowski:2010es} consistent with a $U(2)^5$ flavor symmetry in the fermion sector entering this process. Coefficient names in the model are given in the \verb|UFO| column as implemented in the SMEFT@NLO model \cite{Degrande:2020evl}. See Refs.~\cite{AguilarSaavedra:2018nen} and \cite{Degrande:2020evl} for more details on conventions and the flavor symmetry implementation. 

The list includes gauge operators which modify the Higgs boson couplings, such as the operators $\mathcal{O}_{\varphi W}$ and $\mathcal{O}_{\varphi B}$
which modify the interaction between Higgs bosons and electroweak gauge bosons. Similarly the $\mathcal{O}_{\varphi G}$ operator introduces a direct coupling between the Higgs boson and gluons.The $O_{\varphi d}$ operator generates a wavefunction correction to the
Higgs boson, which rescales all the Higgs boson couplings in a universal manner. The operators $O_{\varphi WB}$ and $O_{\varphi D}$ are the ones
often identified as the $S$ and $T$ oblique parameters.

The 2-fermion operators involve both top and light-quark operators. Light quark degrees of freedom such as  $C_{\varphi q_i}^{(3)}$ and $C_{\varphi q Mi}$ enter in the gluon fusion background by modifying the interactions of the light quarks with the gauge bosons. The corresponding current operators for the leptons enter in the leptonic decay of the vector bosons. Both light quark and lepton operators contribute to EWPO and are therefore expected to be well constrained enough not to be considered in this process.

On the other hand top quark operators are more interesting for this process. These include the chromo-magnetic dipole operator $O_{tG}$, the dimension-six Yukawa operator $O_{t\varphi}$ as well as the electroweak-dipole operators, $O_{tW}$ and $O_{tB}$ and the
current operators,  $O_{\varphi Q}^{(3)}$ and $O_{\varphi t}$.  $O_{tG}$ enters both the signal and background, whilst $O_{t\varphi}$ enters the signal. The weak dipoles and current operators modify only the $gg\to VV$ background. 

In this note we aim to provide instructions on how to produce results for the off-shell Higgs production process 
and its gluon fusion background using the  SMEFT@NLO package and will present some representative results. We hope this will motivate further and detailed studies of this process within the SMEFT. 

\renewcommand{\arraystretch}{1.6}
\begin{center}
\begin{table}    
\begin{tabular}{|p{0.7cm}p{1cm}p{5.8cm}|p{0.7cm}p{1cm}p{5.8cm}|}
      \hline
      \multicolumn{3}{|l}{\emph{Bosonic }}&
      \multicolumn{3}{l|}{} 
      \tabularnewline\hline
      $\mathcal{O}_i$&\code{UFO}& Definition
      &
      $\mathcal{O}_i$&\code{UFO}& Definition
      \tabularnewline\hline
     $\Op{\phi G}$&\code{cpG}&
     $\left(\pdp-\tfrac{v^2}{2}\right)G^{\mu\nu}_{\sss A}\,
                                        G_{\mu\nu}^{\sss A}$
     &
     $\Op{\phi W}$&\code{cpW}&
     $\left(\pdp-\tfrac{v^2}{2}\right)W^{\mu\nu}_{\sss I}\,
                                    W_{\mu\nu}^{\sss I}$
     \tabularnewline
     $\Op{\phi B}$&\code{cpBB}&
     $\left(\pdp-\tfrac{v^2}{2}\right)B^{\mu\nu}\,B_{\mu\nu}$
     &
     $\Op{\phi WB}$&\code{cpWB}&
     $(\phi^\dagger \tau_{\sss I}\phi)\,B^{\mu\nu}W_{\mu\nu}^{\sss I}\,$
     \tabularnewline
     \hline
    
     $\Op{\phi D}$&\code{cpDC}&
     $(\phi^\dagger D^\mu\phi)^\dagger(\phi^\dagger D_\mu\phi)$&
     $\Op{\phi d}$&\code{cdp}&
     $\partial_\mu(\pdp)\partial^\mu(\pdp)$
    \tabularnewline
     \hline
  \end{tabular}\\[3ex]
\begin{tabular}{|p{0.7cm}p{1cm}p{5.8cm}|p{0.7cm}p{1cm}p{5.8cm}|}
    \hline
    \multicolumn{3}{|l}{\emph{2 fermion (chiral flip)}}
    &
    \multicolumn{3}{l|}{}
    \tabularnewline
    \hline
    $\mathcal{O}_i$&\code{UFO}& Definition
    &
    $\mathcal{O}_i$&\code{UFO}& Definition
    \tabularnewline\hline
    $\Op{t\phi}$&\code{ctp}&
    $\left(\pdp-\tfrac{v^2}{2}\right)
    \bar{Q}\,t\,\tilde{\phi} + \text{h.c.}$
    &
    $\Op{tW}$&\code{-}&
    $i\big(\bar{Q}\tau^{\mu\nu}\,\tau_{\sss I}\,t\big)\,
    \tilde{\phi}\,W^I_{\mu\nu}
    + \text{h.c.}$
    \tabularnewline
    $\Op{tG}$&\code{ctG}&
    $ig{\sss S}\,\big(\bar{Q}\tau^{\mu\nu}\,T_{\sss A}\,t\big)\,
    \tilde{\phi}\,G^A_{\mu\nu}
    + \text{h.c.}$
    &
    $\Op{tB}$&\code{-}&
    $i\big(\bar{Q}\tau^{\mu\nu}\,t\big)
    \,\tilde{\phi}\,B_{\mu\nu}
    + \text{h.c.}$
    \tabularnewline
    &&&
    \code{-}&\code{ctW}&
    $\Cp{tW}$
    \tabularnewline
    &&&
    \code{-}&\code{ctZ}&
    $-\sin{\theta_{\sss  W}}\Cp{tB}+\cos{ \theta_{\sss  W}}\Cp{tW}$
    \tabularnewline
    \hline
    \end{tabular}
    \\[3ex]
\begin{tabular}{|p{0.7cm}p{1cm}p{5.8cm}|p{0.7cm}p{1cm}p{5.8cm}|}
    \hline
    \multicolumn{3}{|l}{\emph{2 fermion (current)}}
    &
    \multicolumn{3}{l|}{} 
    \tabularnewline
    \hline
    $\mathcal{O}_i$&\code{UFO}& Definition
    &
    $\mathcal{O}_i$&\code{UFO}& Definition
    \tabularnewline\hline
    $\Opp{\phi l_1}{(1)}$&\code{cpl1}&
    $i\big(\phi^\dagger\lra{D}_\mu\,\phi\big)
      \big(\bar{l}_1\,\gamma^\mu\,l_1\big)$
    &
    $\Opp{\phi l_1}{\sss(3)}$&\code{c3pl1}&
    $i\big(\phi^\dagger\lra{D}_\mu\,\tau_{\sss I}\phi\big)
    \big(\bar{l}_1\,\gamma^\mu\,\tau^{\sss I}l_1\big)$
    \tabularnewline
    $\Opp{\phi l_2}{(1)}$&\code{cpl2}&
    $i\big(\phi^\dagger\lra{D}_\mu\,\phi\big)
      \big(\bar{l}_2\,\gamma^\mu\,l_2\big)$
    &
    $\Opp{\phi l_2}{\sss(3)}$&\code{c3pl2}&
    $i\big(\phi^\dagger\lra{D}_\mu\,\tau_{\sss I}\phi\big)
    \big(\bar{l}_2\,\gamma^\mu\,\tau^{\sss I}l_1\big)$
    \tabularnewline
    $\Opp{\phi l_3}{(1)}$&\code{cpl3}&
    $i\big(\phi^\dagger\lra{D}_\mu\,\phi\big)
      \big(\bar{l}_3\,\gamma^\mu\,l_3\big)$
    &
    $\Opp{\phi l_3}{\sss(3)}$&\code{c3pl3}&
    $i\big(\phi^\dagger\lra{D}_\mu\,\tau_{\sss I}\phi\big)
    \big(\bar{l}_3\,\gamma^\mu\,\tau^{\sss I}l_3\big)$
      \tabularnewline
    \hline
    $\Opp{\phi q_i}{\sss(1)}$&\code{-}&
    $\sum\limits_{\sss i=1,2} i\big(\phi^\dagger\lra{D}_\mu\,\phi\big)
    \big(\bar{q}_i\,\gamma^\mu\,q_i\big)$
    &
    $\Opp{\phi Q}{\sss(1)}$&\code{-}&
    $i\big(\phi^\dagger\lra{D}_\mu\,\phi\big)
    \big(\bar{Q}\,\gamma^\mu\,Q\big)$
    \tabularnewline
    $\Opp{\phi q_i}{\sss(3)}$&\code{-}&
    $\sum\limits_{\sss i=1,2} i\big(\phi^\dagger\lra{D}_\mu\,\tau_{\sss I}\phi\big)
    \big(\bar{q}_i\,\gamma^\mu\,\tau^{\sss I}q_i\big)$
    &
    $\Opp{\phi Q}{\sss(3)}$&\code{-}&
    $i\big(\phi^\dagger\lra{D}_\mu\,\tau_{\sss I}\phi\big)
    \big(\bar{Q}\,\gamma^\mu\,\tau^{\sss I}Q\big)$
    \tabularnewline
    \code{-}&\code{cpq3i}&
    $\Cpp{\phi q_i}{\sss(3)}$
    &
    \code{-}&\code{cpQ3}&
    $\Cpp{\phi Q}{\sss(3)}$
    \tabularnewline
    \code{-}&\code{cpqMi}&
    $\Cpp{\phi q_i}{\sss(1)}-\Cpp{\phi q_i}{\sss(3)}$
    &
    \code{-}&\code{cpQM}&
    $\Cpp{\phi Q}{\sss(1)}-\Cpp{\phi Q}{\sss(3)}$
    \tabularnewline
    \hline
    $\Op{\phi u_i}$&\code{cpu}&
    $\sum\limits_{\sss i=1,2} i\big(\phi^\dagger\,\lra{D}_\mu\,\,\phi\big)
    \big(\bar{u}_i\,\gamma^\mu\,u_i\big)$
    &
    $\Op{\phi d_i}$&\code{cpd}&
    $\sum\limits_{\sss i=1,2(,3)} i\big(\phi^\dagger\,\lra{D}_\mu\,\,\phi\big)
    \big(\bar{d}_i\,\gamma^\mu\,d_i\big)$
    \tabularnewline
    $\Op{\phi t}$&\code{cpt}&
    $i\big(\phi^\dagger\,\lra{D}_\mu\,\,\phi\big)
    \big(\bar{t}\,\gamma^\mu\,t\big)$
    & \multicolumn{3}{l|}{}
     \tabularnewline
    \hline
  \end{tabular}
  \\[3ex]\caption{$CP$-even gauge and 2-fermion operators entering off-shell Higgs production and decay and corresponding gluon fusion background as implemented in the SMEFTatNLO package.}
  \end{table}
\end{center}

\subsection{Generation using SMEFTatNLO}
The SMEFT@NLO implementation of SMEFT operators allows the computation of both signal and background contributions at 1-loop. Taking as an example the $gg\to 4\ell$ process, after importing the model into the MG5\_aMC@NLO code:

\code{import model SMEFTatNLO-NLO}

the following generation commands are needed:
\begin{itemize}
\item  Total contribution

\code{generate g g > e+ e- mu+ mu- NP=2 QCD=2 QED=4 [QCD]}

\item Signal

\code{generate g g > h > e+ e- mu+ mu- NP=2 QCD=2 QED=4 [QCD]}

\item Background

\code{generate g g > e+ e- mu+ mu- /h NP=2 QCD=2 QED=4 [QCD]}

\end{itemize}
The first command (\code{NP=2}) allows to compute all contributions: the SM, the interference of EFT and 
SM ($\mathcal{O}(\Lambda^{-2}$)), and the EFT squared ($\mathcal{O}(\Lambda^{-4}$)) coming from squaring 
amplitudes with single insertions of the operators.  Adding  \code{NP\^{}2==2} to the coupling flags above allows 
the user to obtain only the interference of the EFT and SM contributions, whilst \code{NP\^{}2==4} allows one to 
separately extract the  $\mathcal{O}(\Lambda^{-4})$ contribution. With these commands one can create templates 
which can be combined to extract results for any value of the Wilson coefficients. 

\subsection{Results \label{sec:vryothomresults}}
As a demonstration of how the code works we show in this section results extracted by turning on one coefficient at a time assuming $\Lambda=1$ TeV.  For simplicity we show results for the $gg \to ZZ$ process, keeping the two $Z$'s on-shell.  Results for the $2\to4$ process can also be extracted with the commands shown above.  
\subsubsection{Cross-section results}
As a reference the LO SM prediction for this process extracted with the same setup is: 
\begin{equation*}
    \text{$\sigma_{\rm SM} = 1484 (1)$ fb}
\end{equation*}

\begin{table}[h]
    \centering
    \begin{tabular}{|c|c|c|c|}
    \hline
$\mathcal{O}_i$&\code{UFO}&Squared term (fb)& Interference term (fb)\\[0.5ex] 
\hline\hline
$\mathcal{O}_{\varphi WB}$&\code{cpwb}&$2.797(7)$  &$ 118.9 (3)$ \\

$\mathcal{O}_{\varphi d}$&\code{cdp}&$ 1.273 (3)$ & $0.921 (4)$  \\

$\mathcal{O}_{\varphi W}$&\code{cpw}&$1.162 (3)$ &$16.83 (7) $\\

$\mathcal{O}_{\varphi B}$&\code{cpbb}&$0.1083 (4)$ &$5.17 (1) $ \\

$\mathcal{O}_{\varphi q}^{(3)}$&\code{cpq3i}&$ 23.04 (5)$&$370.0 (7)$  \\

$\mathcal{O}_{\varphi q}^{(-)}$&\code{cpqmi}&$0.1973 (1)$&$34.18 (7)$ 	\\

$\mathcal{O}_{\varphi Q}^{(3)}$&\code{cpq3}&$5.78 (1)$ &$185.1 (2)$ 	\\

$\mathcal{O}_{\varphi Q}^{(-)}$&\code{cpqm}&$1.800 (4)$ &$94.5 (2)$ \\

$\mathcal{O}_{\varphi u}$&\code{cpu}&$0.788 (2)$&$68.07 (4)$ \\

$\mathcal{O}_{\varphi t}$&\code{cpt}&$0.4794 (7)$ &$-1.85 (1) $ \\

$\mathcal{O}_{\varphi d_i}$&\code{cpd}&$0.434 (1)$ &$-50.5 (1)$\\

$\mathcal{O}_{t\varphi}$&\code{ctp}&$0.3245 (6)$&$-0.51(4)$\\

$\mathcal{O}_{tZ}$&\code{ctz}&$0.1546 (3)$ &$-3.53 (1) $ \\

$\mathcal{O}_{tG}$&\code{ctg}&$45.18 (4)$ & $0.47 (6)$ \\

$\mathcal{O}_{\varphi D}$&\code{cpdc}&$0.03983 (3)$ &$8.23 (4)$ \\
\hline
    \end{tabular}
    \caption{Cross-section results for $gg \to ZZ$ for various operator coefficients contributing to the process at 1-loop. Monte Carlo errors on the last digit are shown in the brackets.  The factorisation and renormalisation scale has been set to $M_Z$. \label{Table:xsections}}
\end{table}

In Table \ref{Table:xsections} we show results for the cross-section in fb setting one coefficient to one and setting
 $\Lambda=1$ TeV. These can be rescaled to obtain results for any value of the coefficient. We note that a couple 
 of interference cross-sections (\code{ctG} and \code{ctp})  suffer from large statistical uncertainties. This is due to the fact 
 that the interference changes sign depending on the phase-space region leading to large cancellations. For those 
 contributions a large number of events are required to provide a precise prediction.  We find that for $c=1$ the 
 operators with the largest contributions at the interference level are $\mathcal{O}_{\varphi WB}$, 
 $\mathcal{O}_{\varphi q}^{(3)}$, $\mathcal{O}_{\varphi Q}^{(3)}$ and $\mathcal{O}_{\varphi Q}^{(-)}$. Given that
  all of these operators enter in a lot of other processes one needs to consider these constraints to explore the 
  potential of the off-shell process in providing additional sensitivity. 

Finally it should also be noted that whilst some contributions appear to be suppressed these can become significant if 
the operator is loosely constrained or if the operator can lead to a significant change in the prediction for 
high-energy regions. This motivates examining the behaviour of these contributions at the differential level. 

\subsubsection{Differential distributions}

\begin{figure}
     \centering
     \begin{subfigure}[b]{0.49\textwidth}
         \centering
         \includegraphics[width=\textwidth]{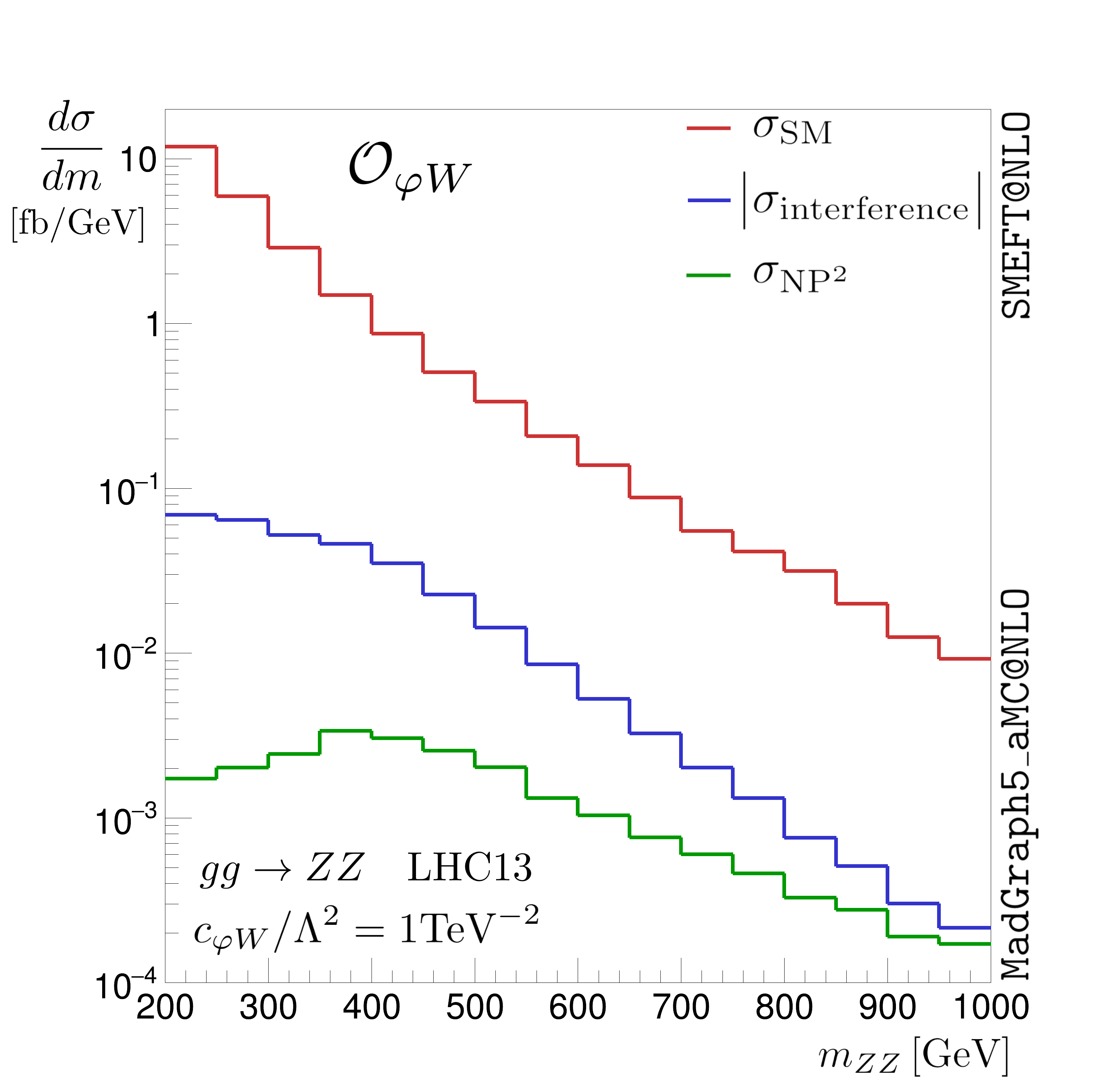}
         \label{fig:cpwb}
     \end{subfigure}
     \hfill
     \begin{subfigure}[b]{0.49\textwidth}
         \centering
         \includegraphics[width=\textwidth]{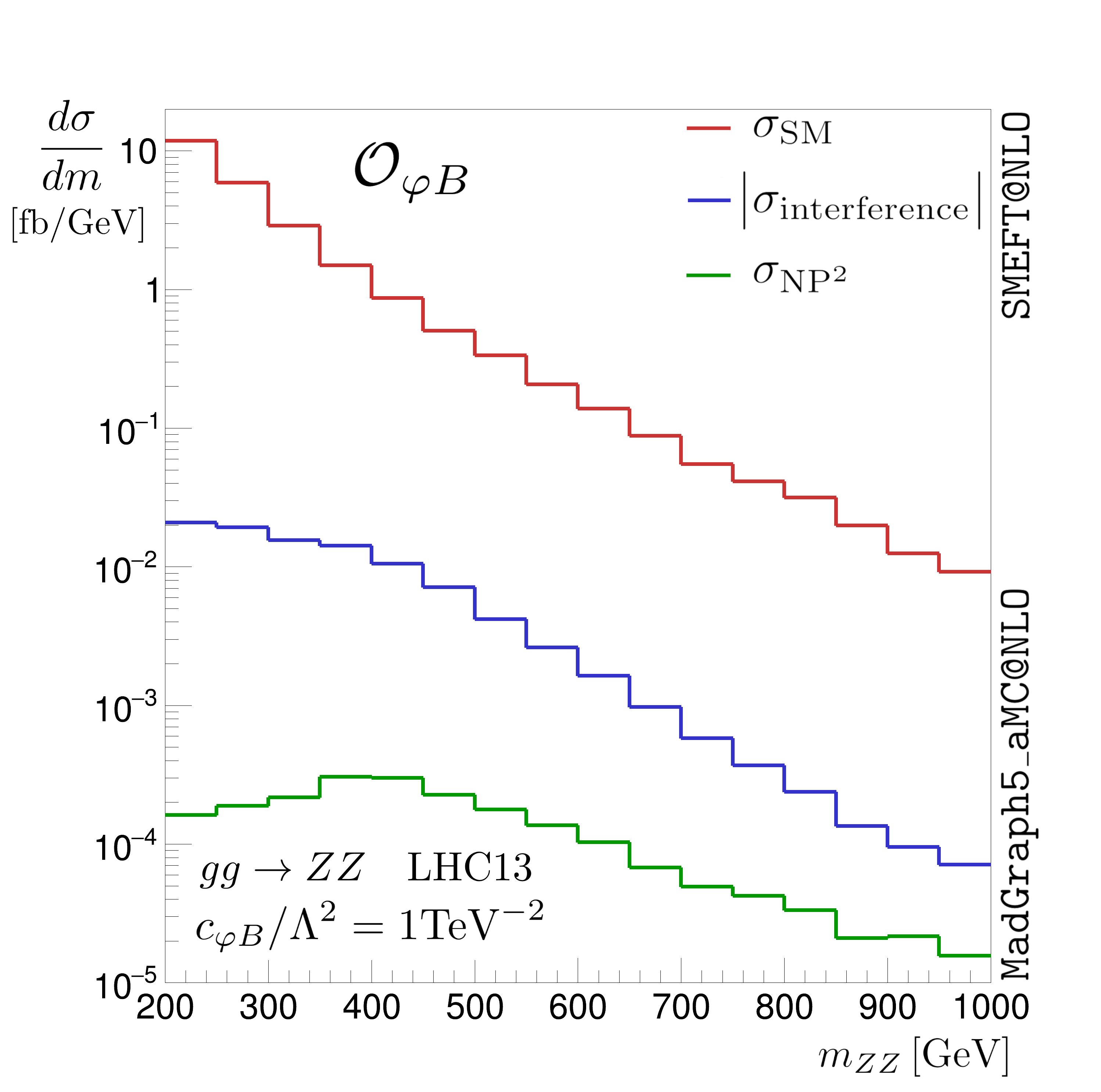}
         \label{fig:cdp}
     \end{subfigure}

             \centering
         \includegraphics[width=0.49\textwidth]{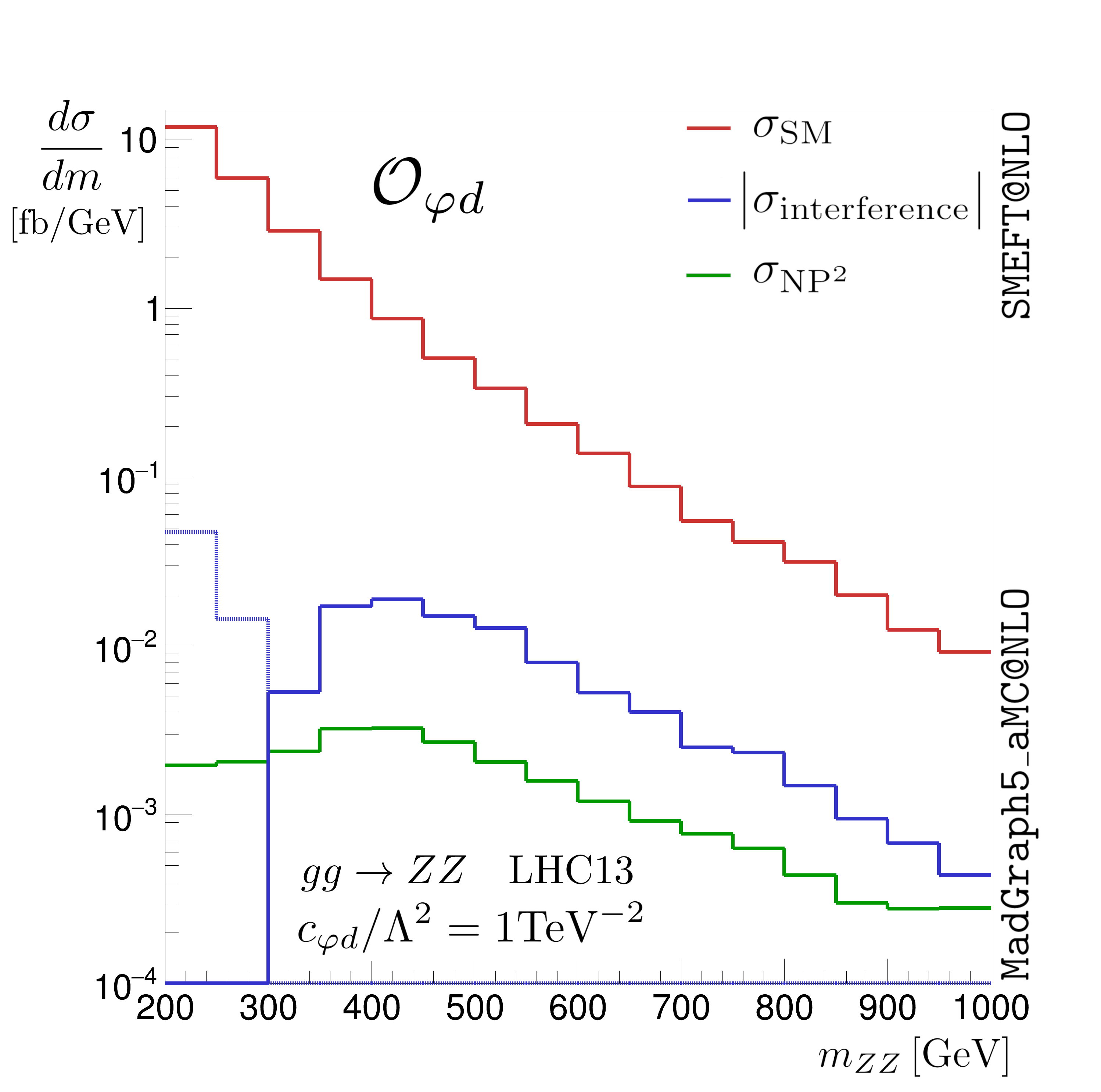}
         \label{fig:cpbb}
        \caption{Differential distributions for Higgs operators, modifying Higgs couplings to the gauge bosons. For the interference, dashed lines denote a negative contribution.         \label{fig:Higgs}}
\end{figure}

In this section we show differential results for a representative sample of the operators discussed above. 
We show the distribution of the
 invariant mass of the $Z$ boson pair. Results are shown separately for the SM, the interference and squared terms. 
 We first show in Fig. \ref{fig:Higgs} results for the gauge operators. We find that their contribution is larger in the tails 
 of the distribution compared to the threshold region. This is particularly true for the squared contributions which are suppressed at low $m(ZZ)$ and enhanced at high $m(ZZ)$. 

\begin{figure}
     \centering
     \begin{subfigure}[b]{0.49\textwidth}
         \centering
         \includegraphics[width=\textwidth]{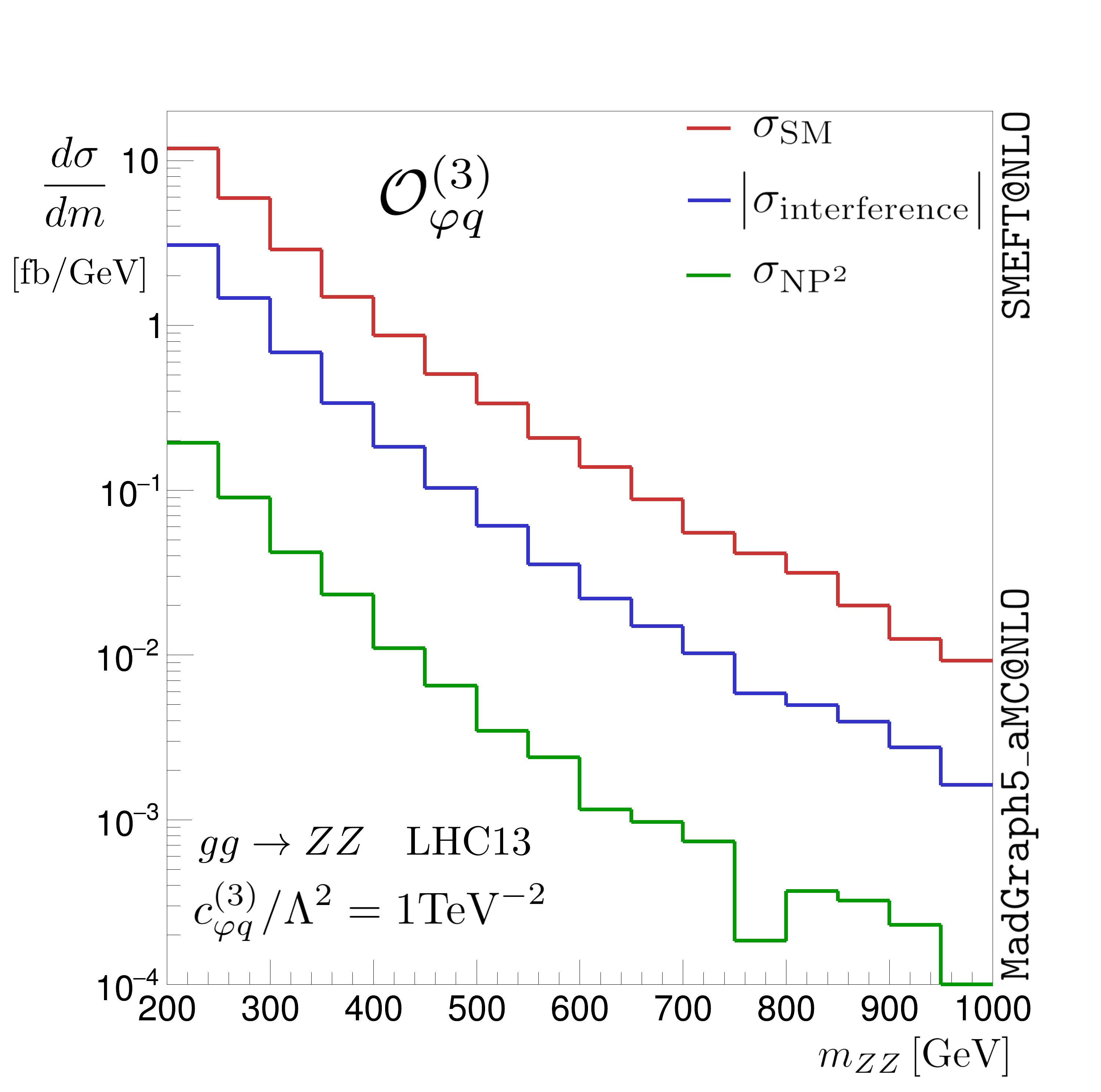}
         \label{fig:cpq3i}
     \end{subfigure}
     \hfill
     \begin{subfigure}[b]{0.49\textwidth}
         \centering
         \includegraphics[width=\textwidth]{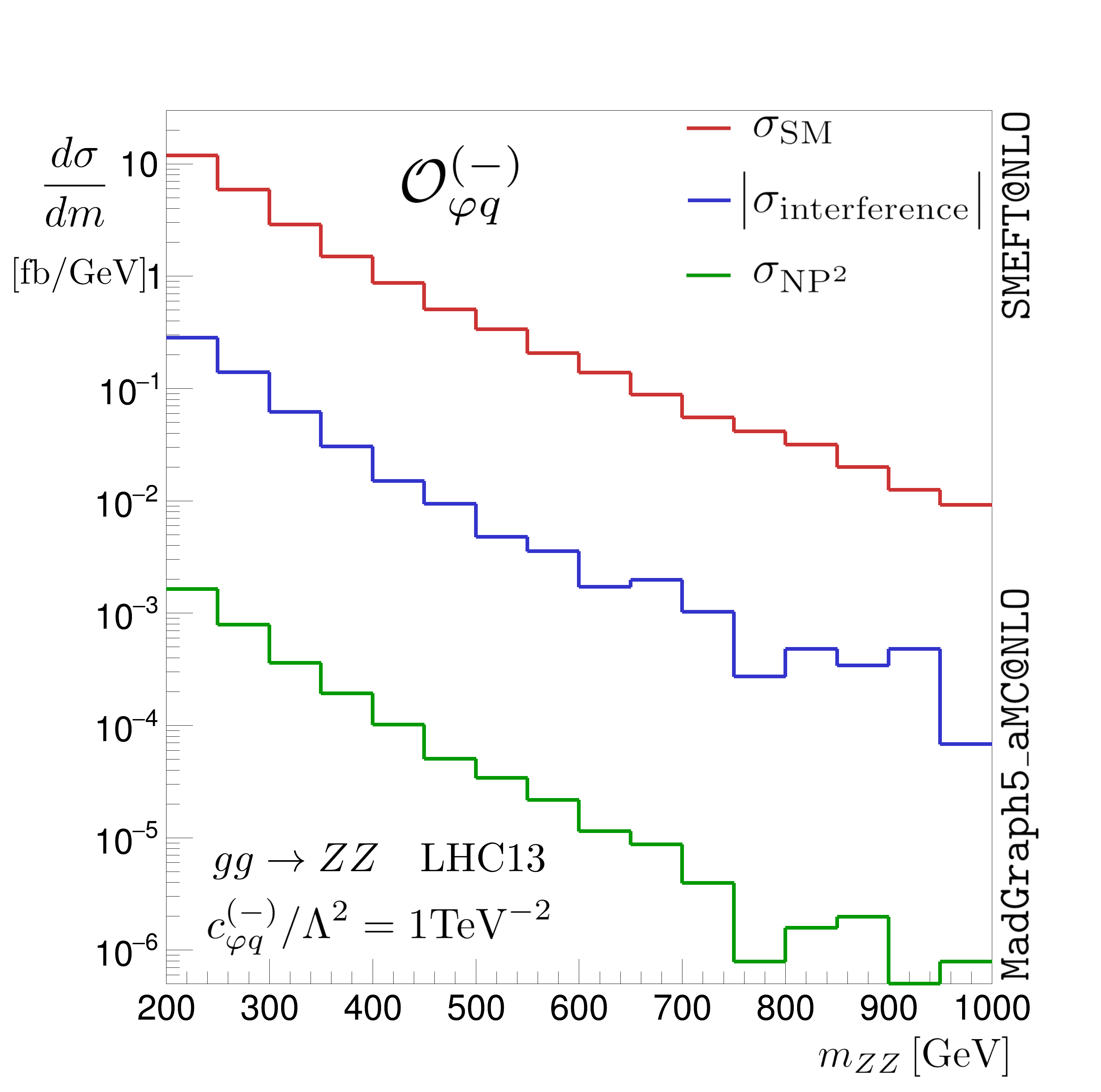}
         \label{fig:cpqmi}
     \end{subfigure}
        \label{fig:fig3}
        
     \centering
     \begin{subfigure}[b]{0.49\textwidth}
         \centering
         \includegraphics[width=\textwidth]{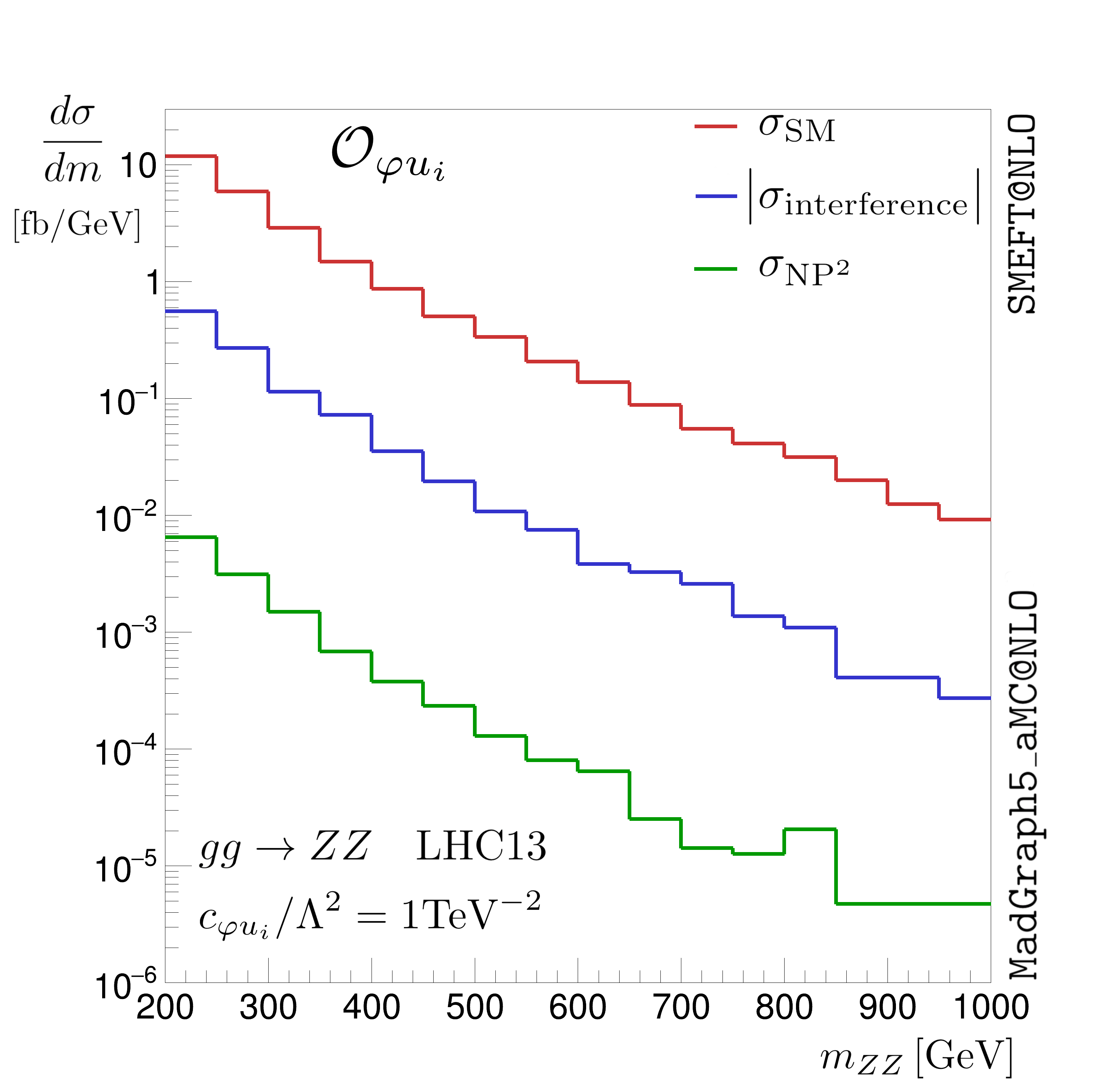}
         \label{fig:cpu}
     \end{subfigure}
     \hfill
     \begin{subfigure}[b]{0.49\textwidth}
         \centering
         \includegraphics[width=\textwidth]{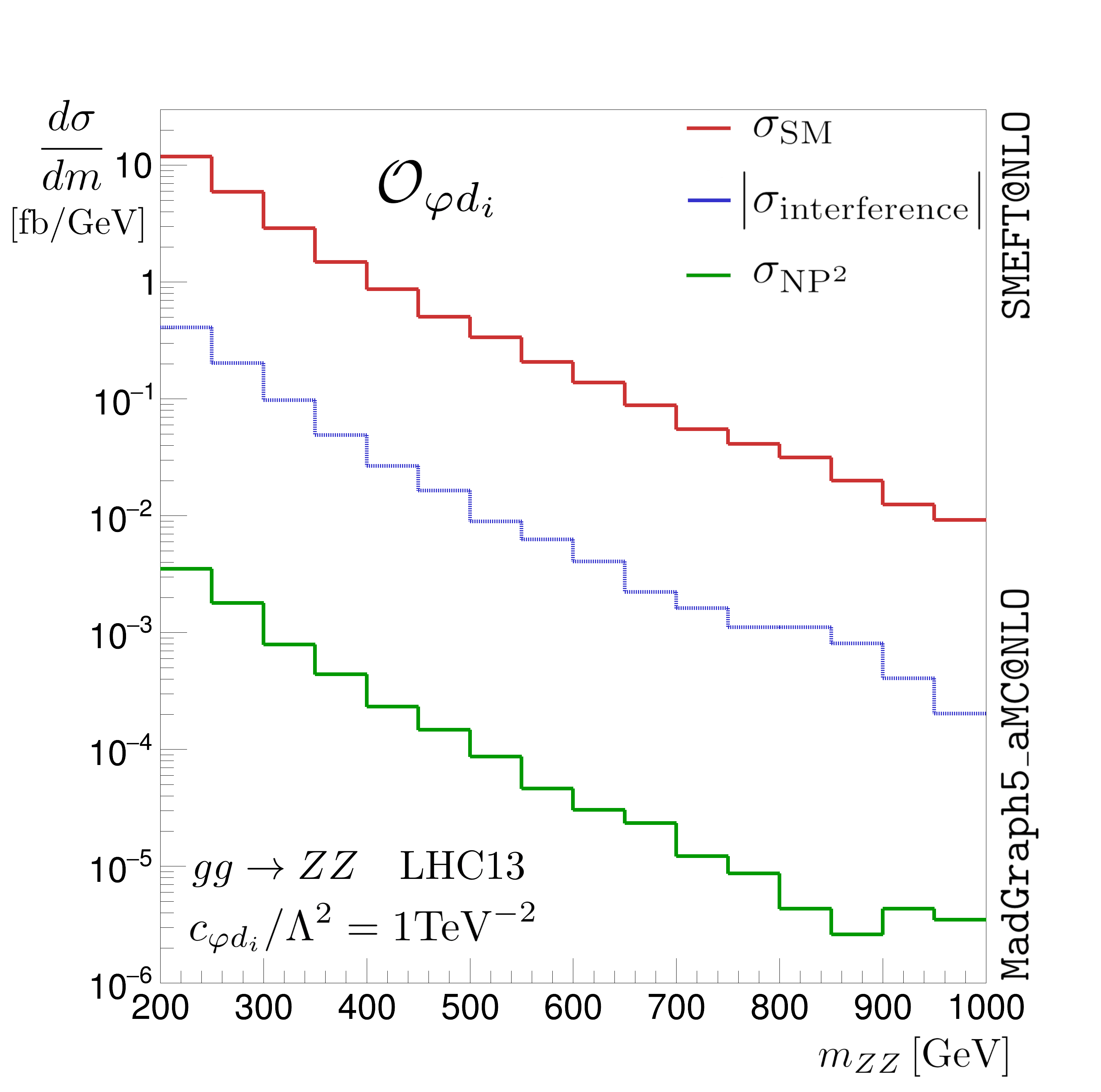}
         \label{fig:cpt}
     \end{subfigure}
 \caption{Differential distributions for light quark operators, modifying light quark couplings to the gauge bosons. For the interference, dashed lines denote a negative contribution.    \label{fig:lightquark}}
\end{figure}

In  Fig.~\ref{fig:lightquark} we show results for the light quark operators. These modify the couplings of up and down 
quarks to the $Z$ boson. These do not show any enhancement in the high-energy tail. The operators receive 
stringent constraints from other processes such as EWPO, Drell-Yan and $q\bar{q}$ initiated diboson production, and therefore are not particularly interesting for the off-shell process. 

\begin{figure}
     \centering
     \begin{subfigure}[b]{0.49\textwidth}
         \centering
         \includegraphics[width=\textwidth]{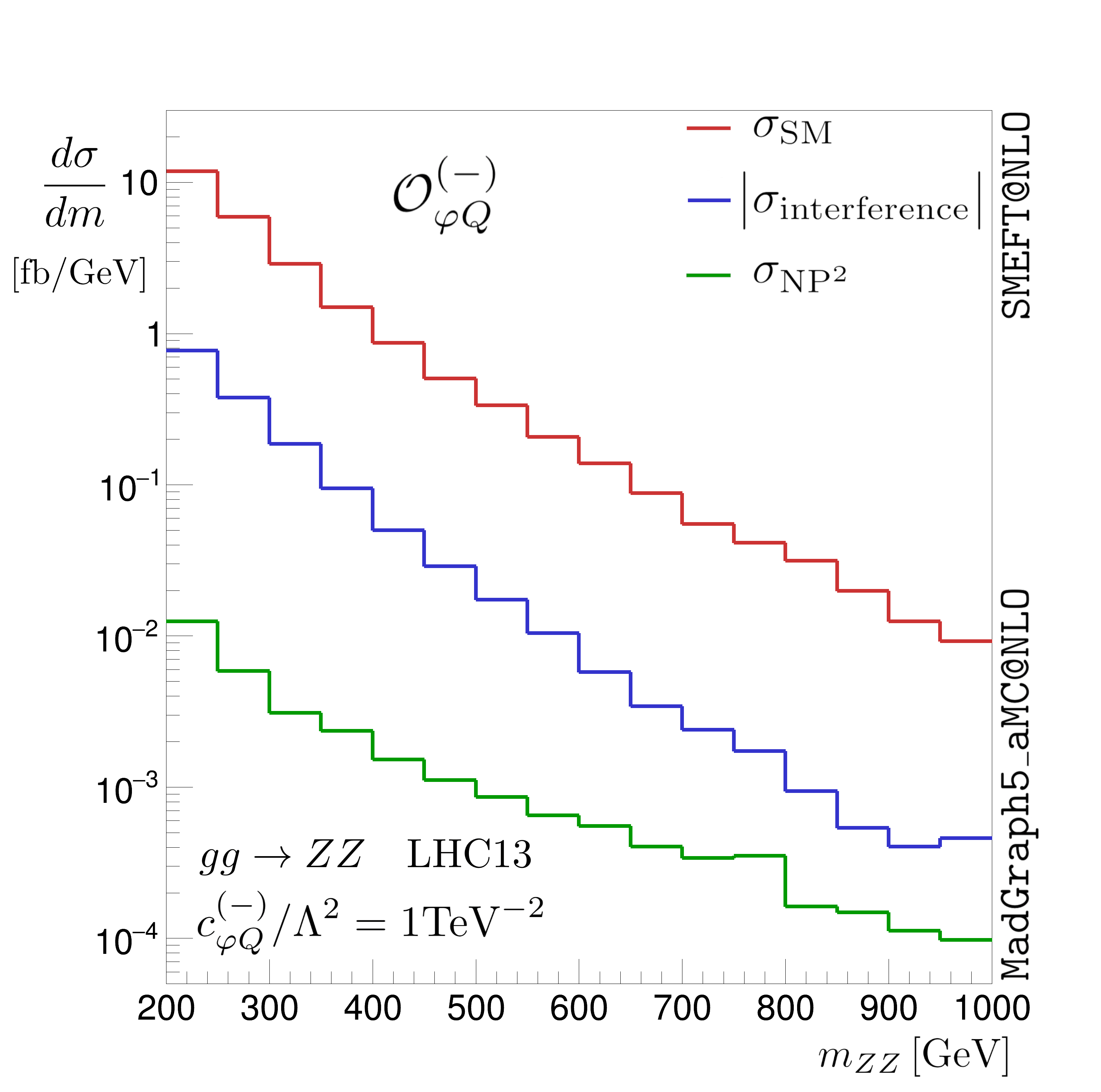}
         \label{fig:cpq3}
     \end{subfigure}
     \hfill
     \begin{subfigure}[b]{0.49\textwidth}
         \centering
         \includegraphics[width=\textwidth]{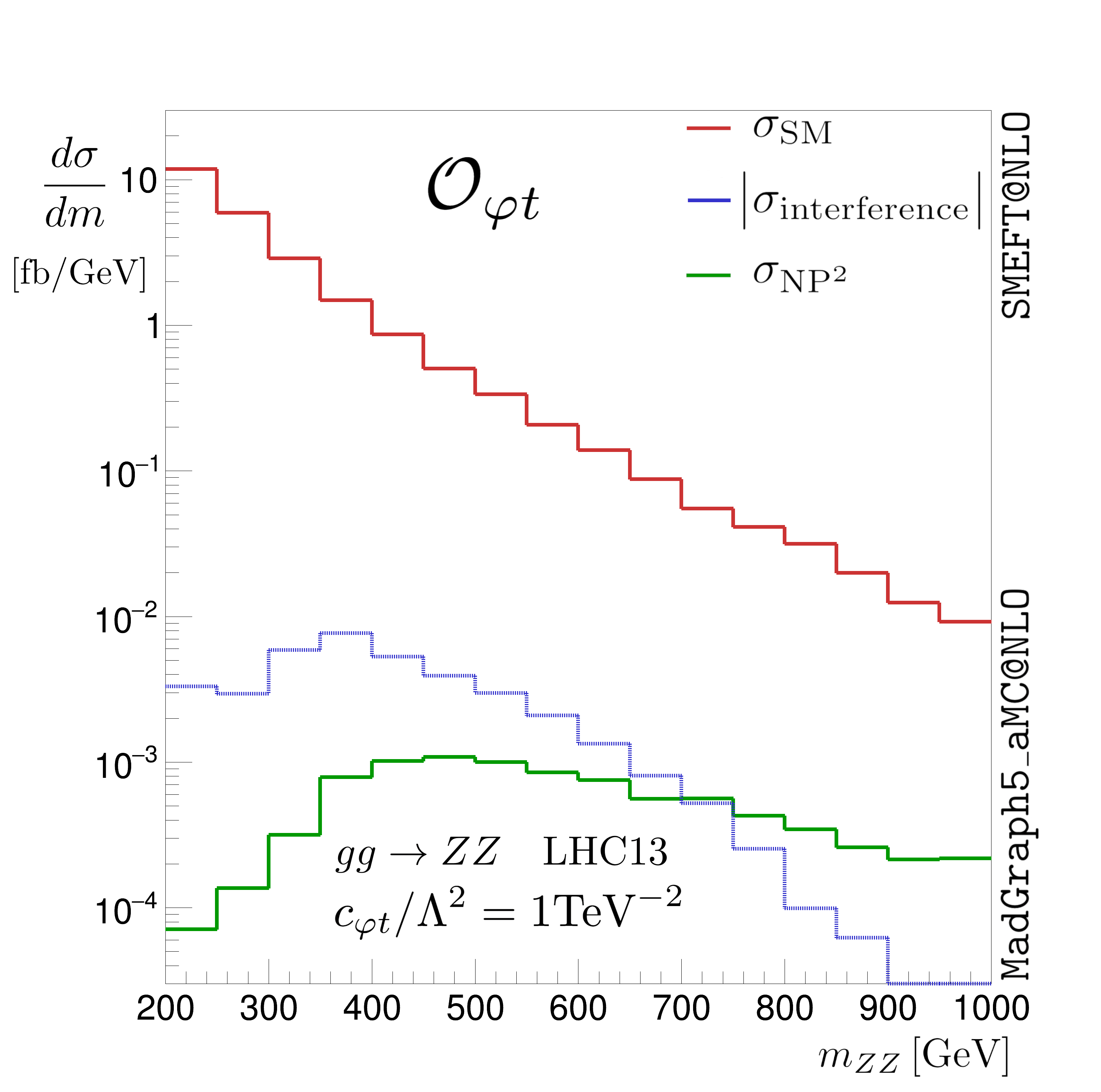}
         \label{fig:cpqm}
     \end{subfigure}
        \label{fig:fig4}
     \centering
             \includegraphics[width=0.49\textwidth]{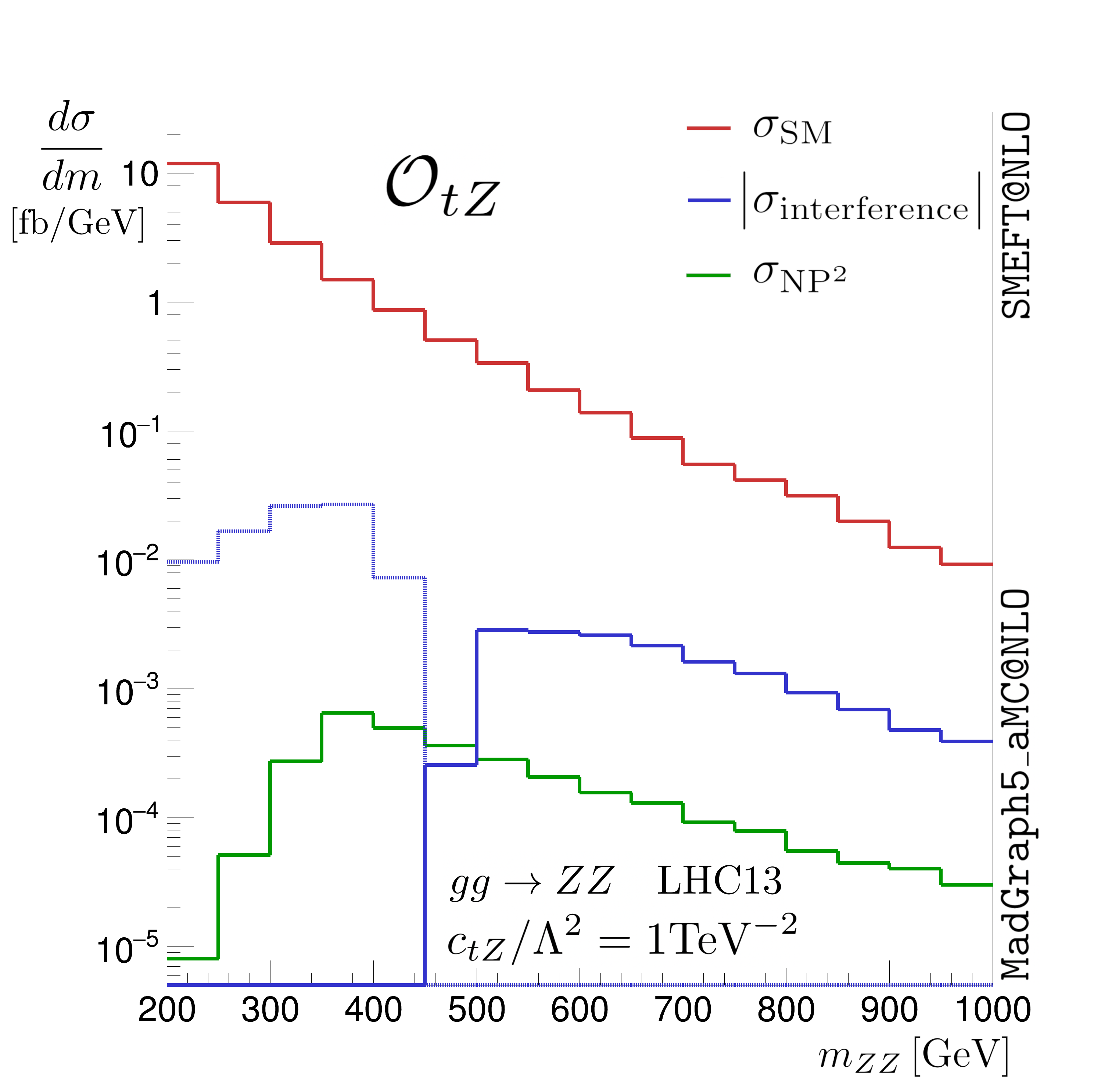}
         \label{fig:ctz}
\caption{Differential distributions for top quark operators, modifying top quark couplings to the $Z$ bosons. For the interference, dashed lines denote a negative contribution.    \label{fig:topquark}}
\end{figure}

In Fig.~\ref{fig:topquark} we present the results for the operators which modify the coupling of the top quark to the Z boson. The operators involving right-handed tops (\code{cpt} and \code{ctZ}) give distributions which are harder than the SM prediction. The typical process to constrain these operators are top production in association with a $Z$ boson, but recent global SMEFT fit results show that these can also be constrained by loop Higgs decays \cite{Ethier:2021bye}.

\begin{figure}
     \centering
     \begin{subfigure}[b]{0.49\textwidth}
         \centering
         \includegraphics[width=\textwidth]{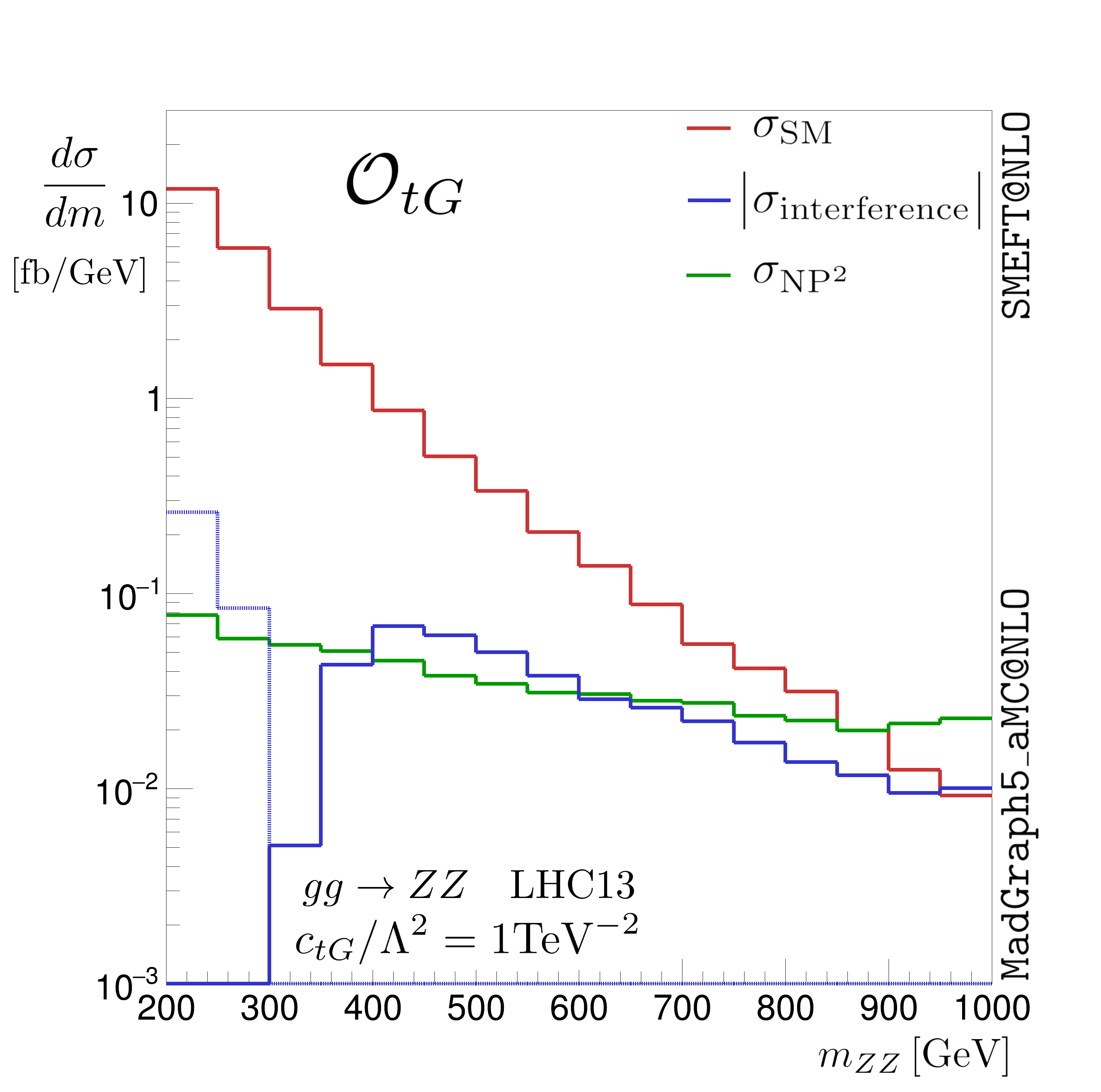}
         \label{fig:cpdc}
     \end{subfigure}
     \hfill
     \begin{subfigure}[b]{0.49\textwidth}
         \centering
         \includegraphics[width=\textwidth]{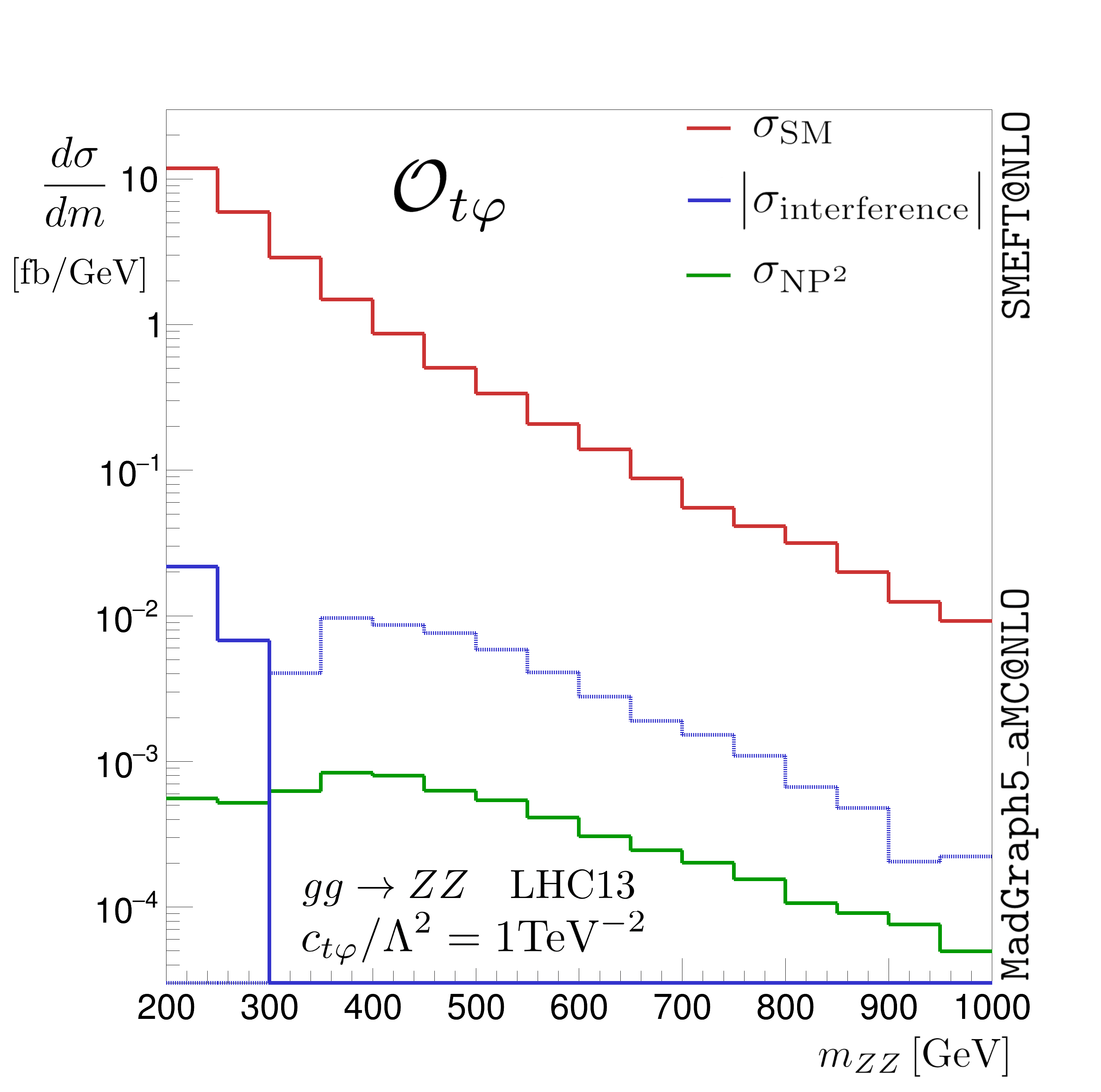}
         \label{fig:ctp}
     \end{subfigure}
        \label{fig:fig7}
        \caption{Differential distributions for top Yukawa and top chromomagnetic dipole moment operators. For the interference, dashed lines denote a negative contribution.    \label{fig:YukChromo}}
\end{figure}

Finally we show in Fig. \ref{fig:YukChromo} the results for the top Yukawa and top chromomagnetic dipole moment operators. We note that the chromomagnetic operator gives a distribution which are significantly harder than the SM one both at the interference and squared level. However as this is receives stringent constraints from top pair production the allowed deviation from the SM prediction is rather small. The top Yukawa operator changes the interaction between the top and Higgs through a rescaling of the top Yukawa coupling. Whilst this operator only rescales the signal, the shape of the $g g \to ZZ $ invariant distribution changes as it comes from a combination of signal, background and their interference.

\subsubsection{Allowed deviations }
\begin{figure}
     \centering
     \begin{subfigure}[b]{0.49\textwidth}
         \centering
         \includegraphics[width=\textwidth]{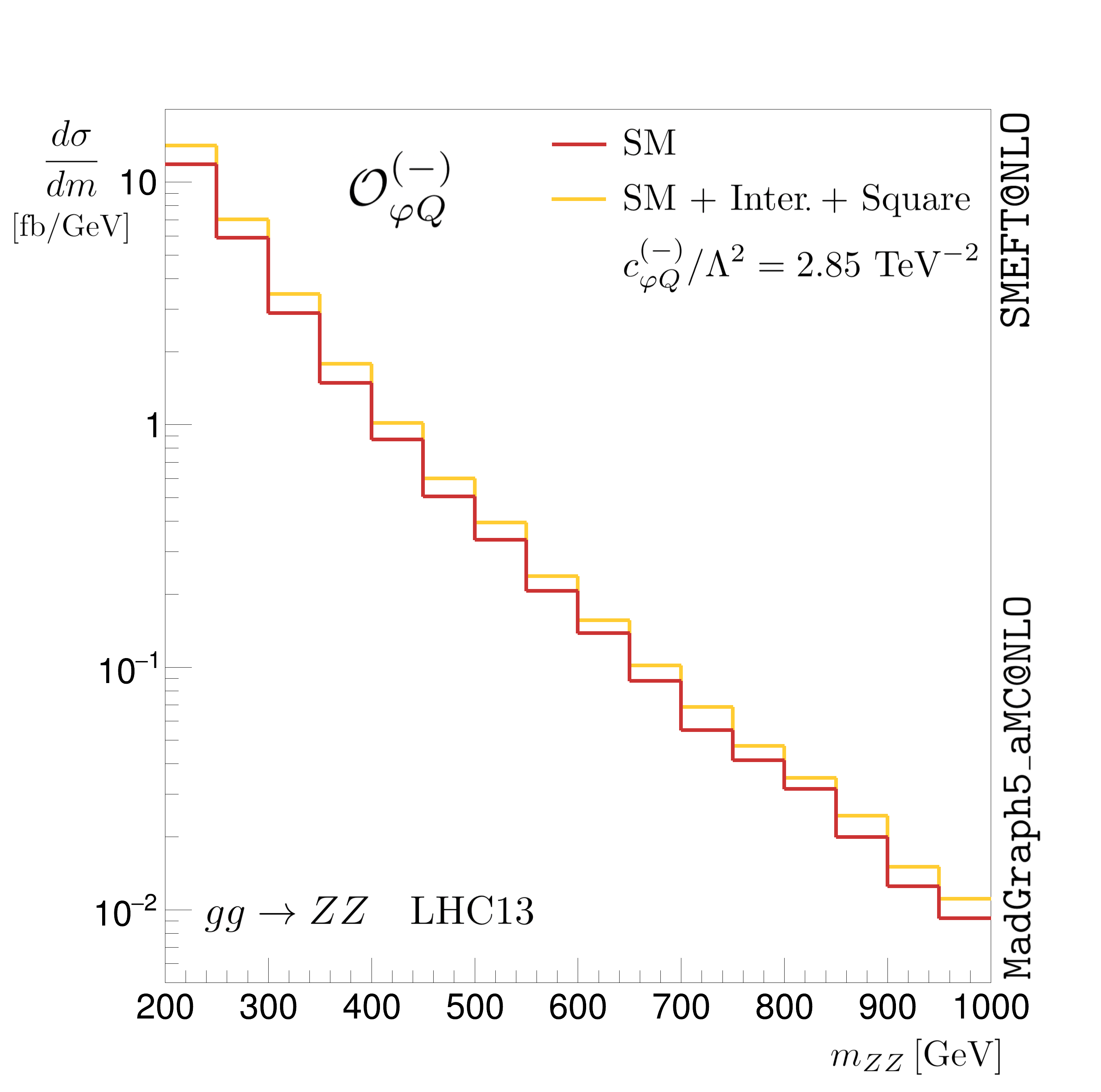}
         \label{fig:cpdc}
     \end{subfigure}
     \hfill
     \begin{subfigure}[b]{0.49\textwidth}
         \centering
         \includegraphics[width=\textwidth]{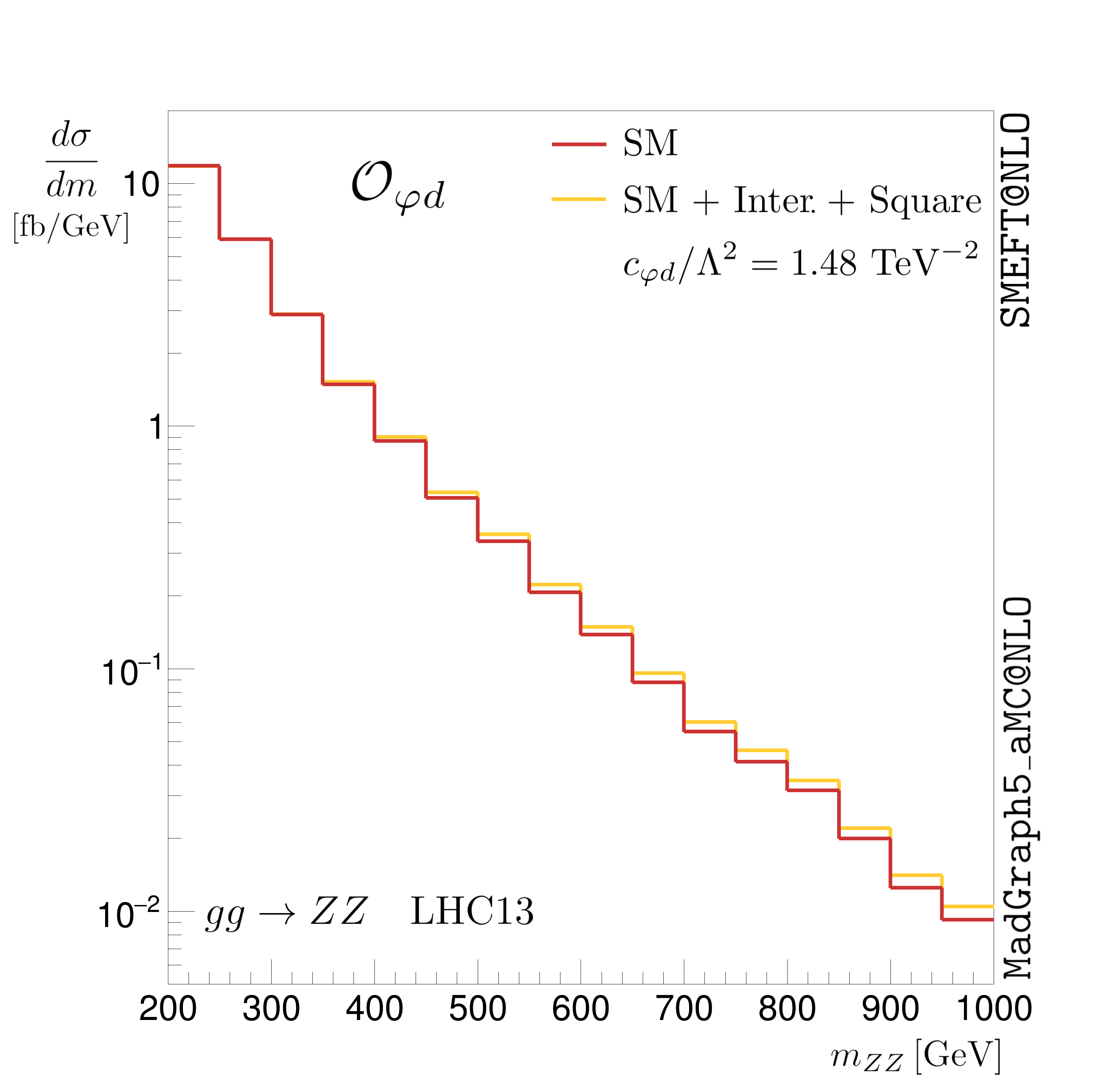}
         \label{fig:ctp}
     \end{subfigure}
        \label{fig:fig7}
     \centering
     \begin{subfigure}[b]{0.49\textwidth}
         \centering
         \includegraphics[width=\textwidth]{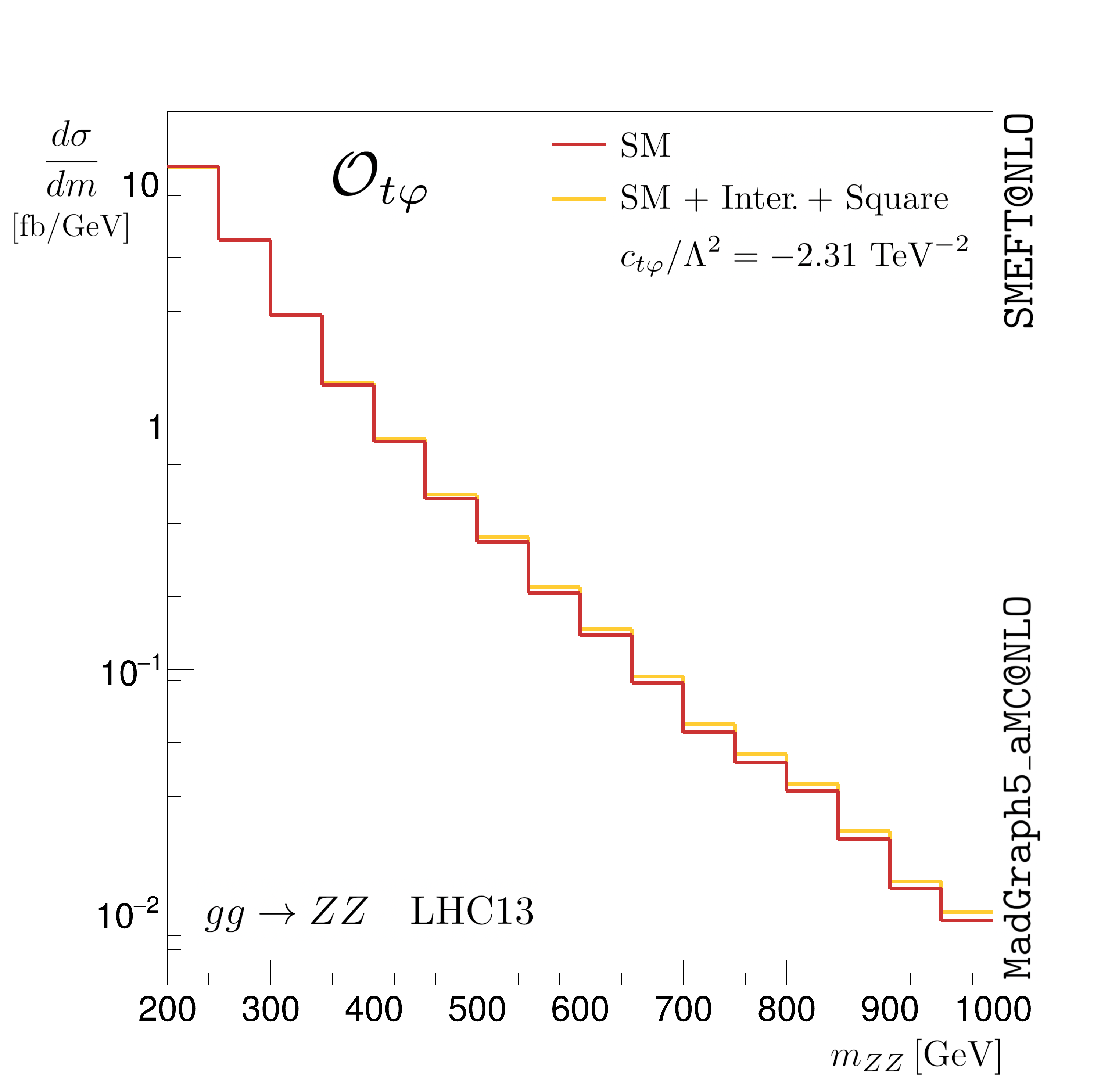}
         \label{fig:cpdc}
     \end{subfigure}
     \hfill
     \begin{subfigure}[b]{0.49\textwidth}
         \centering
         \includegraphics[width=\textwidth]{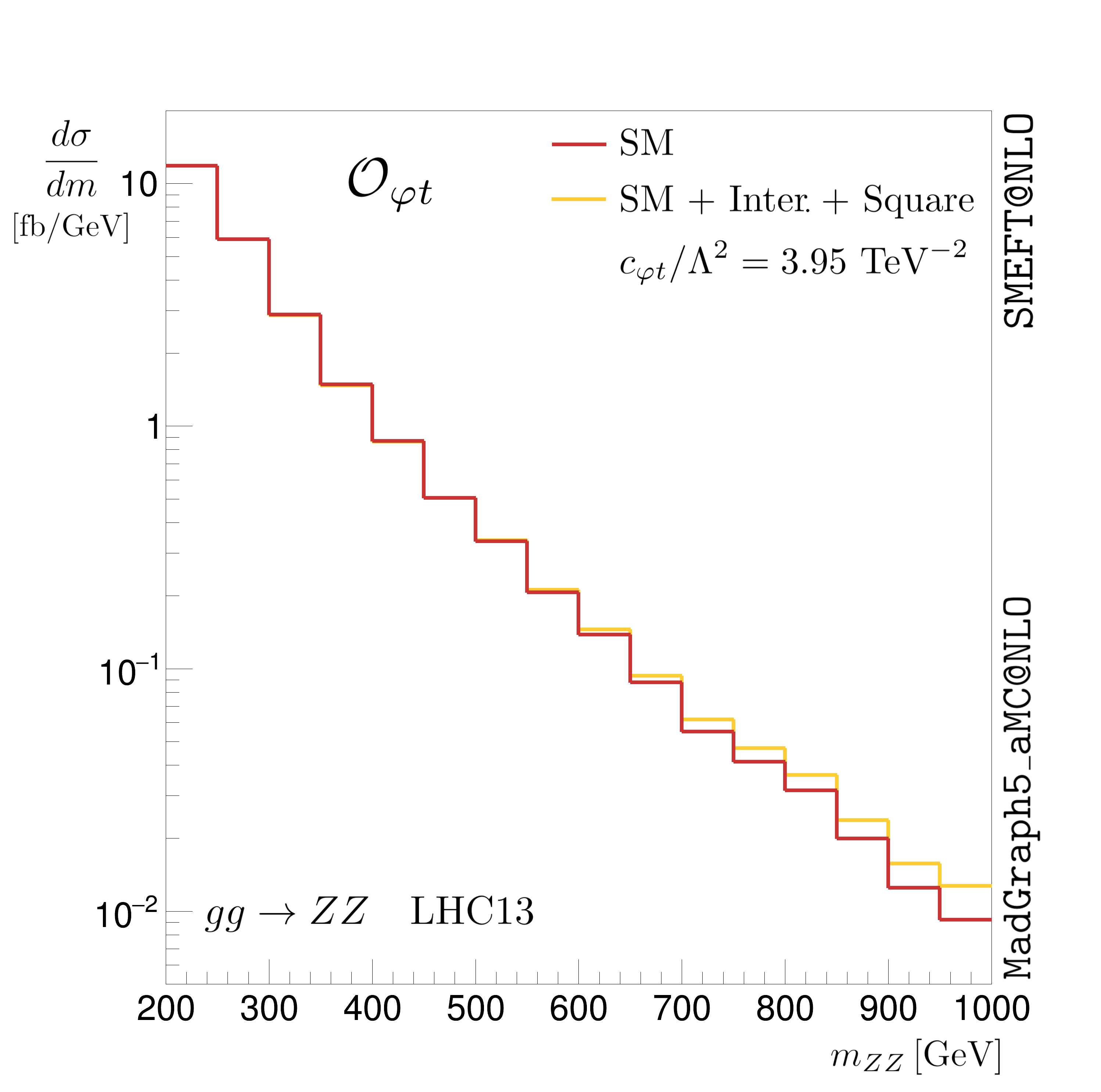}
         \label{fig:ctp}
     \end{subfigure}
 \caption{Differential distributions for maximum allowed values of the coefficients extracted from global SMEFT fits~\cite{Ethier:2021bye}. Both SM-NP interference and NP$^{2}$ terms are included. \label{fig:bounds}}

\end{figure}

To examine the prospects of the off-shell process in providing additional constraints beyond those set in global fits we
 employ the marginalised constraints set in \cite{Ethier:2021bye}. We select the 95\% CL marginalised bounds and extract the corresponding differential distributions. The operators where allowed values of the coefficients can lead to potentially measurable deviations are shown in Fig. \ref{fig:bounds}. Whilst these deviations are not very pronounced they do motivate more detailed studies of this process in particular in the context of potentially breaking degeneracies between operators. 

\clearpage

  \begin{figure}[t]
    \centering
    \includegraphics[width=0.33\textwidth]{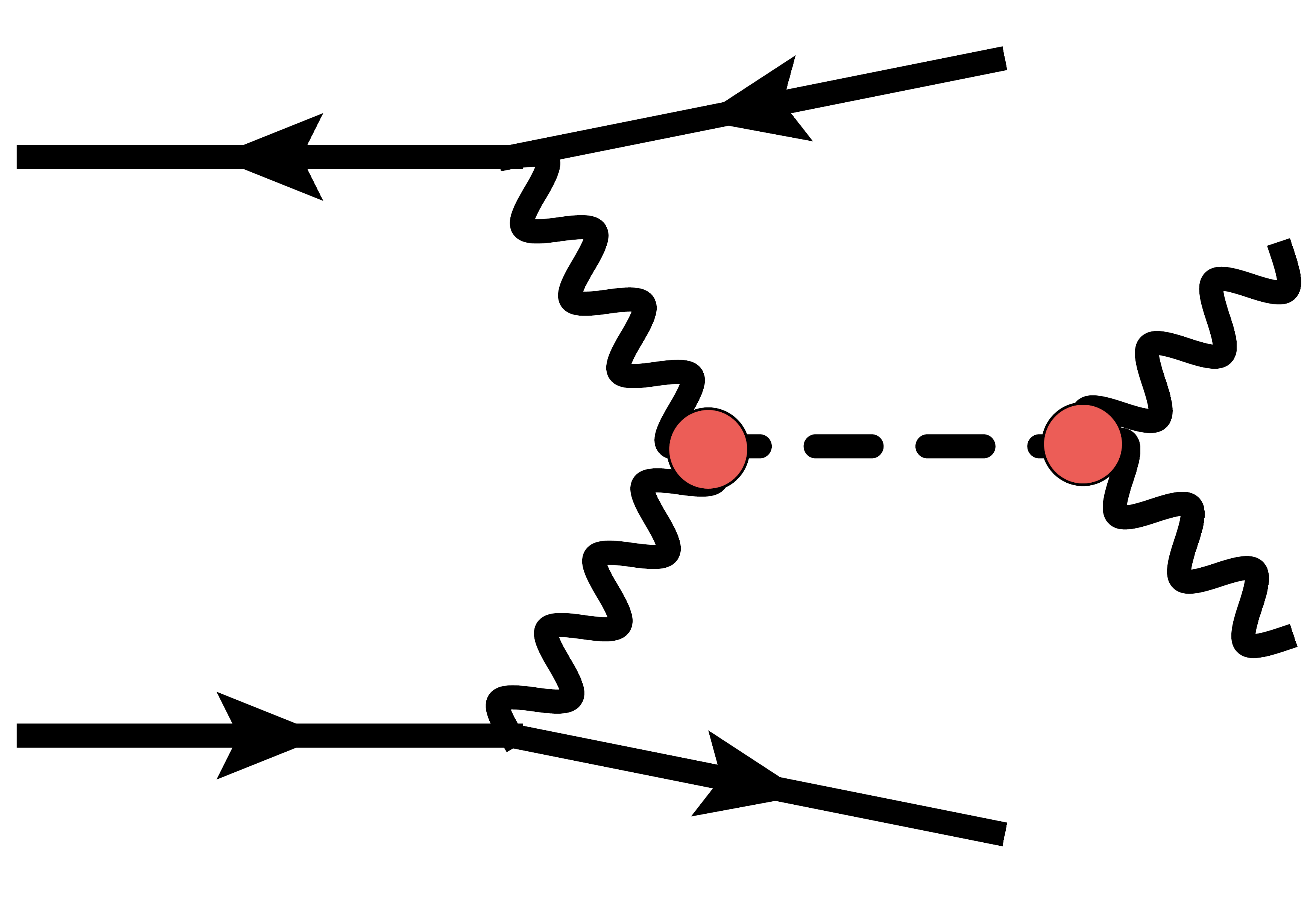}
    \includegraphics[width=0.3\textwidth]{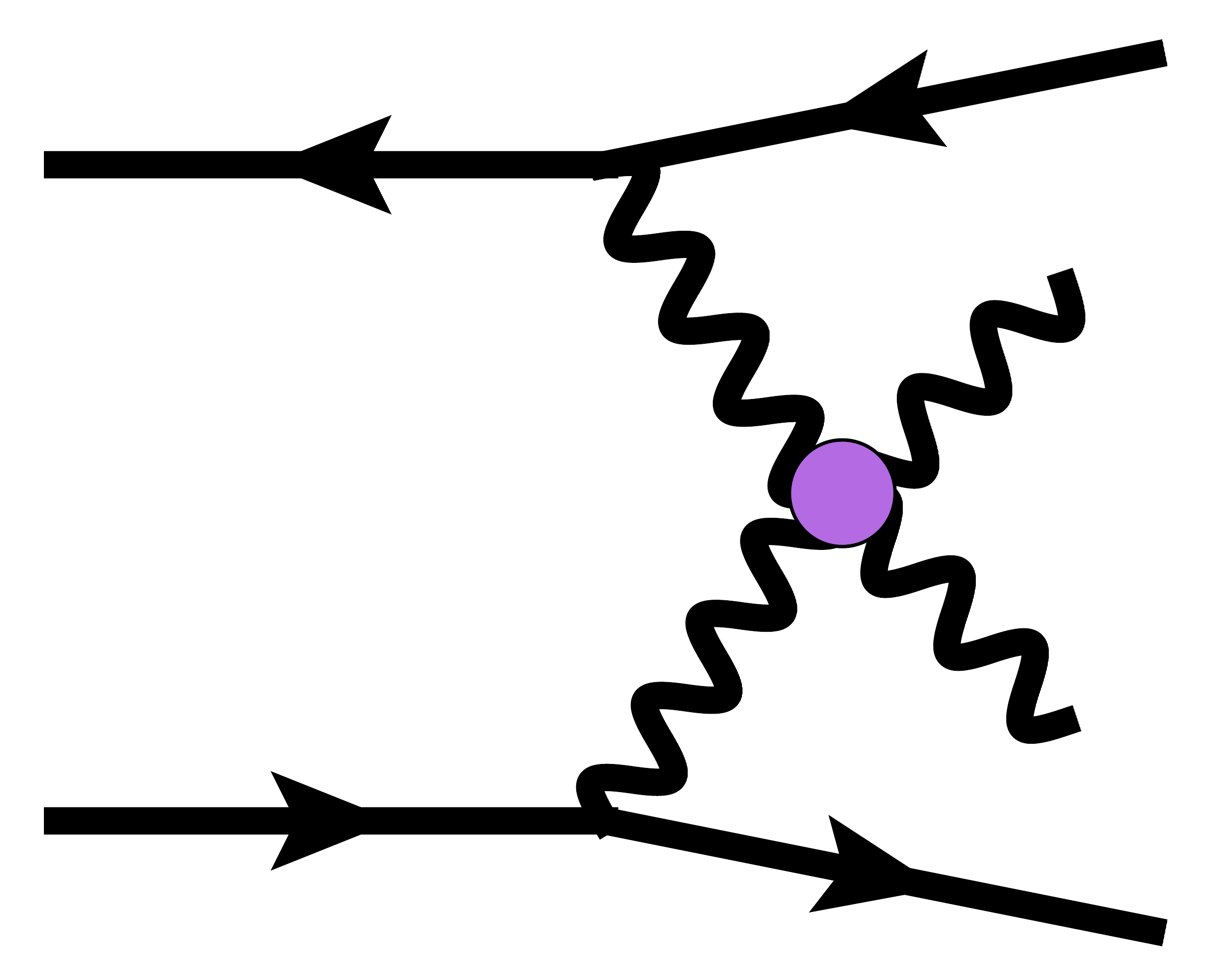}~~
    \includegraphics[width=0.3\textwidth]{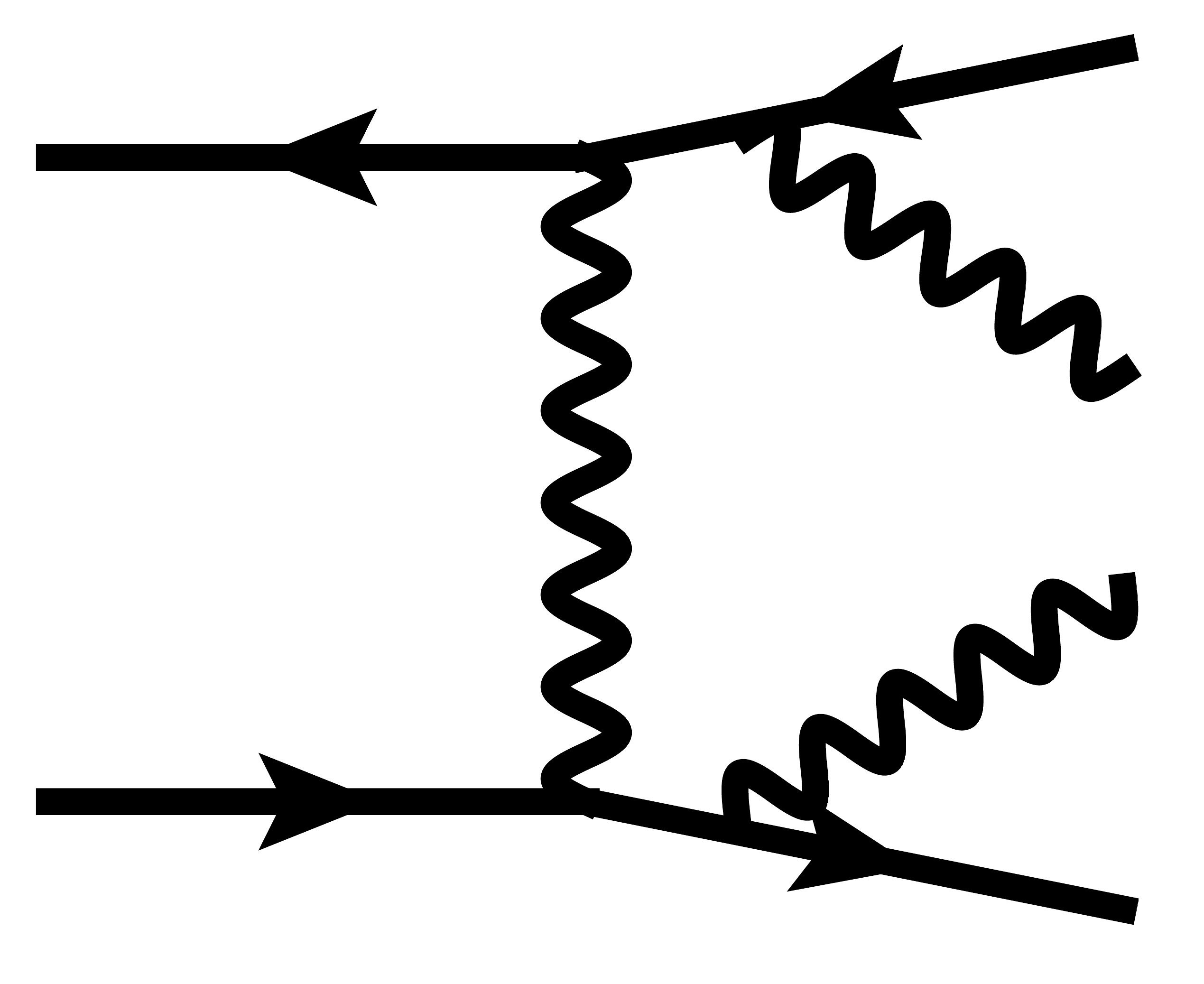}
    \caption{Example Feynman diagrams for electroweak off-shell Higgs production and corresponding background with EFT insertions.}
    \label{fig:ew_diagram}
\end{figure}

\section{Studies with the JHUGen+MCFM framework}
\label{sec:jhugen}
\vspace{-3mm}

The \p{JHUGen} implementation of off-shell Higgs boson production with subsequent decay to $VV\to 4f$
includes interference with background and supports both gluon fusion and electroweak (VBF and $VH$) processes~\cite{Gritsan:2020pib,jhugen}. 
Building on the transparent implementation of standard model matrix elements in \p{MCFM}~\cite{Campbell:2013una,Campbell:2015vwa},
the \p{JHUGen} framework incorporates the general scalar and gauge couplings of the Higgs boson, as well as additional possible states.
The \p{JHUGenLexicon} interface allows for parameterization of EFT effects either in the Higgs (mass eigenstate) or Warsaw 
(weak eigenstate) bases, or directly as modifications of the Higgs boson anomalous interactions with either fermions 
or vector bosons. 

\vspace*{-2mm}
\subsection{Relevant Operators}

Several types of EFT operators affecting Higgs boson physics, 
which appear in Eq.~(\ref{eq:lagrangian}) and are later listed in Eq.~(\ref{eq:DEF_higgs}), are considered.
The typical Feynman diagrams with these operators contributing to the gluons fusion process are
presented in Fig.~\ref{fig:FeynmanDiagrams}, and typical ones contributing to the electroweak off-shell Higgs 
boson production and corresponding background are shown in Fig.~\ref{fig:ew_diagram}. Therefore,
the operators affecting the Higgs boson signal can be classified as follows:
\begin{itemize}
\item  Operators affecting the $HVV$ vertex either in the $H\to VV$ decay or in electroweak production
of the Higgs boson ($VV\to H$, $V\to VH$):
 $\delta c_z, \ c_{z \Box},  \ c_{zz},     \ c_{\gamma \gamma}, \ c_{z \gamma}, \  c_{gg},  \tilde c_{zz},  \  \tilde c_{\gamma \gamma}, \  \tilde c_{z \gamma}$
\item   Operators affecting the $H$gg vertex in gluon fusion (point-like interactions):
 $c_{gg}, \tilde c_{gg}$
\item  Operators affecting Yukawa interaction in the gluon fusion loop:  
CP-odd $\tilde\kappa_t, \tilde\kappa_b$, 
and CP-even $\kappa_t, \kappa_b$, where the latter are equivalent to $\delta y_u, \delta y_d$ in Eq.~(\ref{eq:DEF_higgs})
\item  Operators with a new heavy fermion $Q$ in the gluon fusion loop, which reproduce $c_{gg}$ and $\tilde c_{gg}$
in the limit of $m_Q\to \infty$
\end{itemize}
Moreover, both gluon fusion and electroweak production of the Higgs boson in the off-shell regime require modeling 
of the background processes and their interference with the Higgs boson signal. 
These background processes may be modified by EFT effects. 
Therefore, the following types of EFT operators can also be considered:
\begin{itemize}
\item  Operators which allow for modification of the vector and axial-vector $Zff$ couplings,
either in the $Z$ decay to fermions or through the connection of the $Z$ to the fermion in the gluon fusion loop
in the $gg\to VV\to 4f$ background process
\item  Operators affecting the triple ($d^{\gamma WW}, d^{ZWW}, d_i^\gamma, d_i^Z$)
and quartic ($d^{\gamma\gamma WW}$, $d^{\gamma ZWW}$, $d^{ZZWW}$, $d^{WWWW}$) boson couplings
in the electroweak background production of the $VV\to 4f$ final state in association with jets
\end{itemize}
The former set of operators are not considered in the gluon fusion continuum process yet\cite{jhugen}. 
In the latter case, the triple and quartic electroweak boson couplings, with an example shown in the middle digram of Fig.~\ref{fig:ew_diagram},
are related to the $HVV$ vertices through SMEFT symmetry considerations, while additional terms beyond those listed 
in Eq.~(\ref{eq:DEF_higgs}) and unrelated to any of the Higgs boson contributions are not considered in the current version 
of the framework\cite{jhugen}, which is focused on EFT effects in the Higgs boson couplings. 

It is convenient to use the mass eigenstate basis to parameterize off-shell Higgs boson production due to the 
$Z^*$ or $H^*$ going off-shell in the $H^{(*)}\to Z^{(*)}Z\to 4f$ decay 
(or equivalently $W^{(*)}W\to 4f$). 
Such $H^*$ off-shell enhancement relative to on-shell production would not appear with intermediate 
photons $H^{(*)}\to \gamma^* V\to 4f$ generated by the operators
$c_{\gamma \gamma}, \ c_{z \gamma},  \tilde c_{\gamma \gamma}, \  \tilde c_{z \gamma}$.
As such, these operators can be neglected in the off-shell region, but they are still considered in 
the simulation because of their contribution to on-shell Higgs boson production. 

  \begin{figure}[b!]
    \centering
    \includegraphics[width=0.48\textwidth]{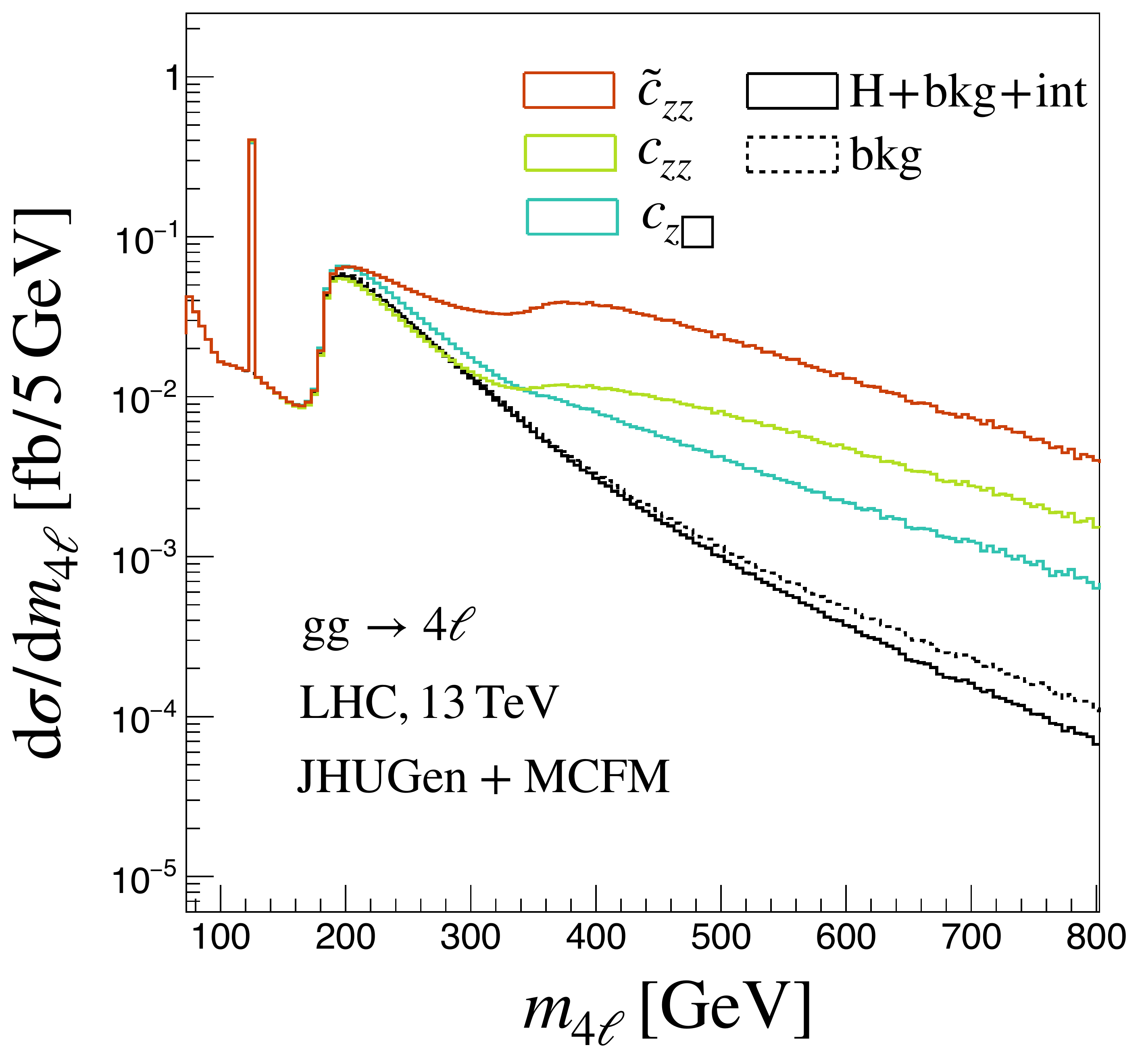}
    \includegraphics[width=0.50\textwidth]{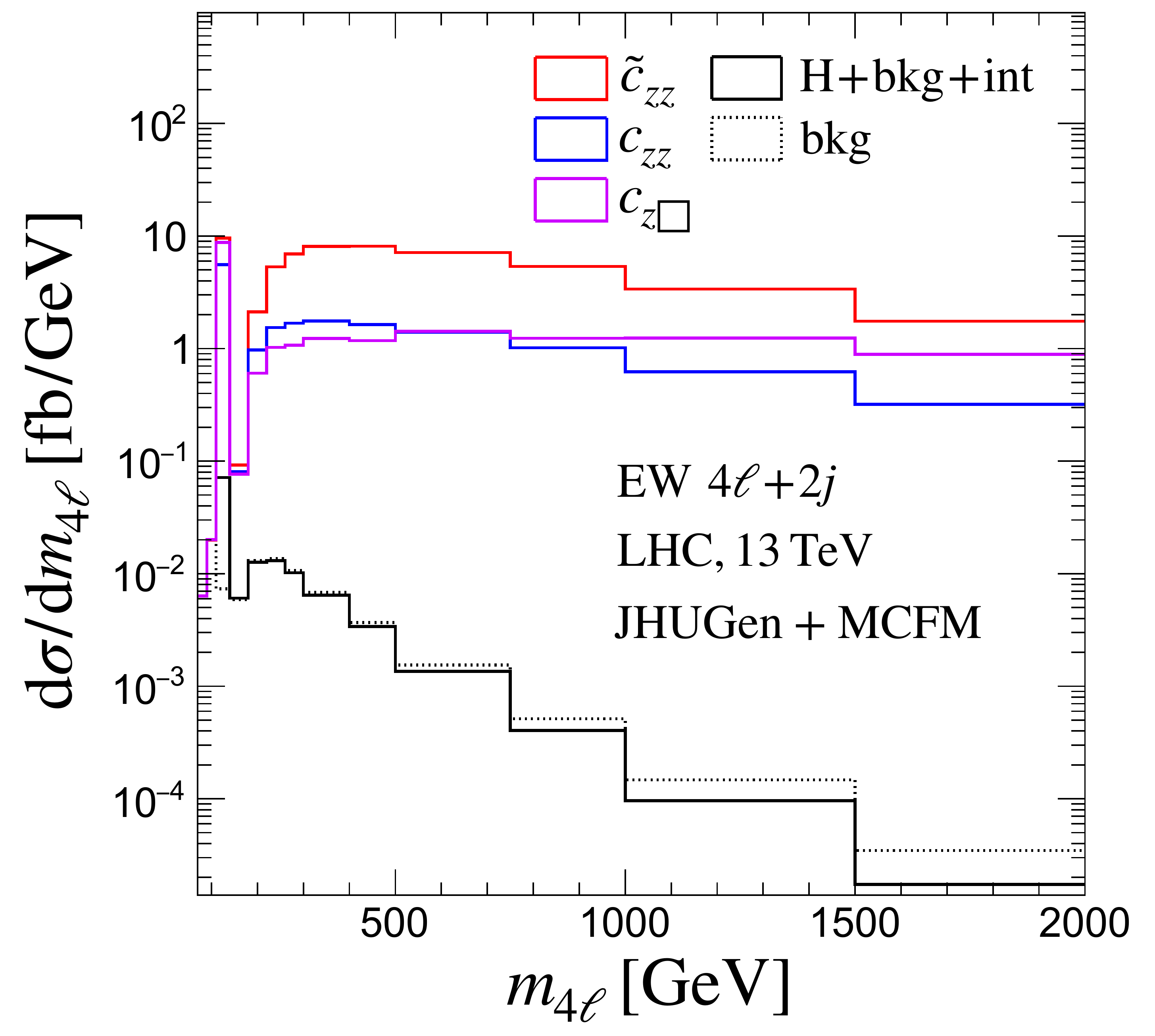}
    \caption{The four-lepton $m_{4\ell}$ invariant mass distributions for gluon fusion (left) and 
    electroweak production in association with two jets (right) at the LHC with a 13 TeV pp collision energy,
    where the histograms describing the electroweak process (right) were originally prepared for Ref.~\cite{Gritsan:2020pib}.
    The total SM production (``H+bkg+int'') and background-only (``bkg'') components are shown in black.
    Three operators $c_{z \Box}$ (magenta), $c_{zz}$ (blue), and $\tilde c_{zz}$ (red) are shown in color, 
    and they are introduced in place of the SM interaction with their strength constrained to reproduce the
    SM cross section of the on-shell Higgs boson signal production through gluon fusion. 
    }
    \label{fig:jhugen-offshell}
\end{figure}

\vspace*{-3mm}
\subsection{Differential Distributions and Expected Constraints}
\vspace*{-2mm}

Differential distributions of the remaining operators describing $HVV$ interactions are illustrated in Fig.~\ref{fig:jhugen-offshell},
where effects of the SM and the $c_{z \Box}$, $c_{zz}$, and $\tilde c_{zz}$ operators are shown
for both gluon fusion and electroweak production. 
Both production and decay vertices are modified in the electroweak process. 
The tensor structure of the $\delta c_z$ operator is identical to that of the SM coupling, and therefore
its impact corresponds to a simple rescaling of the SM signal yield. 

\begin{figure}[t]
    \centering
    \includegraphics[width=0.49\textwidth]{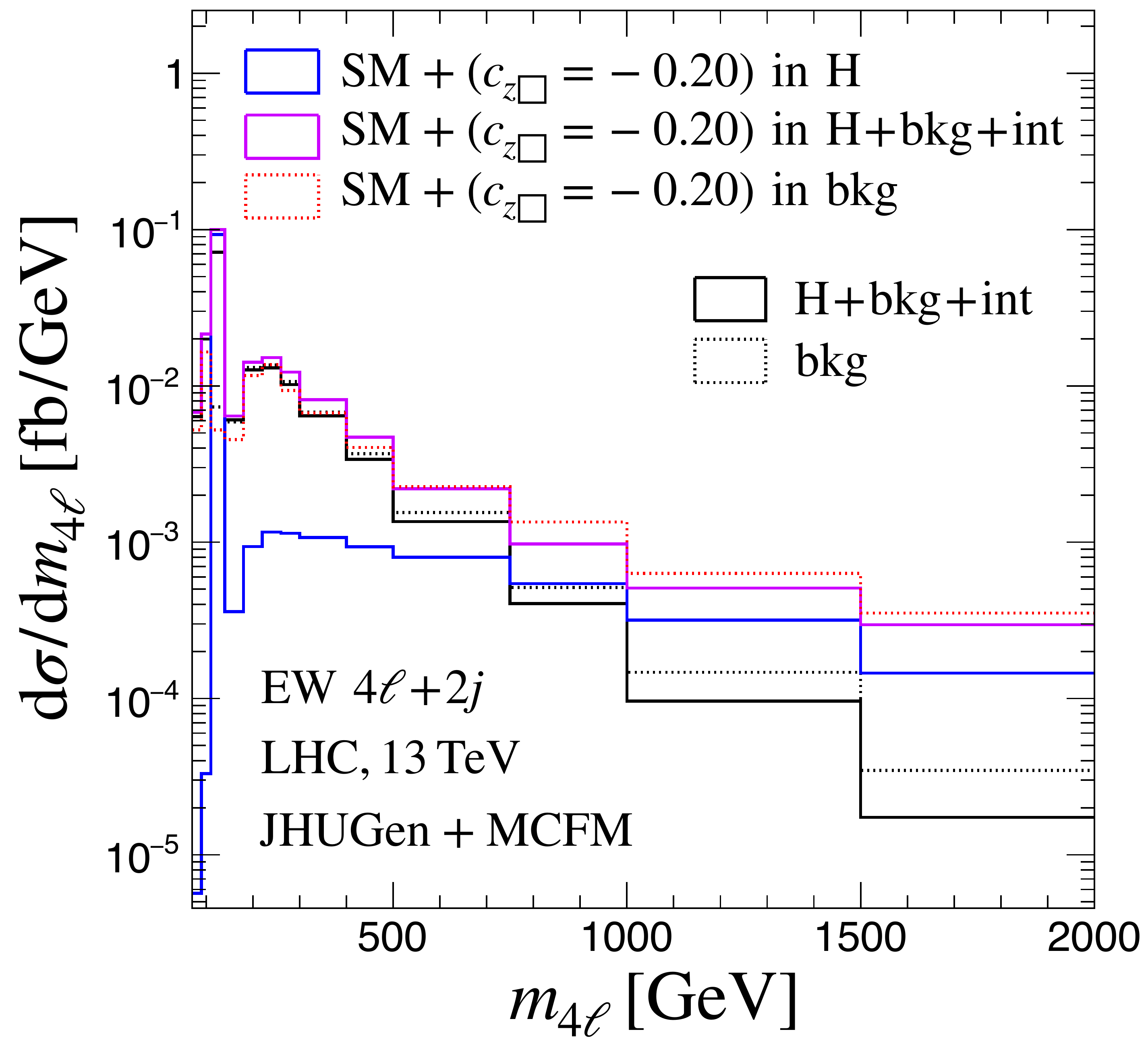}
    \includegraphics[width=0.49\textwidth]{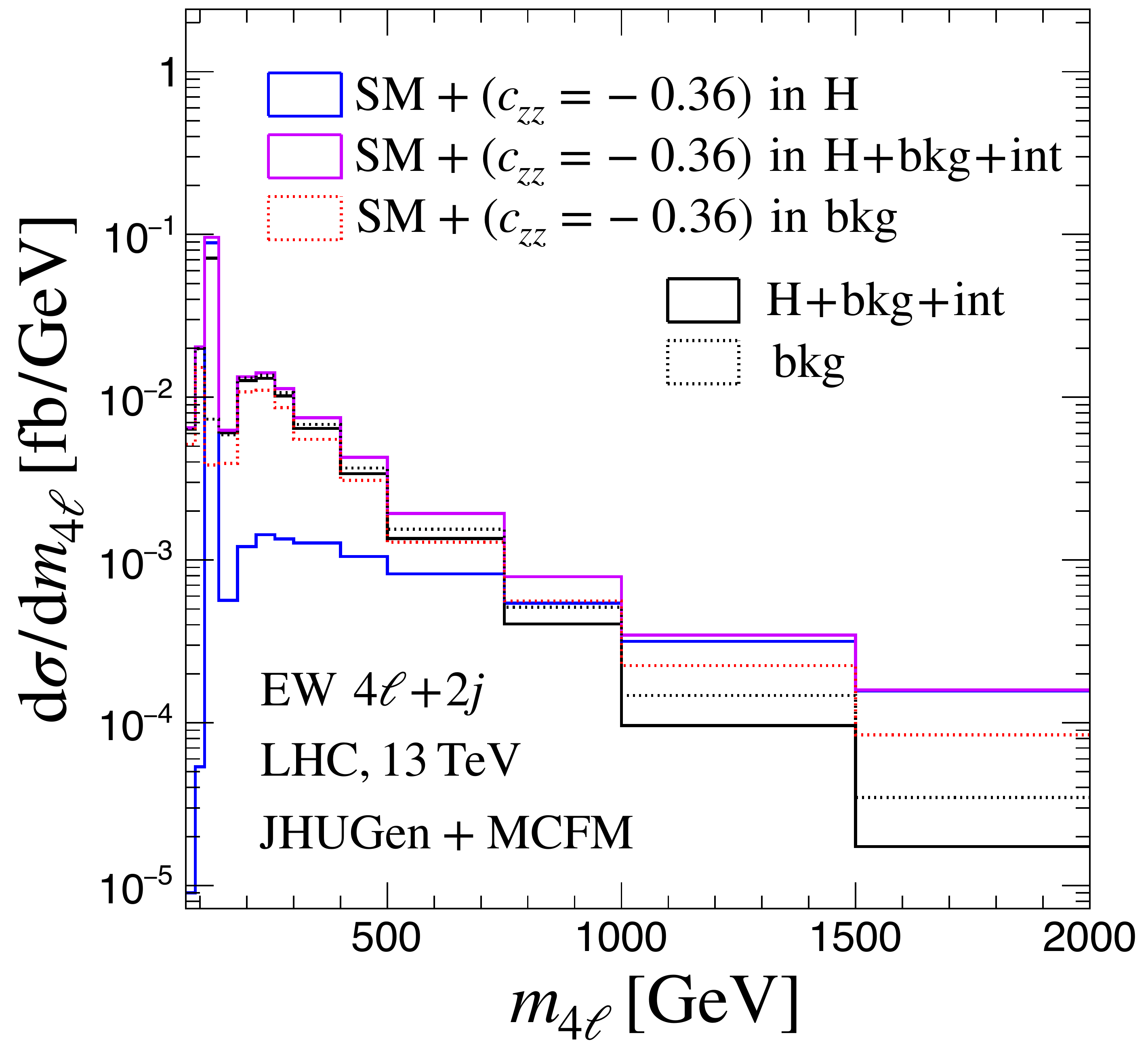}
    \caption{The four-lepton $m_{4\ell}$ invariant mass distributions for 
    electroweak production in association with two jets at the LHC with a 13 TeV pp collision energy,
    where the histograms were originally prepared for Ref.~\cite{Gritsan:2020pib}.
    The $c_{z \Box}=-0.20$ (left) and  $c_{zz}=-0.36$ (right) contributions in addition to the SM diagrams 
    to either signal-only (``H"), background-only (``bkg''), or the full process including their interference (``H+bkg+int'') are modelled.
    The total SM production and background-only components are shown in black.
    }
    \label{fig:jhugen-ew}
\end{figure}

In Fig.~\ref{fig:jhugen-ew}, small modifications of $c_{z \Box}=-0.20$ and  $c_{zz}=-0.36$ operators are considered.
The size of these operators is chosen to be of the scale of expected experimental constraints 
from $H^*$ off-shell data at LHC~\cite{CMS:2019ekd,CMS:2022ley}.
Modifications of the triple and quartic boson couplings in electroweak production are included in the full process 
with interference of the signal and background diagrams. 

\begin{figure}[b]
    \centering
    \includegraphics[width=0.49\textwidth]{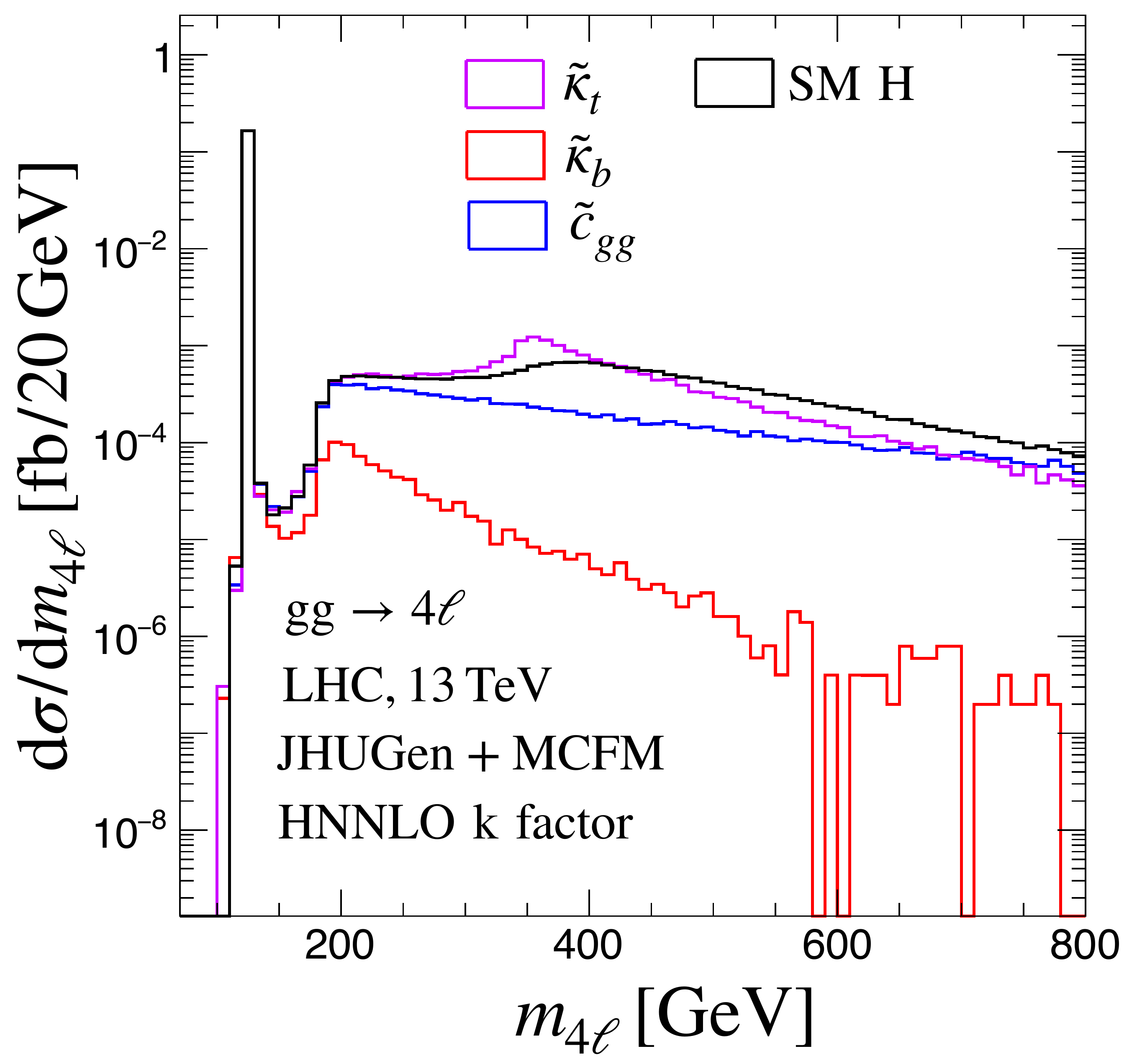}
    \includegraphics[width=0.49\textwidth]{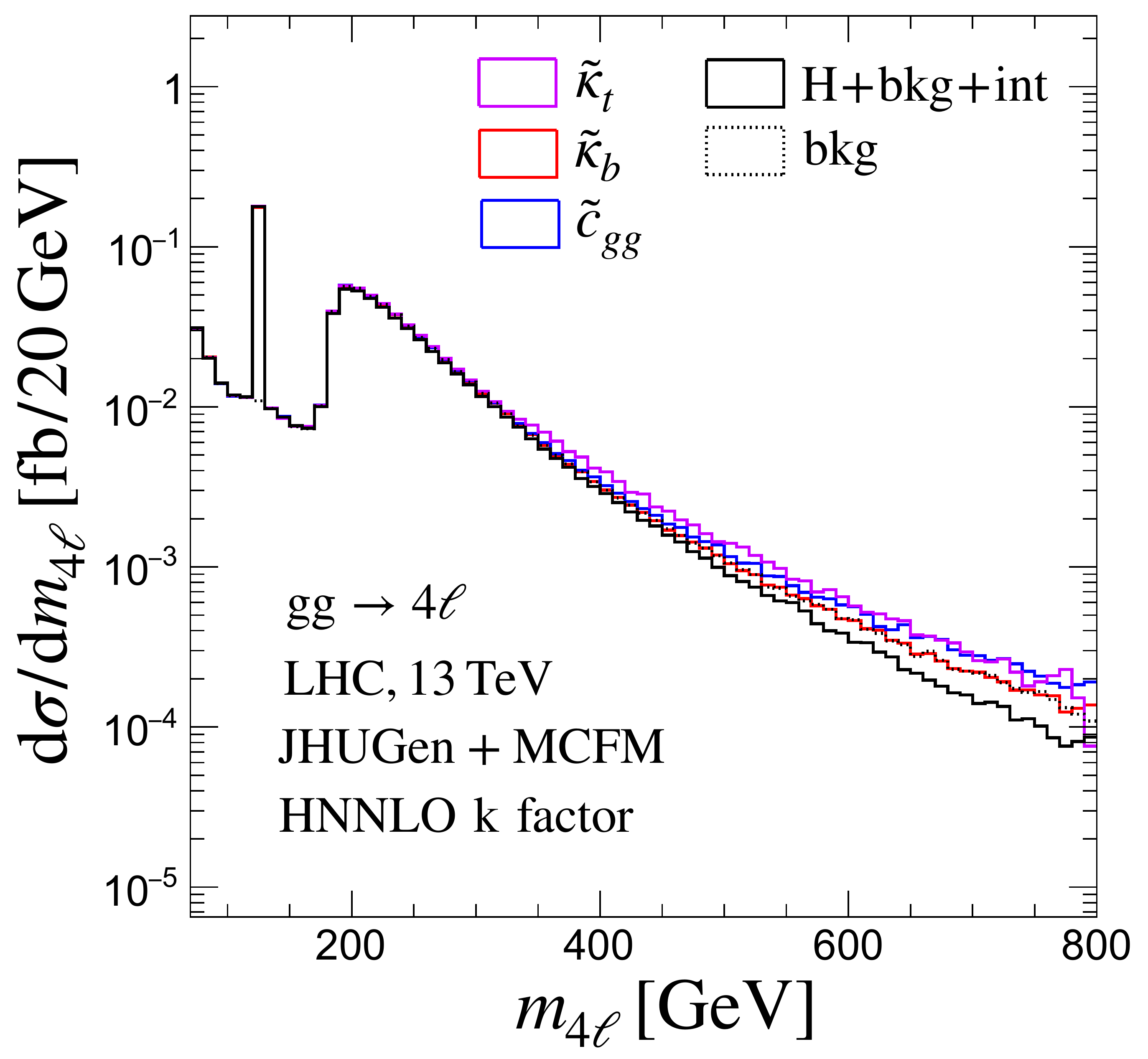}
    \caption{The four-lepton $m_{4\ell}$ invariant mass distributions for gluon fusion process at the LHC with a 13 TeV pp collision,
    where the histograms were originally prepared for Ref.~\cite{Gritsan:2020pib}.
    The left panel illustrates the differences in the Higgs signal-only component,
    where the three CP-odd operators $\tilde\kappa_t$ (magenta), $\tilde\kappa_b$ (red), and $\tilde c_{gg}$ (blue) are shown in color, 
    and they are introduced in place of the SM process with their strength constrained to reproduce 
    cross section of the SM Higgs boson production in gluon fusion (black) in the on-shell region. 
    The right panel illustrates the differences with the background amplitude included (``H+bkg+int''),
    and the background-only (``bkg'') component is shown as a dashed histogram.
    The $m_{4\ell}$-dependent NNLO QCD k~factor~\cite{deFlorian:2016spz} is applied for illustration purpose.  
     }
    \label{fig:jhugen-gg}
\end{figure}

Effects of CP-odd Yukawa couplings or point-like interaction in gluons fusion are illustrated in Fig.~\ref{fig:jhugen-gg}. 
Effects of other operators can be found in Ref.~\cite{Gritsan:2020pib}.
The ranges of allowed deviations of the above operators from the SM with the present Higgs boson 
off-shell data from LHC and using this framework may be found in Refs.~\cite{CMS:2015chx,CMS:2019ekd,CMS:2022ley}.
  
\chapter[Summary of the Higgs basis parametrization of the SMEFT]{Summary of the Higgs basis parametrization of the SMEFT\footnote{contributed by A.\ Falkowski}\label{ch:higgsbasis}}

\section{Pep talk}
\label{sec:pep} 

We consider the extension of the Standard Model (SM) by dimension-6 operators $Q_i$ invariant under the SM gauge symmetries \cite{Buchmuller:1985jz, Grzadkowski:2010es}:
\begin{equation}
\label{eq:Lsmeft}
  \mathcal L_\text{SMEFT} =
  \mathcal L_\text{SM} +
  \sum_i C_i \,Q_i \, . 
\end{equation}
The $Q_i$ set is assumed to be complete and non-redundant, that is to form a {\em basis}. 
Such sets are known to consist of 2499 distinct operators; explicit constructions include the 
Warsaw basis~\cite{Grzadkowski:2010es} and the SILH basis~\cite{Contino:2013kra}.
The parameters $C_i$ are called the {\em Wilson coefficients}. 
Together with the parameters of the SM Lagrangian, they make the parameter space of the the SMEFT. 
In our conventions the Wilson coefficients $C_i$ have dimensions 
$[\rm mass]^{-2}$ and they count as $\cO(\Lambda^{-2})$ in the EFT expansion.
Operators with dimensions higher than six, as well as dimension-5 operators are ignored in this discussion.

The idea behind the Higgs basis~\cite{deFlorian:2016spz} was to create a new parameterization of the space of dimension-6 SMEFT operators satisfying the following properties: 
\begin{enumerate}
\item  Each Wilson coefficient has a simple physical interpretation; 
\item  Higgs observables at leading order are affected by a minimal set of Wilson coefficients; 
\item  Large correlations between the Higgs and electroweak constraints are avoided. 
\end{enumerate}
These features are very helpful for global analyses targeting the multi-parameter space of the SMEFT, see e.g.~\cite{Boselli:2017pef,Durieux:2017rsg,Falkowski:2017pss,Grojean:2018dqj}.
The Higgs basis is a realization of the idea first laid out in~\cite{Gupta:2014rxa}, 
and its Wilson coefficients are what is called {\em primary effects} in that reference.

\section{Definition of the Higgs basis}
\label{sec:def} 

In Ref.~\cite{deFlorian:2016spz} the Higgs basis was introduced as follows. 
One started with the SMEFT Lagrangian including dimension-6 operators in the SILH basis~\cite{Contino:2013kra}. 
From it, a Lagrangian of mass eigenstates after electroweak symmetry breaking was derived. 
That Lagrangian was brought to a more convenient form by a series of fields and couplings redefinitions. 
Finally, the Higgs basis was introduced  by identifying a set of independent linear combinations of SILH Wilson coefficients that fully characterizes the mass eigenstate Lagrangian.  
This algorithm permitted to construct  a 1-to-1 linear map connecting the Higgs basis and SILH basis Wilson coefficients.  

In this note we follow a different route. 
We define the Higgs basis via a map $M_{W \to H}$:\footnote{%
Alternatively, rather than defining  the Higgs basis via a rotation of Wilson coefficients we could  define it as linear combinations of gauge-invariant dimension-6 operators. 
At the operator level, a definition equivalent to the one in \rEq{eq:DEF_map} would be  
$Q_{\rm HB} =  M_{W \to H}^{-1 \, T}  Q_{\rm WB}$, where $Q_{\rm WB}$ is the complete set of dimension-6 operators in the Warsaw basis.}
\begin{equation}
\label{eq:DEF_map}
\vec c_{\rm HB}  = M_{W \to H} \vec C_{\rm WB}. 
\end{equation} 
Here, $\vec c_{\rm HB}$  is a 2499-dimensional vector of Wilson coefficients in the Higgs basis to be defined shortly.
By convention, all components of  $\vec c_{\rm HB}$ are dimensionless.  
On the other hand, $\vec C_{\rm WB}$  is a 2499-dimensional vector of Wilson coefficients in the Warsaw basis of dimension-6 operators~\cite{Grzadkowski:2010es}. 
If one uses SMEFT beyond tree level, one needs to specify at which scale the map in \rEq{eq:DEF_map} is defined;   in our conventions,  \rEq{eq:DEF_map} holds at the scale $m_Z$. 
Finally,  $M_{W \to H}$ is a $2499 \times 2499$-dimensional invertible matrix. 
It was obtained in Ref.~\cite{Falkowski:hdr} and is quoted below.\footnote{%
With respect to that reference, we adapted sign, naming, and flavor conventions to match those of WCxf~\cite{Aebischer:2017ugx}. 
See \rApp{app:notation} for more details. } 
As we shall see shortly, it depends on the parameters $g_L$, $g_Y$, and $g_s$, which are the SM gauge couplings  at  the scale $m_Z$, and on $v$ and $\lambda$, which are the Higgs VEV and self-coupling.  
Their central values are 
\begin{equation}
\label{eq:DEF_couplings}
g_s = 1.2172, \quad  
g_L =  0.6485,  \quad 
 g_Y =  0.3580, \quad 
 v  =  246.22\GeV , \quad 
 \lambda = 0.1291 .  
\end{equation}  
and their errors can be ignored for the present purpose. 

We are ready to write down the map $M_{W \to H}$. 
The vector $\vec c_{\rm HB}$ in \rEq{eq:DEF_map}  contains the following Wilson coefficients:  
\begin{itemize}
\item The Higgs  couplings
\begin{equation} 
\label{eq:DEF_higgs} 
 {\color{blue} \delta c_z, \ c_{z \Box},  \ c_{zz},     \ c_{\gamma \gamma}, \ c_{z \gamma}, \  c_{gg}, \ \delta \lambda_3 } ,  
 \ {\color{blue}  \tilde c_{zz},  \  \tilde c_{\gamma \gamma}, \  \tilde c_{z \gamma}, \  \tilde c_{gg}, \   } 
\,  \delta y_u, \   \delta y_d,  \ \delta y_e , 
\end{equation} 
where the blue parameters are  real, and $\delta y$ are $3 \times 3$ complex matrices in the flavor space. 
They are related to the Wilson coefficients in the Warsaw basis as 
\begin{eqnarray}
\label{eq:DEF_higgsMap}
v^{-2} \delta c_{z} & = & 
C_{\varphi\Box}   - {1 \over 4} C_{\varphi D}
 -{3 \over 2}  \Delta_{G_F}, 
 \nnl 
v^{-2}  c_{z\Box} &= &    {1 \over 2   g_L^2} \left (
C_{\varphi D}   
 +2  \Delta_{G_F}   \right ),   
\nnl 
v^{-2}  c_{gg} &= & {4 \over  g_s^2} C_{\varphi G},  
\nnl 
v^{-2}  c_{\gamma \gamma} &= &  4  \left ( {1  \over   g_L^2} C_{\varphi W} + { 1 \over   g_Y^2} C_{\varphi B} - {1 \over   g_L   g_Y}  C_{\varphi WB} \right )  , 
\nnl  
v^{-2}  c_{zz} &= &    4 \left (  {  g_L^2 C_{\varphi W} +    g_Y^2 C_{\varphi B} +     g_L   g_Y  C_{\varphi WB} \over (  g_L^2 +   g_Y^2)^2} \right ),
\nnl 
v^{-2} c_{z\gamma} &= &     4 \left ( { C_{\varphi W} -   C_{\varphi B} 
-  {  g_L^2 -    g_Y^2 \over 2   g_L   g_Y} C_{\varphi WB} \over   g_L^2 +   g_Y^2} \right )  , 
\nnl 
v^{-2}  \tilde c_{gg} &= & {4 \over  g_s^2} C_{\varphi   \tilde G},  
\nnl 
v^{-2}    \tilde c_{\gamma \gamma} &= &  4  \left ( {1  \over   g_L^2} C_{\varphi   \tilde  W} + { 1 \over   g_Y^2} C_{\varphi   \tilde  B} 
- {1 \over   g_L   g_Y}  C_{\varphi W   \tilde  B} \right )  , 
\nnl  
v^{-2}    \tilde c_{zz} &= &    4 \left (  {  g_L^2 C_{\varphi   \tilde W} +    g_Y^2 C_{\varphi   \tilde B} +     g_L   g_Y  C_{\varphi W   \tilde B} \over (  g_L^2 +   g_Y^2)^2} \right ),
\nnl 
v^{-2}   \tilde c_{z\gamma} &= &     4 \left ( { C_{\varphi   \tilde W} -   C_{\varphi   \tilde B} 
-  {  g_L^2 -    g_Y^2 \over 2   g_L   g_Y} C_{\varphi W   \tilde B} \over   g_L^2 +   g_Y^2} \right )  , 
\nnl 
v^{-2}  \delta \lambda_3   &=& 
   -  {1 \over \lambda } C_{\varphi }    + 3 C_{\varphi \Box}  -  {3 \over 4}  C_{\varphi D}  
 -   {1 \over 2} \Delta_{G_F}  , 
 \nnl 
 v^{-2} [\delta y_f]_{JK}   & =  & 
 - {v  \over \sqrt {2 m_{f_J} m_{f_K}}}  [C_{f \varphi}^\dagger]_{JK}  +  \delta_{JK} \left  (
c_{\varphi \Box}  - {1 \over 4} C_{\varphi D}  - {1 \over 2}  \Delta_{G_F}  
   \right ) ,  
\end{eqnarray} 
where $\Delta_{G_F} = [C^{(3)}_{\varphi l}]_{11} + [C^{(3)}_{\varphi l}]_{22} -  {1 \over 2}[C_{ll}]_{1221}$.  
The interpretation of the Higgs basis Wilson coefficients on the left-hand-side of \rEq{eq:DEF_higgsMap} 
as  certain Higgs boson couplings will become clear in  \rSec{sec:newvar}. 
\item 
The vertex corrections 
\begin{equation} 
\label{eq:DEF_dg}
 \delta g^{Ze}_L, \ \delta g^{Ze}_R,  \  \delta g^{Zu}_L,  \ \delta g^{Zu}_R,   \  \delta g^{Zd}_L,  \ \delta g^{Zd}_R,  \ \delta g^{W \ell}_L,   \ \delta g^{Wq}_R, 
\end{equation}
which are all  $3 \times 3$ Hermitian matrices in the flavor space, 
except for  $ \delta g^{Wq}_R$ which is a general complex matrix. 
They are related to the Wilson coefficients in the Warsaw basis as 
\begin{eqnarray}
\label{eq:DEF_dgMap}
v^{-2}  \delta g^{W \ell}_L & = &   C^{(3)}_{\varphi l} + f(1/2,0) - f(-1/2,-1),  
\nnl 
v^{-2} \delta g^{Z\ell}_L & = &    - {1 \over 2} C^{(3)}_{\varphi l} - {1\over 2} C_{\varphi l}^{(1)}+   f(-1/2, -1) , 
\nnl 
v^{-2}  \delta g^{Z\ell }_R & = &  - {1\over 2} C_{\varphi e}^{(1)}   +  f(0, -1) ,
\nnl 
v^{-2}  \delta g^{Z u}_L & = &   {1 \over 2}  C^{(3)}_{\varphi q} - {1\over 2} C_{\varphi q}^{(1)}   + f(1/2,2/3) ,  
\nnl
v^{-2}  \delta g^{Zd}_L & = &   
 - {1 \over 2}  C^{(3)}_{\varphi q} - {1\over 2} C_{\varphi q}^{(1)}   + f(-1/2,-1/3), 
\nnl
v^{-2}  \delta g^{Zu}_R & = &    - {1\over 2} C_{\varphi u}   +  f(0,2/3), 
\nnl
v^{-2}  \delta g^{Zd}_R & = &    - {1\over 2} C_{\varphi d}  +  f(0,-1/3), 
\nnl 
v^{-2} \delta g^{Wq}_R & = &    {1 \over 2 } C_{\varphi ud},
\end{eqnarray} 
where 
\begin{equation} 
\label{eq:DEF_fdeltag}
f(T^3,Q)  \equiv  \bigg \{  
-   Q  {g_L  g_Y \over  g_L^2 -  g_Y^2} C_{\varphi WB} 
 -  \left ( 
{1 \over 4} C_{\varphi D}  +  {1 \over 2 } \Delta_{G_F}  \right )  \left ( T^3 + Q { g_Y^2 \over  g_L^2 -   g_Y^2} \right )  \bigg \} {\bf 1} . 
\end{equation}
\item The $F^3$ couplings  
\begin{equation}
\label{eq:DEF_f3}
 \lambda_z, \  \tilde \lambda_z,  \ \lambda_{g}, \ \tilde  \lambda_{g}  , 
\end{equation} 
which are all real. 
They are related to the Wilson coefficients in the Warsaw basis as  
\begin{eqnarray} 
\label{eq:DEF_f3Map}
v^{-2} \lambda_z   &= &  {3 \over 2} g_L C_{W}, \qquad  v^{-2} \tilde \lambda_z   =  {3 \over 2} g_L C_{\tilde W}, 
\nnl 
v^{-2} \lambda_g  &= & {C_{G} \over g_s^3}, \qquad  v^{-2} \tilde \lambda_g  = {C_{\tilde G} \over g_s^3}.   
\end{eqnarray} 
\item 
The dipole couplings  
\begin{eqnarray} 
\label{eq:DEF_dipole}
&  
d_{G u}, \ d_{G d},   \ d_{A e}, \  d_{A u}, \ d_{A d}, \  d_{Z e}, \ d_{Z u}, \ d_{Z d},    
\end{eqnarray} 
which are all complex  $3 \times 3$ matrices. 
They are related to the Wilson coefficients in the Warsaw basis as 
\begin{eqnarray}
\label{eq:DEF_dipoleMap}
v^{-2} d_{Gf}  &=& - {16 \over g_s^2}  C_{fG}^*,
\nonumber \\ 
v^{-2} d_{A f}   &=& - {16 \over g_L^2}   \left ( \eta_f C_{fW}^*  + C_{fB}^*  \right ),
\nonumber \\  
v^{-2} d_{Z f}  &=& -  16  \left (  \eta_f  {1 \over g_L^2 + g_Y^2} C_{fW}^*  - {g_Y^2 \over g_L^2 (g_L^2 + g_Y^2)}   C_{fB}^*  \right ),
\nonumber \\ 
v^{-2} d_{Wf} &= & - {16 \over g_L^2} C_{fW}^*,   
\end{eqnarray} 
where $\eta_u = +1$, $\eta_{d,e} = -1$. 
\item Four-fermion couplings 
\begin{eqnarray} 
\label{eq:DEF_4f}
& c_{ll}, \  c_{qq}^{(1)} , \ c_{qq}^{(3)} , \  c_{l q}^{(1)}  , \ c_{l q}^{(3)} ,  \ 
c_{ee} ,  \  c_{uu} , \  c_{dd} , \  c_{eu} , \  c_{ed} , \  c_{ud}^{(1)} , \  c_{ud}^{(3)}   
\nnl & 
 c_{l e} , \  c_{l u} , \  c_{ld} , \  c_{qe} , \  c_{qu}^{(1)}  , \  c_{qu}^{(8)}  , \  c_{qd}^{(1)}   , \  c_{qd}^{(8)} ,    
  \ c_{ledq}, \ c_{quqd}^{(1)},  \ c_{quqd}^{(8)} , \  c_{lequ}^{(1)} , \ c_{lequ}^{(3)}  . 
 \nnl 
\end{eqnarray}
which are all 4-index tensors in the flavor space. 
They are trivially related to the Wilson coefficients in the Warsaw basis as  
\begin{equation}
\label{eq:DEF_4fMap}
v^{-2} c_i  = C_i.   
\end{equation} 
\end{itemize} 
The full set of Higgs basis Wilson coefficients is displayed in  Eqs.~(\ref{eq:DEF_higgs}), (\ref{eq:DEF_dg}), (\ref{eq:DEF_f3}), (\ref{eq:DEF_dipole}), and (\ref{eq:DEF_4f}). 
The map $M_{W \to H}$ is completely specified by Eqs.~(\ref{eq:DEF_higgsMap}), (\ref{eq:DEF_dgMap}), (\ref{eq:DEF_f3Map}), (\ref{eq:DEF_dipoleMap}), and (\ref{eq:DEF_4fMap}).  
The physical interpretation of the Higgs basis Wilson coefficients will be clarified in \rSec{sec:newvar}.

\section{Lagrangian for mass eigenstates}
\label{sec:lme} 

In order to derive physical predictions of the SMEFT 
the first step is to recast its Lagrangian in~\rEq{eq:Lsmeft} in terms of the mass eigenstates after electroweak symmetry breaking. 
In the Warsaw basis this exercise was completed in~\cite{Dedes:2017zog}, 
where all the interactions vertices with the corresponding Feynman rule were given. 
To derive the mass eigenstate Lagrangian in the Higgs basis  one could for example borrow the interaction terms from that reference and translate the couplings  to the Higgs basis using the map in  Eqs.~(\ref{eq:DEF_higgsMap}), (\ref{eq:DEF_dgMap}), (\ref{eq:DEF_f3Map}), (\ref{eq:DEF_dipoleMap}), and (\ref{eq:DEF_4fMap}).\footnote{
In practice, we rederive all interactions using a custom-made computer code. }  
Below I quote the part of the mass eigenstate Lagrangian most relevant for the LHC and Higgs phenomenology.  
 
By definition of the mass eigenstate basis, 
the kinetic terms for the Higgs, $W$, $Z$ bosons, photons, gluons, and fermions  
 are diagonal and canonically normalized: 
\begin{equation} 
\cL \supset 
 {1 \over 2} \partial_\mu \hat h \partial_\mu  \hat h 
 - {1 \over 2} W_{\mu \nu}^+  W_{\mu \nu}^- -   {1 \over 4} Z_{\mu \nu} Z_{\mu \nu} 
-   {1 \over 4} A_{\mu \nu} A_{\mu \nu}   - {1 \over 4} G_{\mu \nu}^a  G_{\mu \nu}^a
+  \sum_{f \in u,d,e,\nu}  i \bar f  \gamma_\mu \partial_\mu  f  . 
\end{equation}  
Above, we mark the Higgs boson field $\hat h$ because later we will switch to a more convenient variable to describe this particle. 
The mass terms for the Higgs, $W$, $Z$ bosons, and fermions are also diagonal:  
\begin{equation} 
\cL \supset 
- {1 \over 2} m_h^2 h^2 + m_W^2 W_{\mu}^+  W_{\mu}^-  + {1\over 2} m_Z^2  Z_\mu Z_\mu 
-  \sum_{f \in  u,d,e}  m_f  \bar f  f ,  
\end{equation} 
where in the Higgs basis 
\begin{eqnarray}
\label{eq:HL_masses}
m_h^2 & = &  2 \hat \lambda \hat v^2 \bigg [  
1 - {3 \hat g_L^4 \over \hat g_L^2 - \hat g_Y^2 } c_{z \Box} 
- {3 \hat g_L^2 \hat g_Y^2 \over \hat g_L^2 - \hat g_Y^2 } c_{z z}
+ {3 \hat g_L^4 \hat g_Y^4 \over (\hat g_L^2 - \hat g_Y^2)  (\hat g_L^2 + \hat g_Y^2)^2 } c_{\gamma \gamma}
+ {3 \hat g_L^2 \hat g_Y^2 \over \hat g_L^2 + \hat g_Y^2 } c_{z \gamma}  
\nnl  & & 
- {5 \over 2 } \delta c_z  + {3 \over 2} \delta \lambda_3 - 3 \Delta
\bigg ], 
\nnl 
m_W^2 & = & {\hat g_L^2 \hat v^2 \over 4} \bigg [ 
{\hat g_L^2 \over 2 } c_{z z}
+ {\hat g_L^2 \hat g_Y^4 \over 2 (\hat g_L^2 + \hat g_Y^2)^2 } c_{\gamma \gamma}
+ { \hat g_L^2 \hat g_Y^2 \over \hat g_L^2 + \hat g_Y^2 } c_{z \gamma}  
\bigg ]  
\nnl 
m_Z^2 & = & {(\hat g_L^2 + \hat g_Y^2 ) \hat v^2 \over 4} \bigg [ 
\hat g_Y^2 c_{z \gamma}
- {\hat g_Y^2 (\hat g_L^2 + \hat g_Y^2 ) \over \hat g_L^2 - \hat g_Y^2 } c_{z \Box}
+  {\hat g_L^2 \hat g_Y^4 \over \hat g_L^4 - \hat g_Y^4}  c_{\gamma \gamma}
\nnl && 
+  { \hat g_L^6 -  \hat g_L^4  \hat g_Y^2 -  3 \hat g_L^2  \hat g_Y^4 -   \hat g_Y^6 
\over 2 \hat g_L^2 (\hat g_L^2 - \hat g_Y^2) } c_{z z}
-  \left ( 1 + {\hat g_Y^2 \over \hat g_L^2}  \right  )  \Delta 
\bigg ]  
\end{eqnarray} 
and $\Delta \equiv \delta g^{We}_L + \delta g^{W\mu}_L  - {1\over 2} [c_{ll}]_{1221}$. 
Above $\hat g_L$ and $\hat g_Y$ are the  $SU(2) \times U(1)$ gauge couplings in the SM Lagrangian 
$\cL_{\rm SM}$ in \rEq{eq:Lsmeft},  
$\hat v$ is the vacuum expectation value of the Higgs field, $\langle H^\dagger H \rangle = \hat v^2/2$, 
and $\hat \lambda$ is the quartic Higgs coupling in the SM Lagrangian. 
Since dimension-6 operators affect the input observables from which these couplings are determined in the SM context, in SMEFT one {\em cannot} assume the hatted couplings have the numerical values in \rEq{eq:DEF_couplings}. 
In fact, their numerical values vary as a function of dimension-6 Wilson coefficients.

The gluon couplings to matter are given by 
\begin{equation}
\label{eq:HL_gluon}
\cL \supset - \hat g_s \left (1+ {\hat g_s^2\over 4} c_{gg} \right )  
G_\mu^a  \sum_{f \in u,d}  \bar f  \gamma_\mu T^a f  . 
\end{equation} 
The photon couplings to matter are given by 
\begin{equation}
\label{eq:HL_photon}
\cL \supset -  {\hat g_L \hat g_Y \over \sqrt{\hat g_L^2 + \hat g_Y^2 } } \left ( 
1+  { \hat g_L^2 \hat g_Y^2 \over 4 (\hat g_L^2 + \hat g_Y^2) } c_{\gamma \gamma}  
\right )  A_\mu  \sum_{f \in u,d,e} Q_f \bar f  \gamma_\mu  f  . 
\end{equation} 
The $Z$ boson couplings to charged leptons and quarks are given by 
\begin{equation}
\cL \supset  - \sqrt{\hat g_L^2 + \hat g_Y^2} Z_\mu  
\sum_{f \in u, d,e}  \bar f   \gamma_\mu \left (T^3_f - \hat s_\theta^2 Q_f +  \hat \delta g^{Zf}  \right)  f 
\end{equation}  
where $\hat s_\theta = \hat g_Y/\sqrt{\hat g_L^2 + \hat g_Y^2 } $ and 
\begin{equation}
 \hat \delta g^{Zf}   = \delta g^{Zf} 
 + \left ( T^3_f + {\hat g_Y^2 \over \hat g_L^2 - \hat g_Y^2 }Q_f \right )  
 \left ( {\hat g_L^2 \over 2}   c_{z \Box} +  {\hat g_L^2 + \hat g_Y^2 \over 4} c_{zz} \right ) 
 - Q_f  {\hat g_L^2  \hat g_Y^2 \over 2 (  \hat g_L^2 - \hat g_Y^2) (  \hat g_L^2 + \hat g_Y^2)^2 } c_{\gamma \gamma} . 
\end{equation} 
 The contact interactions between fermions and the Higgs and Z bosons are given by 
\begin{equation}
\label{eq:HL_contactVertex}
\cL \supset  - \sqrt{\hat g_L^2 + \hat g_Y^2} \left ( {2 \hat h \over \hat v}  + {\hat h^2 \over \hat v^2} \right )   Z_\mu  
\sum_{f \in u, d,e,\nu}  \bar f   \gamma_\mu \hat \delta g^{hZf}  f 
\end{equation}  
where 
\begin{equation}
 \hat \delta g^{hZf}   = \delta g^{Zf} 
 + {\hat g_L^2 \over 2}   \left ( T^3_f + {\hat g_Y^2 \over \hat g_L^2 - \hat g_Y^2 }Q_f \right )    c_{z \Box} 
-  Q_f  {\hat g_L^2  \hat g_Y^2 \over 2 (  \hat g_L^2 - \hat g_Y^2)  }
\left ( c_{zz} 
+ { \hat g_L^2 - \hat g_Y^2 \over \hat g_L^2 + \hat g_Y^2} c_{z \gamma} 
+ { \hat g_L^2 \hat g_Y^2 \over   ( \hat g_L^2 + \hat g_Y^2)^2 } c_{\gamma \gamma} 
\right ) . 
\end{equation} 
The cubic Higgs boson self-interactions are given by 
\begin{equation}
\label{eq:HL_h3}
\cL \supset   - \hat \lambda \hat v (1 + \hat \delta \lambda_3) \hat h^3 
+ { \hat \lambda_3^{(2)}  \over \hat v}  \hat h (\partial_\mu \hat h)^2 , 
\end{equation}
where 
\begin{eqnarray}
\hat \delta \lambda_3 & = & 
- {11 \hat g_L^4 \over 2 (\hat g_L^2 - \hat g_Y^2) } c_{z \Box} 
- {11 \hat g_L^2 \hat g_Y^2 \over 2 (\hat g_L^2 - \hat g_Y^2) } c_{z z}
+ {11 \hat g_L^4 \hat g_Y^4 \over 2 (\hat g_L^2 - \hat g_Y^2)  (\hat g_L^2 + \hat g_Y^2)^2 } c_{\gamma \gamma}
+ {11 \hat g_L^2 \hat g_Y^2 \over 2 (\hat g_L^2 + \hat g_Y^2) } c_{z \gamma}  
\nnl  & & 
- {9 \over 2 } \delta c_z  + {5 \over 2} \delta \lambda_3 - {11 \over 2} \Delta, 
\nnl 
\hat \lambda_3^{(2)} &=  &
 - {3 \hat g_L^4 \over \hat g_L^2 - \hat g_Y^2 } c_{z \Box} 
- {3 \hat g_L^2 \hat g_Y^2 \over \hat g_L^2 - \hat g_Y^2 } c_{z z}
+ {3 \hat g_L^4 \hat g_Y^4 \over (\hat g_L^2 - \hat g_Y^2)  (\hat g_L^2 + \hat g_Y^2)^2 } c_{\gamma \gamma}
+ {3 \hat g_L^2 \hat g_Y^2 \over \hat g_L^2 + \hat g_Y^2 } c_{z \gamma}  
\nnl  & & 
- 2  \delta c_z  - 3 \Delta . 
\end{eqnarray} 
The SM-like Higgs boson couplings to massive gauge bosons are given by 
\begin{equation} 
\cL \supset  {\hat h \over \hat v} \left [ 
 (1 + \hat \delta c_w)   {\hat g_L^2 \hat v^2 \over 2} W_\mu^+ W_\mu^-
  +   (1 + \hat \delta c_z)  {(\hat g_L^2 + \hat g_Y^2) \hat v^2 \over 4} Z_\mu Z_\mu
  \right ],  
  \end{equation} 
  where 
  \begin{eqnarray}
\hat \delta c_w  & = & \delta c_z 
+  {3 \hat g_L^4 \over 2 (\hat g_L^2 - \hat g_Y^2) } c_{z \Box} 
 + { \hat g_L^2( \hat g_L^2 + 2 \hat g_Y^2) \over 2 (\hat g_L^2 - \hat g_Y^2) } c_{z z}
- {\hat g_L^2 \hat g_Y^4 (2 \hat g_L^2 + \hat g_Y^2) \over 2 (\hat g_L^2 - \hat g_Y^2)  (\hat g_L^2 + \hat g_Y^2)^2 } c_{\gamma \gamma}
- { \hat g_L^2 \hat g_Y^2 \over 2 (\hat g_L^2 + \hat g_Y^2) } c_{z \gamma}  
\nnl  & & 
+ {3 \over 2}  \Delta , 
\nnl   
\hat \delta c_z  & = & \delta c_z 
+  {(3 \hat g_L^4 - 4  \hat g_L^2  \hat g_Y^2 ) \over 2 (\hat g_L^2 - \hat g_Y^2) } c_{z \Box} 
 + { \hat g_L^4 -\hat g_L^2 \hat g_Y^2 - \hat g_Y^4    \over 2 (\hat g_L^2 - \hat g_Y^2) } c_{z z}
+ {\hat g_L^4 \hat g_Y^4  \over 2 (\hat g_L^2 - \hat g_Y^2)  (\hat g_L^2 + \hat g_Y^2)^2 } c_{\gamma \gamma}
+ { \hat g_L^2 \hat g_Y^2 \over 2 (\hat g_L^2 + \hat g_Y^2) } c_{z \gamma}  
\nnl  & & 
- {1 \over 2}  \Delta .
 \end{eqnarray} 
 The 2-derivative Higgs boson couplings to gauge bosons are given by 
 \begin{eqnarray}  
\cL & \supset &   {\hat h \over \hat v}  \bigg \{ 
 c_{gg} {\hat g_s^2 \over 4} G_{\mu \nu}^a G_{\mu \nu}^a   
 +  \tilde c_{gg} {\hat g_s^2 \over 4} G_{\mu \nu}^a \tilde G_{\mu \nu}^a   
 + c_{ww}  {\hat g_L^2 \over  2} W_{\mu \nu}^+  W_{\mu\nu}^-  
 + \tilde c_{ww}  {\hat g_L^2 \over  2} W_{\mu \nu}^+   \tilde W_{\mu\nu}^- 
\nnl  && 
+ c_{\gamma \gamma} {\hat g_L^2 \hat g_Y^2 \over 4 (\hat g_L^2 + \hat g_Y^2)} A_{\mu \nu} A_{\mu \nu} 
+ \tilde c_{\gamma \gamma} {\hat g_L^2 \hat g_Y^2 \  \over 4 (\hat g_L^2 + \hat g_Y^2)} A_{\mu \nu} \tilde A_{\mu \nu}  
+ c_{z \gamma} {\hat g_L \hat g_Y \over  2} Z_{\mu \nu} A_{\mu\nu} 
+ \tilde c_{z \gamma} {\hat g_L \hat g_Y \over  2} Z_{\mu \nu} \tilde A_{\mu\nu}
\nnl  &&
+ c_{zz} {\hat g_L^2 + \hat g_Y^2 \over  4} Z_{\mu \nu} Z_{\mu\nu}
+ \tilde c_{zz}  {\hat g_L^2 + \hat g_Y^2  \over  4} Z_{\mu \nu} \tilde Z_{\mu\nu}, 
\bigg \} , 
\end{eqnarray} 
where 
\begin{eqnarray}
c_{ww} & = & c_{zz} +  {2 \hat g_Y^2 \over \hat g_L^2 + \hat g_Y^2} c_{z \gamma} 
 + { \hat g_Y^4 \over (\hat g_L^2 + \hat g_Y^2)^2 } c_{\gamma \gamma} , 
\nnl 
\tilde c_{ww} & = & \tilde  c_{zz} +  {2 \hat g_Y^2 \over \hat g_L^2 + \hat g_Y^2} \tilde  c_{z \gamma} 
 + { \hat g_Y^4 \over (\hat g_L^2 + \hat g_Y^2)^2 } \tilde  c_{\gamma \gamma} . 
\end{eqnarray}

\section{New variables}
\label{sec:newvar} 

The Lagrangian displayed in \rSec{sec:lme} is perfectly valid, and can be used to calculate physical predictions of SMEFT. 
However, it has a number of inconvenient features. 
In this subsection, we use equations of motion and field and couplings redefinitions to bring it into a more convenient form, where the physical interpretation of the Higgs basis Wilson coefficients is more transparent. 
We stress that this is purely cosmetic: the physical observables up to $\cO(1/\Lambda^2)$ in the EFT expansion are exactly the same, whether calculated with the fields and couplings in \rSec{sec:lme} or the ones in \rSec{sec:newvar}. 

Below we enumerate the redefinitions and briefly discuss motivations for each of them. 

{\bf \#1} In the SM, the hatted couplings $\hat g_s$,   $\hat g_L$,  $\hat g_Y$,  $\hat \lambda$, and the VEV $\hat v$  can be directly related to input observables. 
On that basis they can be assigned well-defined numerical values with error intervals. 
That is no longer true in the SMEFT. 
As can be seen in \rEq{eq:HL_masses} and \rEq{eq:HL_photon}, 
the presence of dimension-6 operators complicates the relation between the SM couplings and traditional input observables such as $m_Z$ or $\alpha$.  
For this reason it is convenient to introduce a new (unhatted) set of of couplings, related to the original couplings by 
\begin{eqnarray}  &  
\label{eq:UHL_shiftG}
\hat  g_s =  g_s \left (1 + \delta g_s \right ), \quad  \hat g_L =  g_L \left (1 + \delta g_L \right ), 
 \quad \hat g_Y = g_Y \left (1 + \delta g_Y \right ),
 & \nnl &   
 \hat v =  v \left (1 + \delta v \right ),   \quad  
 \hat  \lambda =  \lambda \left ( 1 + \delta \lambda \right ). 
\end{eqnarray} 
We choose the shifts as 
\begin{eqnarray} 
\delta g_s & = &  - {g_s^2\over 4} c_{gg}, 
 \nnl 
\delta g_L & = & 
- { g_L^4 \over 2 (g_L^2 -  g_Y^2) } c_{z \Box} 
-  { g_L^2(g_L^2 +  g_Y^2) \over 4 (g_L^2 - g_Y^2) } c_{z z}
+ {g_L^2  g_Y^4 \over 4 (g_L^4 -  g_Y^4) } c_{\gamma \gamma},
\nnl 
\delta g_Y & = & 
 { g_L^2 g_Y^2\over 2 (g_L^2 -  g_Y^2) } c_{z \Box} 
+ { g_Y^2(g_L^2 +  g_Y^2)  \over 4 (g_L^2 - g_Y^2) } c_{z z}
- {g_L^4  g_Y^2 \over 4 (g_L^4 -  g_Y^4) } c_{\gamma \gamma},
\nnl 
\delta v & = & 
{ g_L^4 \over 2 (g_L^2 -  g_Y^2) } c_{z \Box} 
+ { g_L^2   g_Y^2 \over 2 (g_L^2 - g_Y^2) } c_{z z}
- {g_L^4  g_Y^4 \over 2 (g_L^2 -  g_Y^2) (g_L^2 +  g_Y^2)^2 } c_{\gamma \gamma}
- { g_L^2   g_Y^2 \over 2 (g_L^2 + g_Y^2) } c_{z \gamma}
+ {1\over 2 } \Delta, 
\nnl 
\delta \lambda & = & 
{ 2 g_L^4 \over g_L^2 -  g_Y^2 } c_{z \Box} 
+ { 2 g_L^2   g_Y^2 \over g_L^2 - g_Y^2 } c_{z z}
- {2 g_L^4  g_Y^4 \over  (g_L^2 -  g_Y^2) (g_L^2 +  g_Y^2)^2 } c_{\gamma \gamma}
- { 2 g_L^2   g_Y^2 \over g_L^2 + g_Y^2 } c_{z \gamma}
\nnl && + {5 \over 2} \delta c_z  - {3 \over 2} \delta \lambda_3 
+ 2 \Delta. 
 \end{eqnarray} 
Recall that $\Delta \equiv \delta g^{We}_L + \delta g^{W\mu}_L  - {1\over 2} [c_{ll}]_{1221}$. 
In these new variables, the mass terms in \rEq{eq:HL_masses} simplify 
\begin{eqnarray}
\label{eq:UHL_masses}
m_h^2  =   2 \lambda v^2 , \qquad 
m_Z^2  =  {(g_L^2 + g_Y^2 ) v^2 \over 4}, 
\qquad 
m_W^2  =  {g_L^2 v^2 \over 4} \left ( 1 + \Delta  \right )  . 
\end{eqnarray} 
Furthermore, the gluon and photon couplings to matter in \rEq{eq:HL_gluon} and \rEq{eq:HL_photon} simplify as 
\begin{equation}
\cL \supset -  g_s G_\mu^a  \sum_{f \in u,d}  \bar f  \gamma_\mu T^a f  
 -  {g_L g_Y \over \sqrt{g_L^2 + g_Y^2 } }  A_\mu  \sum_{f \in u,d,e} Q_f \bar f  \gamma_\mu  f   . 
\end{equation} 
Finally, one can show that the Fermi constant measured in muon decay is related via $G_{F} = {1 \over \sqrt 2 v^2}$ to the parameter $v$ in \rEq{eq:UHL_shiftG}. 
All in all, the new parameter set $g_s$, $g_L$, $g_Y$, $v$, $\lambda$ introduced in \rEq{eq:UHL_shiftG}, 
at tree level,  is related to the input observables $\alpha_s$, $\alpha$, $m_Z$, $G_F$, $m_h$ in the same way as the corresponding parameters in the SM. 
With these new parameters the mass eigenstate Lagrangian simplifies considerably,   
and moreover they can be assigned the numerical values displayed in \rEq{eq:DEF_couplings}, which are independent of the dimension-6 Wilson coefficients up to $\cO({1 \over 16 \pi^2 \Lambda^2})$  corrections. 

{\bf \#2}
The cubic Higgs boson interactions in \rEq{eq:HL_h3} come in two forms:
one the familiar SM-like cubic coupling, and the other a 2-derivative interaction. 
However, the latter can be eliminated by a suitable choice of variables, 
after which its effect is absorbed into the former, and into other couplings involving two Higgs bosons and gauge bosons and/or fermions.  
To this end one can perform the following non-linear redefinition of the Higgs field: 
\begin{equation}
\label{eq:UHL_deltaH}
\hat h = h +  \delta_h \left ( h^2 + {1\over 3} h^3 \right ) , 
\end{equation} 
where 
\begin{eqnarray} 
\delta_h  &=& 
 {3 g_L^4 \over 2(g_L^2 -  g_Y^2) } c_{z \Box} 
+ { 3 g_L^2   g_Y^2 \over 2 (g_L^2 - g_Y^2) } c_{z z}
- {3 g_L^4  g_Y^4 \over  2 (g_L^2 -  g_Y^2) (g_L^2 +  g_Y^2)^2 } c_{\gamma \gamma}
- { 3 g_L^2   g_Y^2 \over 2 (g_L^2 + g_Y^2) } c_{z \gamma}
 \nnl  && 
+ \delta c_z +   {3 \over 2}  \Delta. 
\end{eqnarray}  
This eliminates {\em all} derivative Higgs boson self-interactions from the Lagrangian. 
In the new variable $h$ the self-interactions take the form 
\begin{equation}
\label{eq:UHL_lhself}
\cL \supset 
 - \lambda  v \left (1 + \delta \lambda_3 \right )   h^3  
 -  {\lambda \over 4} \left (1 + \delta \lambda_4 \right )  h^4 
 -  \lambda_5  {\lambda \over v} h^5  -  \lambda_6 {\lambda \over v^2} h^6 ,   
\end{equation}  
where 
\begin{equation}
\delta \lambda_4  = 6 \delta \lambda_3 - {4 \over 3} \delta c_z,
\quad
\lambda_5  = {3 \over 4} \delta \lambda_3 - {1 \over 4} \delta c_z,
\quad 
\lambda_6  = {1 \over 8} \delta \lambda_3 - {1 \over 24 } \delta c_z. 
\end{equation} 
We stress that, although the field redefinition in \rEq{eq:UHL_deltaH} changes the Lagrangian,  it 
does {\em not}  change on-shell S-matrix elements. 
More generally, on-shell S-matrix elements, whether tree- or loop-level, are not affected by general field redefinitions, even non-linear ones or non-gauge-invariant ones, as long as they satisfy certain minimal conditions~\cite{Arzt:1993gz}.   
Therefore, amplitudes for all Higgs boson production processes will be the same whether calculated with the Lagrangian in \rSec{sec:lme} using the field $\hat h$, or with the Lagrangian in this subsection using the field $h$.  

{\bf \#3} 
In the new variables introduced in this subsection, the $Z$ boson couplings  to charged leptons and quarks also simplify:  
\begin{equation}
\label{eq:UHL_vertexZ}
\cL \supset  - \sqrt{g_L^2 + g_Y^2} Z_\mu  
\sum_{f \in u, d,e}  \bar f   \gamma_\mu \left (T^3_f -  s_\theta^2 Q_f +  \delta g^{Zf}  \right)  f , 
\end{equation}  
where $s_\theta = g_Y/\sqrt{g_L^2 + g_Y^2}$. 
That is to say, the Wilson coefficients $\delta g^{Zf}$ in the Higgs basis, cf. \rEq{eq:DEF_dg}, are interpreted as {\em vertex corrections} to the $Z$ boson couplings as compared to the SM prediction.  
There is a similar kind of interaction in \rEq{eq:HL_contactVertex} which differs from \rEq{eq:UHL_vertexZ} by the presence of additional Higgs boson fields. 
Since it is already $\cO(\Lambda^{-2})$, it remains the same in the new variables, except for the trivial relabeling 
$\hat X \to X$. 
It is however possible to combine the vertex correction in  \rEq{eq:UHL_vertexZ} and  the Higgs interactions in  \rEq{eq:UHL_vertexZ} into one compact expression: 
\begin{equation}
\label{eq:UHL_vertexCombined}
\cL \supset  - \sqrt{g_L^2 + g_Y^2} Z_\mu  \left (1 + {h \over v} \right)^2  
\sum_{f \in u, d,e}  \bar f   \gamma_\mu  \delta g^{Zf}   f ,  
\end{equation} 
The motivation to do so is the following. 
The interactions in \rEq{eq:HL_contactVertex} are certainly relevant for the LHC Higgs phenomenology (in particular in the $H \to Z Z^*$ channel) and must be taken into account to correctly assess the parameter space. 
On the other hand, there are strong model independent constraints on the vertex corrections 
$\delta g^{Zf}$~\cite{Falkowski:2017pss}, at the level of $\cO(10^{-3})$ for the leptonic vertex correction.  
Such strongly suppressed vertex corrections will not be relevant for LHC Higgs phenomenology, where typical accuracy is  $\cO(10^{-1})$. 
Eliminating from  \rEq{eq:HL_contactVertex}  all terms {\em not} proportional to the vertex correction  $\delta g$ 
 offers users an option to ignore this class of interactions in the LHC context.  
In practice, the elimination can be achieved by adding to the Lagrangian the terms 
\begin{eqnarray}
\label{eq:UHL_eom}
\cL_{\rm eom} &  = & 
\left ( {2 h \over v} + {h^2 \over v^2} \right )
\bigg \{ 
x_{ZB} Z_\mu  \left [  \partial_\nu B_{\nu \mu} +  {i g_Y \over 2} H^\dagger  \overleftrightarrow {D_\mu} H  +  g_Y j_\mu^Y \right ] 
\nnl  &  & 
+ x_{ZW} Z_\mu  \left [ 
D_\nu W_{\nu \mu}^3  +  {i  \over 2} g_L H^\dagger \sigma^3  \overleftrightarrow {D_\mu} H +  g_L  j_\mu^3 \right ] 
 \nnl  &  &
+ \sum_{i=1}^2 x_{W} W_\mu^i  \left [ 
D_\nu W_{\nu \mu}^i  +  {i  \over 2} g_L H^\dagger \sigma^i  \overleftrightarrow {D_\mu} H +  g_L  j_\mu^i \right ] 
\bigg \} . 
\end{eqnarray}  
where $j_{\mu}^a$ and $j_{\mu}^Y$ are the fermionic currents coupled to the $SU(2)\times U(1)$ gauge bosons in the SM, and 
\begin{eqnarray}
x_{ZB} &= &  
 - { g_L^2 g_Y \sqrt{g_L^2 + g_Y^2} \over 2(g_L^2 -  g_Y^2) } c_{z \Box} 
-  { g_L^2   g_Y \sqrt{g_L^2 + g_Y^2}  \over 2 (g_L^2 - g_Y^2) } c_{z z}
+ { g_L^4  g_Y^3 \over  2 (g_L^2 -  g_Y^2) (g_L^2 +  g_Y^2)^{3/2} } c_{\gamma \gamma}
+ {  g_L^2   g_Y \over 2 \sqrt{g_L^2 + g_Y^2} } c_{z \gamma}, 
\nnl 
x_{ZW} &= &  
 - { g_L^3  \sqrt{g_L^2 + g_Y^2} \over 2(g_L^2 -  g_Y^2) } c_{z \Box} 
-  {  g_L g_Y^2 \sqrt{g_L^2 + g_Y^2}  \over 2 (g_L^2 - g_Y^2) } c_{z z}
+ { g_L^3  g_Y^4 \over  2 (g_L^2 -  g_Y^2) (g_L^2 +  g_Y^2)^{3/2} } c_{\gamma \gamma}
+ {  g_L   g_Y^2 \over 2 \sqrt{g_L^2 + g_Y^2} } c_{z \gamma},
\nnl 
x_{W} & = & 
  - { g_L^4  \over 2(g_L^2 -  g_Y^2) } c_{z \Box} 
-  {  g_L^2  g_Y^2  \over 2 (g_L^2 - g_Y^2) } c_{z z}
+ { g_L^4  g_Y^4 \over  2 (g_L^2 -  g_Y^2) (g_L^2 +  g_Y^2)^{2} } c_{\gamma \gamma}
+ {  g_L^2   g_Y^2 \over 2 (g_L^2 + g_Y^2) } c_{z \gamma} . 
\end{eqnarray} 
Note that each term in square brackets in \rEq{eq:UHL_eom} vanishes due to the SM equations of motion. 
Therefore adding it to the Lagrangian does not change S-matrix elements, whether at tree or at loop level.  
The role of \rEq{eq:UHL_eom} is to eliminate all $h^n V \bar f f$ contact interactions that are not proportional to the vertex corrections $\delta g$. 
Of course, the eliminated interactions do not vanish, but re-emerge in a different (more transparent) form. 
In this case, their effect on single Higgs processes is taken over by 2-derivative Higgs boson interactions with electroweak gauge bosons:  
\begin{equation}
\cL \supset 
c_{z \Box} g_L^2 Z_\mu \partial_\nu Z_{\mu \nu} 
+ c_{\gamma \Box} g_L g_Y Z_\mu \partial_\nu A_{\mu \nu}
+ c_{w \Box} g_L^2 \left (W_\mu^- \partial_\nu W_{\mu \nu}^+ + \hc \right )  , 
\end{equation} 
where $c_{z \Box}$ is already one of the Higgs basis Wilson coefficients, and the other two parameters can be expressed by the Wilson coefficients as  
\begin{eqnarray} 
 c_{w \Box}  & = &  {1 \over g_L^2 - g_Y^2} \left [ 
 g_L^2 c_{z \Box} + g_Y^2 c_{zz} 
 - {g_Y^2 (g_L^2 - g_Y^2) \over g_L^2 + g_Y^2} c_{z \gamma }
- { g_L^2  g_Y^4 \over  (g_L^2 +  g_Y^2)^2 } c_{\gamma \gamma}
\right ],  
\nnl 
 c_{\gamma \Box}  & = &  {1 \over g_L^2 - g_Y^2} \left [ 
 2 g_L^2 c_{z \Box} + (g_L^2 + g_Y^2) c_{zz} 
 - (g_L^2 - g_Y^2) c_{z \gamma }
- { g_L^2  g_Y^2 \over  g_L^2 +  g_Y^2 } c_{\gamma \gamma}
\right ]  . 
\end{eqnarray}

\vspace{1cm}

The change of variables and transformations described in  {\bf \#1},  {\bf \#2}, {\bf \#3} leads to a more convenient form of the mass eigenstate Lagrangian, 
in which the interpretation of various Higgs basis Wilson coefficients is more transparent. 
The Lagrangian in these variables is the one introduced in Ref.~\cite{deFlorian:2016spz}.  
However, we stress again that applying these transformations is a question of taste.  
Using the original variables and mass eigenstate Lagrangian from \rSec{sec:lme} would lead to the same amplitudes for all physical processes up to  $\cO(\Lambda^{-2})$, that is up to the maximum order for which our EFT is defined. 

\section{Final Lagrangian}
\label{sec:leff} 

In this section we summarize the SMEFT Lagrangian in the Higgs basis, rewritten in the variables introduced in \rSec{sec:newvar}.
This is the same Lagrangian as the one in Ref.~\cite{deFlorian:2016spz}, 
up to a change in sign and CKM conventions  to match those in  WCxF.  
The idea is to list here all terms that may be relevant for the current Higgs analyses at the LHC.  
The complete Lagrangian is available in a custom-made computer code, and any additional terms can be obtained on request. 

\subsubsection{Kinetic and mass terms}

The kinetic terms are diagonal and canonically normalized: 
\begin{equation} 
\cL \supset 
 {1 \over 2} \partial_\mu  h \partial_\mu  h 
 - {1 \over 2} W_{\mu \nu}^+  W_{\mu \nu}^- -   {1 \over 4} Z_{\mu \nu} Z_{\mu \nu} 
-   {1 \over 4} A_{\mu \nu} A_{\mu \nu}   - {1 \over 4} G_{\mu \nu}^a  G_{\mu \nu}^a
+  \sum_{f \in u,d,e,\nu}  i \bar f  \gamma_\mu \partial_\mu  f  . 
\end{equation}  
Here $h$, $W_\mu^\pm$, $Z_\mu$, $A_\mu$, $G_\mu^a$ and $f$ are respectively Higgs boson, $W$ boson, $Z$ boson, photon, gluon, and fermion fields. 

The mass terms are given by 
\begin{equation} 
\cL \supset 
- {1 \over 2} m_h^2 h^2 + m_W^2 W_{\mu}^+  W_{\mu}^-  + {1\over 2} m_Z^2  Z_\mu Z_\mu 
-  \sum_{f \in  u,d,e}  m_f  \bar f  f ,  
\end{equation} 
where  
\begin{eqnarray}
\label{eq:LEFF_masses}
m_h  & = &     \sqrt{2 \lambda} v ,
\nnl 
m_Z  & = &   {\sqrt{ g_L^2 + g_Y^2} v \over 2}, 
\nnl 
m_W  &= &   {g_L v \over 2} \left ( 1 + \delta m_w  \right )  ,  
\qquad \delta m_w = {1 \over 2} \delta g^{We}_L + {1 \over 2} \delta g^{W\mu}_L  - {1\over 4} [c_{ll}]_{1221}, 
\nnl 
m_f  & = &  {Y_f v \over \sqrt 2} \left ( 1 + \delta m_f \right ), 
\qquad  \delta m_{f_J} = {1\over 2} [\delta y_f]_{JJ} -   {1\over 2} \delta c_z . 
\end{eqnarray} 
Note that the neutrinos are treated as massless.

\subsubsection{Single Higgs couplings}

The single Higgs couplings to matter are given by 
\begin{eqnarray}
\label{eq:LEFF_higgs}
\cL & \supset& {h \over v} \left [ 
 (1 + \delta c_w)   {g_L^2 v^2 \over 2} W_\mu^+ W_\mu^-
  +   (1 + \delta c_z)  {(g_L^2 + g_Y^2) v^2 \over 4} Z_\mu Z_\mu
\right . \nn &- & \left . 
\sum_{f \in u,d,e} \sum_{IJ}    
\sqrt{m_{f_I} m_{f_J}} 
\left [  \left (\delta_{IJ} + [\delta y_f]_{IJ} \right ) \bar f_L  f_R  + \hc   \right ] 
\right . \nn & & \left .
+ c_{ww}  {g_L^2 \over  2} W_{\mu \nu}^+  W_{\mu\nu}^-  + \tilde c_{ww}  {g_L^2 \over  2} W_{\mu \nu}^+   \tilde W_{\mu\nu}^- 
+ c_{w \Box} g_L^2 \left (W_\mu^- \partial_\nu W_{\mu \nu}^+ + \hc \right )  
\right . \nn & & \left . 
+  c_{gg} {g_s^2 \over 4} G_{\mu \nu}^a G_{\mu \nu}^a   
+ c_{\gamma \gamma} {g_L^2 g_Y^2 \over 4 (g_L^2 + g_Y^2) }  A_{\mu \nu} A_{\mu \nu} 
+ c_{z \gamma} {g_L g_Y  \over  2} Z_{\mu \nu} A_{\mu\nu} 
+ c_{zz} {g_L^2 + g_Y^2 \over  4} Z_{\mu \nu} Z_{\mu\nu}
\right . \nn & & \left . 
+c_{z \Box} g_L^2 Z_\mu \partial_\nu Z_{\mu \nu} 
+ c_{\gamma \Box} g_L g_Y Z_\mu \partial_\nu A_{\mu \nu}
\right . \nn & & \left . 
+  \tilde c_{gg} {g_s^2 \over 4} G_{\mu \nu}^a \tilde G_{\mu \nu}^a  
+ \tilde c_{\gamma \gamma} {g_L^2 g_Y^2 \over 4 (g_L^2 + g_Y^2) } A_{\mu \nu} \tilde A_{\mu \nu} 
+ \tilde c_{z \gamma} {g_L g_Y \over  2} Z_{\mu \nu} \tilde A_{\mu\nu}
+ \tilde c_{zz}  {g_L^2 + g_Y^2  \over  4} Z_{\mu \nu} \tilde Z_{\mu\nu}
\right ].  
\nnl 
 \end{eqnarray}
Most of the EFT parameters above are identical to the  Higgs basis Wilson coefficients, as defined in \rSec{sec:def}. 
The remaining EFT parameters can be expressed by the Wilson coefficients as  
\begin{eqnarray} 
\label{eq:LEFF_cwHB}
\delta c_w & =& 
\delta c_z  + 4 \delta m_w, \qquad 
 \delta m_w \equiv  {1 \over 2} \delta g^{We}_L + {1 \over 2} \delta g^{W\mu}_L  - {1\over 4} [c_{ll}]_{1221},
\nnl 
c_{ww} & = & c_{zz} +  {2  g_Y^2 \over  g_L^2 +  g_Y^2} c_{z \gamma} 
 + { g_Y^4 \over (g_L^2 + g_Y^2)^2 } c_{\gamma \gamma} , 
\nnl 
\tilde c_{ww} & = & \tilde  c_{zz} +  {2  g_Y^2 \over g_L^2 +  g_Y^2} \tilde  c_{z \gamma} 
 + {  g_Y^4 \over ( g_L^2 + g_Y^2)^2 } \tilde  c_{\gamma \gamma} , 
 \nnl 
 c_{w \Box}  & = &  {1 \over g_L^2 - g_Y^2} \left [ 
 g_L^2 c_{z \Box} + g_Y^2 c_{zz} 
 - {g_Y^2 (g_L^2 - g_Y^2) \over g_L^2 + g_Y^2} c_{z \gamma }
- { g_L^2  g_Y^4 \over  (g_L^2 +  g_Y^2)^2 } c_{\gamma \gamma}
\right ]  , 
\nnl 
 c_{\gamma \Box}  & = &  {1 \over g_L^2 - g_Y^2} \left [ 
 2 g_L^2 c_{z \Box} + (g_L^2 + g_Y^2) c_{zz} 
 - (g_L^2 - g_Y^2) c_{z \gamma }
- { g_L^2  g_Y^2 \over  g_L^2 +  g_Y^2 } c_{\gamma \gamma}
\right ]  . 
\end{eqnarray}

\subsubsection{Higgs self-interactions} 
 
 The Higgs boson self-interactions take the form 
 \begin{equation}
\label{eq:LEFF_lhself}
\cL \supset 
 - \lambda  v \left (1 + \delta \lambda_3 \right )   h^3  
 -  {\lambda \over 4} \left (1 + \delta \lambda_4 \right )  h^4 
 -  \lambda_5  {\lambda \over v} h^5  -  \lambda_6 {\lambda \over v^2} h^6 ,   
\end{equation}  
where $\delta \lambda_3$ is one of the Wilson coefficients in the Higgs basis, 
and the remaining EFT parameters can be expressed by the Wilson coefficients as  
\begin{equation}
\delta \lambda_4  = 6 \delta \lambda_3 - {4 \over 3} \delta c_z,
\quad
\lambda_5  = {3 \over 4} \delta \lambda_3 - {1 \over 4} \delta c_z,
\quad 
\lambda_6  = {1 \over 8} \delta \lambda_3 - {1 \over 24 } \delta c_z. 
\end{equation} 
 
\subsubsection{Gauge interactions} 
 
The gauge interactions have the form 
\begin{eqnarray}
\label{eq:LEFF_lvff}
\hspace{-1cm}
\cL & \supset&
- {g_L g_Y \over \sqrt{g_L^2 + g_Y^2} } A_\mu \sum_{f \in u,d,e} Q_f   \bar f  \gamma_\mu  f
-  g_s G_\mu^a \sum_{f \in u,d} \bar f  \gamma_\mu T^a  f , 
\nnl  & - &  
{g_L \over \sqrt 2}  \left ( 
W_\mu^+  \bar \nu_L  \gamma_\mu ( {\bf I} + \delta g^{W\ell}_L)  e_L  
+ W_\mu^+ \bar u_L \gamma_\mu (V_{\rm CKM } + \delta g^{Wq}_L ) d_L 
+  W_\mu^+  \bar u_R \gamma_\mu \delta g^{Wq}_R  d_R  + \hc \right )
\nn &- & \sqrt{g_L^2 + g_Y^2} Z_\mu   \left [  
   \sum_{f \in u, d,e,\nu}  \bar f_L  \gamma_\mu ( T^3_f - s_\theta^2 Q_f  +   \delta g^{Zf}_L)  f_L
+    \sum_{f \in u, d,e}   \bar f_R  \gamma_\mu (  - s_\theta^2 Q_f  +  \delta g^{Zf}_R) f_R   \right ], 
\nnl
\end{eqnarray}
where all the departures from the SM couplings are parametrized by the vertex corrections $\delta g^{Vf}$. 
Note that, by construction, there are no vertex corrections to photon and gluon couplings.    
Most of the vertex corrections are Wilson coefficients of the Higgs basis, as defined in \rSec{sec:def}. 
The remaining EFT parameters can be expressed by the Wilson coefficients as  
\begin{eqnarray}
\label{eq:LEFF_dgrelations}
\delta g^{Z\nu}_L  &=& \delta g^{W\ell}_L  +  \delta g^{Z e}_L,  
\nnl 
\delta g^{Wq}_L  &=& V_{\rm CKM}^\dagger \delta g^{Zu}_L V_{\rm CKM} -   \delta g^{Zd}_L .  
\end{eqnarray}

\subsubsection{Dipole interactions} 

The dipole-type interactions take the form 
\begin{eqnarray}
\label{eq:LEFF_dipole}
\cL & \supset& 
- {1 + h/v \over 4 v}  \left [ 
g_s \sum_{f \in u,d}    {\sqrt{m_{f_I} m_{f_J}} \over v} 
\bar f_I  \sigma_{\mu \nu} T^a   [d_{Gf}]_{IJ}  P_L  f_J  G_{\mu\nu}^a 
\right . \nnl && \left .  
+ {g_L g_Y \over  \sqrt{g_L^2 + g_Y^2}  }  \sum_{f \in u,d,e}   {\sqrt{m_{f_I} m_{f_J}} \over v} 
\bar f_I   \sigma_{\mu \nu}  [d_{A f}]_{IJ}  P_L   f_J   A_{\mu\nu} 
\right . \nnl && \left . 
+ \sqrt{g_L^2 + g_Y^2}  \sum_{f \in u,d,e}   {\sqrt{m_{f_I} m_{f_J}} \over v} 
\bar f_I    \sigma_{\mu \nu}  [d_{Zf}]_{IJ}  P_L  f_J   Z_{\mu\nu}
\right . \nnl && \left . 
+ \sqrt {2} g_L    {\sqrt{m_{u_I} m_{u_J}} \over v}   
\bar u_I   \sigma_{\mu \nu}  [d_{Wu}]_{IJ}    P_L   d_J  W_{\mu\nu}^+ 
+    \sqrt {2} g_L    {\sqrt{m_{d_I} m_{d_J}} \over v}  
\bar d_I  \sigma_{\mu \nu}  [d_{Wd}]_{IJ}  P_L   u_J W_{\mu\nu}^-   
   \right . \nnl && \left . 
+ \sqrt {2} g_L  {\sqrt{m_{e_I} m_{e_J}} \over v}   
\bar e_I   \sigma_{\mu \nu}  [d_{We}]_{IJ} P_L  \nu_J  W_{\mu\nu}^- 
+ \hc  
  \right ],    
\end{eqnarray} 
where $\sigma_{\mu\nu}  = {i\over 2} \left ( \gamma_\mu \gamma_\nu -  \gamma_\nu \gamma_\mu \right )$, and  $d_{G f}$, $d_{A f}$,  $d_{Z f}$,  and $d_{W f}$  are complex $3 \times 3$ matrices.  
Out of these, 
$d_{G d}$,  $d_{A f}$, and  $d_{Z f}$  are already Wilson coefficients of the Higgs basis, as defined in \rSec{sec:def}. 
The remaining dipole parameters can be expressed by the Wilson coefficients as  
\begin{equation}
\eta_f d_{Wf} = d_{Zf} + {g_Y^2 \over g_L^2 + g_Y^2} d_{Af}, 
\end{equation} 
where $\eta_u = +1$, $\eta_{d,e} = -1$. 

\subsubsection{Contact Higgs-gauge-fermion interactions} 

The contact Higgs-gauge-fermion interactions have the form
\begin{eqnarray} 
\label{eq:LEFF_hvff}
\cL & \supset&
- \sqrt 2 g_L \left ( {h \over v} + {h^2 \over 2 v^2} \right ) \left ( 
W_\mu^+  \bar \nu_L \gamma_\mu  \delta g^{W\ell}_L  e_L  
+ W_\mu^+ \bar u_R  \gamma_\mu \delta g^{Wq}_L  d_R  
+  W_\mu^+  \bar u_R  \gamma_\mu \delta g^{Wq}_R  d_R  + \hc \right )
\nn &-& 2 \sqrt{g_L^2 + g_Y^2}   \left ( {h \over v} + {h^2 \over 2 v^2} \right )  Z_\mu   \left [  
   \sum_{f \in u, d,e,\nu}  \bar f_L \gamma_\mu  \delta g^{Zf}_L  f_L 
+    \sum_{f \in u, d,e}  \bar f_R  \gamma_\mu   \delta g^{Zf}_R f_R   \right ]. 
\nnl  
\end{eqnarray} 
The EFT parameters describing these interactions are completely fixed by the vertex corrections discussed in the previous subsection.

\subsubsection{Triple gauge couplings} 

The triple gauge couplings of electroweak gauge bosons are customarily parametrized as~\cite{Hagiwara:1993ck} 
 \begin{eqnarray}   
\label{eq:LEFF_tgc}
\hspace{-1cm}
 \cL  & \supset &  - i {g_L \over \sqrt{g_L^2 + g_Y^2} } \bigg \{ 
     g_Y \left ( W_{\mu \nu}^+ W_\mu^-  -  W_{\mu \nu}^- W_\mu^+ \right ) A_\nu  
  +  g_L  g_{1,z} \left ( W_{\mu \nu}^+ W_\mu^-  -  W_{\mu \nu}^- W_\mu^+ \right ) Z_\nu 
\nnl & +  & 
 g_Y \kappa_\gamma W_\mu^+W_\nu^-   A_{\mu\nu} 
+ g_L \kappa_z W_\mu^+W_\nu^- Z_{\mu\nu}
+g_Y  \tilde \kappa_\gamma  W_\mu^+W_\nu^-   \tilde A_{\mu\nu}
+ g_L  \tilde \kappa_z  W_\mu^+W_\nu^- \tilde Z_{\mu\nu}
  \nnl   &+  &      
    { \lambda_\gamma  g_Y \over m_W^2 }  W_{\mu \nu}^+W_{\nu \rho}^- A_{\rho \mu} 
+  { \lambda_z  g_L \over m_W^2 }   W_{\mu \nu}^+W_{\nu \rho}^- Z_{\rho \mu}    
+  { \tilde \lambda_\gamma  g_Y  \over m_W^2 }  W_{\mu \nu}^+W_{\nu \rho}^-  \tilde A_{\rho \mu}  
 +   {  \tilde\lambda_z  g_L \over m_W^2 }   W_{\mu \nu}^+W_{\nu \rho}^-  \tilde Z_{\rho \mu}  
 \bigg \} . 
  \nnl 
\end{eqnarray} 
Above, only $\lambda_z$ and $\tilde \lambda_z$   are Wilson coefficients in the Higgs basis, as defined in \rSec{sec:def}. 
The remaining EFT parameters can be expressed by the Wilson coefficients as  
\begin{eqnarray}
\label{eq:LEFF_tgcpar}
g_{1,z} & = &  1 + {1 \over 2 (g_L^2 - g_Y^2) } \left [ 
-  g_L^2 (g_L^2 + g_Y^2) c_{z \Box} 
 -  g_Y^2 (g_L^2 + g_Y^2) c_{zz} 
 +  g_Y^2 (g_L^2 - g_Y^2)c_{z \gamma }
+  { g_L^2  g_Y^2 \over  g_L^2 +  g_Y^2 } c_{\gamma \gamma} \right ], 
\nnl 
\kappa_z & = &  1 + {1 \over  g_L^2 - g_Y^2 } \left [ 
-  {g_L^2 (g_L^2 + g_Y^2) \over 2}  c_{z \Box} 
 -  g_L^2  g_Y^2 c_{zz} 
 +  {g_L^2 g_Y^2 (g_L^2 - g_Y^2) \over g_L^2 + g_Y^2} c_{z \gamma }
+  { g_L^4  g_Y^4 \over  (g_L^2 +  g_Y^2)^2 } c_{\gamma \gamma} \right ], 
\nnl 
\kappa_\gamma & = &  1 + {g_L^2 \over 2  } \left [ 
 c_{zz} 
-  {g_L^2 - g_Y^2 \over g_L^2 + g_Y^2} c_{z \gamma }
-   { g_L^2  g_Y^2 \over  (g_L^2 +  g_Y^2)^2 } c_{\gamma \gamma} 
\right ] ,
\nnl 
\tilde \kappa_\gamma & = & {g_L^2 \over 2  } \left [ 
 \tilde c_{zz} 
-  {g_L^2 - g_Y^2 \over g_L^2 + g_Y^2} \tilde c_{z \gamma }
-   { g_L^2  g_Y^2 \over  (g_L^2 +  g_Y^2)^2 } \tilde c_{\gamma \gamma} 
\right ] ,
\nnl 
\tilde \kappa_z & = &  - {g_Y^2 \over 2  } \left [ 
 \tilde c_{zz} 
-  {g_L^2 - g_Y^2 \over g_L^2 + g_Y^2} \tilde c_{z \gamma }
-   { g_L^2  g_Y^2 \over  (g_L^2 +  g_Y^2)^2 } \tilde c_{\gamma \gamma} 
\right ] ,
\nnl 
\lambda_\gamma & = & \lambda_z, \qquad \tilde \lambda_\gamma  =  \tilde \lambda_z  . 
\end{eqnarray} 
One can verify that these expressions lead to the usual dimension-6 SMEFT relations between the triple gauge couplings: 
$\kappa_z  =   g_{1,z}   - {g_Y^2 \over g_L^2} (\kappa_\gamma -1)$,  
$\tilde  \kappa_z  =    - {g_Y^2 \over g_L^2} \tilde \kappa_\gamma$.

\subsubsection{Double Higgs couplings to matter} 

The  interactions between two Higgs bosons and two other SM fields  are given by 
\begin{eqnarray} 
\label{eq:LEFF_lhh}
\hspace{-2cm}
 \cL  & \supset &  
h^2   \bigg \{ 
 \left (1 +   \delta c_z^{(2)} \right ) {g_L^2 + g_Y^2 \over 8}  Z_\mu Z_\mu  
+   \left (1 +  \delta c_w^{(2)} \right )  {g_L^2 \over 4} W_\mu^+  W_\mu^-
\nnl & - & 
 {1 \over 2 v^2} \sum_{f} \sqrt{m_{f_J} m_{f_K}} 
\left [ \bar f_{J}  [y_f^{(2)}]_{JK}   P_L f_{K} + {\mathrm h.c.} \right ] 
\nonumber \\  &+& 
 {1 \over 8 v^2}  \bigg  [ c_{gg}^{(2)}  g_s^2 G_{\mu \nu}^a G_{\mu \nu}^a  +   \tilde c_{gg}^{(2)}  g_s^2 G_{\mu \nu}^a \tilde G_{\mu \nu}^a  \bigg ] 
\nonumber \\  &+& 
 {1 \over 8 v^2}   \left ( 2 c_{ww}^{(2)} g_L^2 W_{\mu \nu}^+ W_{\mu \nu}^-   
 + c_{zz}^{(2)}  (g_L^2 + g_Y^2) Z_{\mu \nu} Z_{\mu \nu}   
 +  2 c_{z\gamma}^{(2)} g_L g_Y Z_{\mu \nu} A_{\mu \nu}   
  +  c_{\gamma \gamma}^{(2)} {g_L^2 g_Y^2 \over g_L^2 + g_Y^2 }   A_{\mu \nu} A_{\mu \nu} 
 \right )
\nonumber \\  &+& 
{1 \over 8 v^2} \left ( 
2   \tilde c_{ww}^{(2)} g_L^2 W_{\mu \nu}^+ \tilde W_{\mu \nu}^-   
+   \tilde c_{zz}^{(2)} (g_L^2 + g_Y^2)Z_{\mu \nu}  \tilde Z_{\mu \nu}   
+  2   \tilde c_{z\gamma}^{(2)} g_L g_Y Z_{\mu \nu} \tilde A_{\mu \nu}    
+    \tilde c_{\gamma \gamma}^{(2)} {g_L^2 g_Y^2 \over g_L^2 + g_Y^2 }  A_{\mu \nu} \tilde A_{\mu \nu}  \right ) 
\nonumber \\ & + &  
 {1 \over 2 v^2} \left  ( g_L^2 c_{w \Box}^{(2)} (W_\mu^+ \partial_\nu W_{\mu \nu}^- + W_\mu^- \partial_\nu W_{\mu \nu}^+ ) 
+ g_L^2  c_{z \Box}^{(2)} Z_\mu \partial_\nu Z_{\mu \nu} 
+ g_L g_Y  c_{\gamma \Box}^{(2)} Z_\mu \partial_\nu A_{\mu \nu}  \right )
\bigg \} .
\end{eqnarray}  
The parameters above are related to the Wilson coefficients in the Higgs basis as 
\begin{eqnarray}
\delta c_z^{(2)} &=& 4 \delta c_z, \qquad   \delta c_w^{(2)} = 4 \delta c_z + 6 \Delta, \quad 
\nonumber \\ 
\,  [y_f^{(2)}]_{JK}   & = &  3  [\delta y_f]_{JK} -  \delta c_z \, \delta_{JK}, 
\nonumber \\ 
c_{vv}^{(2)}  & =& c_{vv}, \qquad \tilde c_{vv}^{(2)}  = \tilde  c_{vv}, \qquad v \in \{g,w,z,\gamma \},   
\nonumber \\
c_{v\Box}^{(2)}  & =& c_{v\Box},  \qquad v \in \{w,z,\gamma \}, 
 \end{eqnarray}
 where the expressions of $c_{ww}$, $\tilde c_{ww}$, $c_{z\Box}$ and $ c_{\gamma \Box}$ 
 in terms of the Higgs basis Wilson coefficients are given in \rEq{eq:LEFF_cwHB}.

\section{Discussion}
\label{sec:disc} 

We close this note with a number of scattered comments. 
\begin{itemize}
\item 
Using the Warsaw or Higgs basis is entirely a matter of convenience, and leads to fully equivalent results at $\cO(1/\Lambda^2)$ in the EFT expansion. 
One can verify these statements for any particular process, by calculating it in both bases, and comparing the results using the map in \rEq{eq:DEF_map}.  
Moreover, using the Higgs basis with the Lagrangian in \rSec{sec:lme}, 
or the one with redefined fields and couplings in \rSec{sec:leff} leads to the same  results at $\cO(1/\Lambda^2)$. 
\item 
The Higgs basis is designed to be convenient for the characterization of Higgs processes at the LHC.  
It does not mean it is convenient for {\em any} application. 
One counterexample is diboson production. 
In the Higgs basis, the cubic CP-even  electroweak gauge couplings  are described  by five parameters  
$c_{zz}$, $c_{z\gamma}$, $c_{\gamma \gamma}$, $c_{z \Box}$, and $\lambda_z$, see \rEq{eq:LEFF_tgcpar}. 
In diboson analyses it is more convenient to use the standard TGC parametrization in terms of $g_{1,z}$, $\kappa_\gamma$ and $\lambda_z$, and only a-posteriori translate the results  to the Higgs basis using   \rEq{eq:LEFF_tgcpar}. 
One could in fact construct another basis, call it {\em Higgs-TGC basis}, where 
$\delta g_{1,z} \equiv  g_{1,z}-1$ and $\delta \kappa_\gamma \equiv \kappa_\gamma -1$ are defined as Wilson coefficients, at the expense of two Wilson coefficients from the original Higgs basis, e.g. $c_{zz}$ and $c_{z \Box}$.  
Similarly, for the analysis of $h \to WW^*$ alone, 
one would rather use the $\delta c_w$, $c_{ww}$, $\tilde c_{ww}$, $c_{w \Box}$ variables, 
and only later translate to the Higgs basis using  \rEq{eq:LEFF_cwHB}. 
Again, for this purpose one could also  construct a new basis, call it  {\em Higgs-WW basis}, where  
$\delta c_w$, $c_{ww}$, $\tilde c_{ww}$, $c_{w \Box}$ are the Wilson coefficients, 
at the expense of e.g. $\delta c_z$, $c_{zz}$, $\tilde c_{zz}$,  $c_{z \Box}$. 
 \item 
To reduce the number of free parameters in a SMEFT analysis of Higgs observables one may take advantage of the fact that some combinations of Wilson coefficients are strongly constrained by other precision measurements, notably by the electroweak data from LEP-1. 
A nice feature of the Higgs basis is that it separates the Wilson coefficients affecting only the Higgs observables at tree level (the ones in \rEq{eq:DEF_higgs}) from those affecting also electroweak precision measurements and thus being strongly constrained (the ones in \rEq{eq:DEF_dg}).  
In particular, all leptonic, bottom and charm vertex corrections $\delta g$ are constrained at a level of $10^{-2}$ or better~\cite{Efrati:2015eaa}. 
The current experimental sensitivity at the LHC is not sufficient to probe the effect of these vertex corrections on the Higgs observables, thus for all practical purposes one can simply set these $\delta g$ to zero when analyzing  the LHC Higgs data.
Similarly, $\delta m_w$ is very strongly constrained by $W$ mass measurements, and thus can be set to zero.    
This greatly reduces the number of Wilson coefficients that a typical Higgs analysis has to deal with. 
\item 
Some caution regarding  the point above has to be exercised, however.
First, not all vertex corrections and dipole couplings have been strongly constrained by prior non-Higgs measurements. 
For example, the vertex corrections to the $Z t \bar t$ couplings, $\delta g_{L,R}^{Zt}$, for obvious reasons are not constrained by LEP-1, and thus they should not be neglected whenever they contribute to Higgs observables.  
Moreover, for some of the observables the effect of strongly constrained parameters may be amplified at the LHC.
One such case was identified in \cite{Zhang:2016zsp}. 
In the case of diboson production at the LHC, the effect of the light quark vertex correction is enhanced  by the factor $s/v^2$ at high invariant diboson mass $\sqrt{s}$, and is hence not negligible.  
\end{itemize}


\chapter[Short notes on the SMEFT]{Short notes on the SMEFT\footnote{contributed by E.\ Salvioni}\label{ch:shortnotes}}

\section{Higgs basis with additional constraint \label{sec:higgs-basis-constraint}}

Imposing an additional constraint is motivated by the desire to reduce the number of degrees of freedom in the EFT fit as well as a simplification of the mapping between the Wilson coefficients of two SMEFT bases, for instance the Higgs (or \p{JHUGen}) basis and the Warsaw basis.  Constraints that can be related to experimental bounds or an exact, approximate or assumed symmetry of the Lagrangian are preferable to ad hoc constraints.  It is important to keep in mind that such a constraint is not part of the definition of the SMEFT basis, but rather an optional choice made at the analysis level.

The constraint $\delta m = 0$, see (\ref{eq:1}), is directly motivated by
experiment, as $W$ mass measurements imply $|\delta m| < 10^{-3}$.
This constraint is a well motivated choice in the context of Higgs searches.

For example, by choosing to impose $\delta m = 0$ for $gg\to ZZ$ the experimental analysis will constrain the Higgs basis coefficients \{ $c_{gg}$, $\delta c_z$, $c_{z\Box}$, $c_{zz}$ \} in the $CP$-even sector, since the photon does not enter that particular process.  When the resulting likelihood is translated to the Warsaw basis, $\delta m = 0$ will constrain one linear combination of the 5 relevant Warsaw basis operators.

Setting $\delta v = 0$, see (\ref{eq:deltam}), is another possible constraint.  But, experimental constraints on that combination of coefficients are somewhat weaker and more correlated with other electroweak constraints.

Custodial symmetry is not an exact symmetry even in the SM.  However, the custodial limit is often defined by setting the Warsaw basis Wilson coefficient $c_{HD} = 0$.  Some authors also motivate the choice $\delta m = 0$ with custodial symmetry considerations, see \cite{deFlorian:2016spz}, p.\ 299.

\section{Relation between Higgs and Warsaw bases \label{sec:rel-higgs-warsaw}}
Start from the 15 couplings in (II.2.20) of \cite{deFlorian:2016spz}, supplemented by the correction to the $W$ mass $\mathcal{L}\supset 2 \delta m \tfrac{g^2 v^2}{4} W_\mu^+ W^{-\,\mu}$. Consider first the 10 $CP$ even couplings. In a dimension 6 EFT, four coefficients can be expressed as functions of the others, see (II.2.38) there. The remaining $6$ coefficients $\{ c_{gg}, \delta c_z, c_{z\Box}, c_{zz}, c_{z\gamma}, c_{\gamma\gamma} \}$ as well as $\delta m$ are independent, and here we express them in terms of the coefficients of the Warsaw basis, whose relevant operators are defined as
{\footnotesize
\begin{align}
v^2 \mathcal{L}_6^{\rm even} \,=&\, c_{GG} H^\dagger H G_{\mu \nu}^a G^{a\,\mu\nu} + c_{WW} H^\dagger H W_{\mu\nu}^i W^{i\,\mu\nu} + c_{WB} \, H^\dagger \sigma^i H W_{\mu\nu}^i B^{\mu\nu} + c_{BB} H^\dagger H B_{\mu\nu} B^{\mu\nu} \\
 +&\, c_{H\Box}  (H^\dagger H) \Box (H^\dagger H) + c_{HD} |H^\dagger D_\mu H|^2 + \sum_{k = 1,2} i (c_{H\ell}^{(3)})_{kk} (\bar{\ell} \sigma^i \gamma^\mu \ell)_{kk} (H^\dagger \sigma^i \hspace{-1.5mm} \stackrel{\leftrightarrow}{D_\mu} \hspace{-1mm} H )  + (c_{\ell\ell})_{1221} (\bar{\ell} \gamma^\mu \ell \,\bar{\ell} \gamma_\mu \ell)_{1221} \,. \nonumber 
\end{align}
}
Here we follow the notation of \cite{Falkowski:2015wza}, but not exactly: $c_{H\Box} = - c_H - c_T $ and $ c_{HD} = - 4\, c_T $. We do this to work with the proper Warsaw basis, where $c_{H\Box}, c_{HD}$ appear rather than $c_H, c_T$. We also normalize the operators as in the original Warsaw basis, except for setting $\Lambda = v$. The convention is $v \simeq 246$~GeV.

The relations are (see Eqs.~(A.1) and (A.9) to (A.11) of \cite{Falkowski:2015wza})
{\scriptsize
\begin{equation*}
c_{zz} =4\, \frac{ g^2 c_{WW} + g g^{\prime} c_{WB} + g^{\prime\,2} c_{BB}}{(g^2 + g^{\prime\,2})^2} \qquad c_{z\gamma} = 2\, \frac{2 c_{WW} - \frac{g^2 - g^{\prime\,2}}{g g'} \,c_{WB} - 2 c_{BB}}{g^2 + g^{\prime\,2}} \qquad c_{\gamma\gamma} = 4 \Big( \frac{ c_{WW}}{g^2} - \frac{c_{WB}}{g g'}  + \frac{c_{BB}}{g^{\prime\,2}}   \Big)
\end{equation*}

\begin{equation} \label{eq:1}
c_{gg} = \frac{4}{g_s^2} c_{GG} \qquad c_{z\Box} = \frac{c_{HD} + 4 \delta v}{2 g^2} \qquad  \delta c_z = c_{H\Box} - \frac{1}{4} c_{HD} - 3 \delta v \qquad \delta m = \frac{g^2}{g^2 - g^{\prime\,2}} \left\{ - \frac{g^{\prime}}{g} c_{WB} - \frac{1}{4} c_{HD} - \frac{g^{\prime\,2}}{g^2} \delta v  \right\}
\end{equation}
}
where the $\delta v$ combination of Warsaw basis operators is
\begin{equation} \label{eq:deltam}
\delta v \equiv  \frac{1}{2} \Big( (c_{H\ell}^{(3)})_{11} + (c_{H\ell}^{(3)})_{22} \Big) - \frac{1}{4} (c_{\ell\ell})_{1221}\,.
\end{equation}
The above transformations have been explicitly verified using Rosetta.\footnote{Note that for the numerical example of \cite{Falkowski:2015wza} we find that it is $\delta m$ that vanishes, and not $\delta v \approx 0.0329$, contrarily to what was stated there. Thanks to Ken Mimasu for confirming this is a typo in their paper.} They can be inverted algebraically, yielding $\{c_{GG}, c_{WW}, c_{WB}, c_{BB}, c_{HD}, c_{H\Box}, \delta v\}$ as functions of \\
$\{c_{gg}, c_{zz}, c_{z\gamma}, c_{\gamma\gamma}, c_{z\Box},  \delta c_z , \delta m \}$. Custodial symmetry corresponds to $\delta m = 0\,$.

In the $CP$ odd sector, one coefficient can be expressed as function of the others in a dimension 6 EFT, leaving $4$ independent coefficients $\{ \tilde{c}_{zz}, \tilde{c}_{z\gamma}, \tilde{c}_{\gamma\gamma}, \tilde{c}_{gg} \}$. The Warsaw basis Lagrangian is
\begin{align}
v^2 \mathcal{L}_6^{\rm odd} \,=&\, \tilde{c}_{GG} H^\dagger H G_{\mu \nu}^a \widetilde{G}^{a\,\mu\nu} + \tilde{c}_{WW} H^\dagger H W_{\mu\nu}^i \widetilde{W}^{i\,\mu\nu} + \tilde{c}_{WB} \, H^\dagger \sigma^i H \widetilde{W}_{\mu\nu}^i B^{\mu\nu} + \tilde{c}_{BB}  H^\dagger H B_{\mu\nu} \widetilde{B}^{\mu\nu} 
\end{align}
and the transformations are identical to the $CP$ even counterparts (see (A.12) of \cite{LHCHXSWG-INT-2015-001}),
{\scriptsize
\begin{equation}
 \tilde{c}_{zz} = 4\, \frac{ g^2 \tilde{c}_{WW} +  g g^{\prime} \tilde{c}_{WB}  +  g^{\prime\,2} \tilde{c}_{BB}  }{(g^2 + g^{\prime\,2} )^2} \qquad \tilde{c}_{z\gamma} = 2\, \frac{2 \tilde{c}_{WW}  - \frac{g^2 - g^{\prime\,2}}{g g'}\, \tilde{c}_{WB} - 2 \tilde{c}_{BB} }{g^2 + g^{\prime\,2}} \qquad  \tilde{c}_{\gamma\gamma} = 4 \Big( \frac{\tilde{c}_{WW}}{g^2} - \frac{\tilde{c}_{WB}}{g g'} + \frac{\tilde{c}_{BB}  }{g^{\prime\,2}} \Big)
\end{equation}
}
and $\tilde{c}_{gg} = \frac{4}{g_s^2} \tilde{c}_{GG}$. Again these relations can be inverted algebraically.


\chapter[Effective Field Theory calculations and tools]{Effective Field Theory calculations and tools\footnote{contributed by E.\ Vryonidou and A.V.\ Gritsan}\label{ch:tools}}

The success of the Standard Model Effective Theory (SMEFT) programme
at the LHC relies on the availability of public tools for calculations in this framework. Among the most important of these are Monte Carlo (MC) tools for providing realistic predictions for collider processes both for phenomenological studies and experimental analyses. 
In this respect, significant efforts have been made to implement the effects of dimension-6 operators in MC event generators. 
\p{SMEFTsim} \cite{Brivio:2017btx} is a complete implementation of the dimension-6 operators in the Warsaw basis that can be used to simulate arbitrary processes in the \p{SMEFT} at tree level.  The Higgs couplings to $gg$, $\gamma\gamma$ and $Z\gamma$ are also implemented at tree level via effective vertices in the infinite top mass limit.  Other tools for leading order (LO) predictions include an alternative implementation of the Warsaw basis in the $R_{\xi}$ gauge \cite{Dedes:2017zog}, \p{dim6top}, an implementation of top quark operators under various flavour assumptions \cite{AguilarSaavedra:2018nen} and the Higgs Effective Lagrangian (HEL) \cite{Alloul:2013naa} implementation of SILH basis operators. Complementary to SMEFT implementations, there also exist several models of anomalous couplings such as the Higgs Characterisation \cite{Maltoni:2013sma,Demartin:2014fia,Demartin:2015uha} and BSM Characterisation models \cite{Falkowski:2015wza}. These models are all made available in the Universal \p{FeynRules} Output (UFO) format that can be imported into general purpose Monte Carlo tools, such as {\sc MadGraph5\_aMC@NLO} or \p{Sherpa}, to generate events and interface them to parton shower generators (PS). A powerful aspect of this workflow is that, once implemented, the model is generic enough to enable event generation for any desired process.

Implementations of particular processes in the presence of dimension-6 operators exist also in other frameworks. An example is the weak production of Higgs in association with a vector boson in \p{POWHEG} based on the NLO computation of \cite{Mimasu:2015nqa}, the implementation of Higgs pair production in the EFT in \p{Hpair} (including approximate NLO corrections) \cite{Grober:2017gut} and in \p{Herwig} \cite{Goertz:2014qta,Bellm:2015jjp}. Two well-known tools for calculating cross sections for Higgs production via gluon fusion including higher order QCD corrections, \p{HiGlu} \cite{Spira:1995mt,Spira:1996if} and \p{SusHi} \cite{Harlander:2016hcx}, can also include the effects of modified top and bottom quark Yukawas and the dimension-5 Higgs-gluon-gluon operator. The latter code also permits event generation at NLOQCD+PS accuracy via \p{aMCSusHi} \cite{Mantler:2015vba} including modified top and bottom quark Yukawa couplings. 
For a variety of processes with electroweak and Higgs bosons in the final state (VBF H, W and Z production, weak boson pair production, vector-boson-scattering processes, triboson production) the \p{VBFNLO} program \cite{Arnold:2008rz,Arnold:2011wj} provides NLO QCD corrections together with implementations of dimension-6 operators and, in the case of VBS and triboson production, dimension-8 operators.

There are also EFT-specific tools providing a number of useful interfaces and calculations. 
\\\p{eHDECAY} \cite{Djouadi:1997yw,Contino:2014aaa} is a package for the calculation of Higgs boson branching fractions including SMEFT effects parametrised by SILH basis operators. The freedom of basis choice in the SMEFT implies that arbitrarily many equivalent descriptions of the model can be formulated. This has important consequences for the development of EFT tools given that any numerical implementation of EFT effects requires choosing a specific basis.  A SMEFT basis translation tool, \p{Rosetta} \cite{Falkowski:2015wza}, can be used to numerically transform points in parameter space from one basis to another. It adopts the SLHA convention for model parameter specification and provides an interface to Monte Carlo event generation tools through the aforementioned BSMC model. Furthermore, additional interfaces exist to other programs such as \p{eHDECAY}, internal routines testing compatibility of Higgs signal-strength and EW precision measurements as well as providing predictions for di-Higgs production cross sections in the SMEFT. {\textsc Rosetta} provides SMEFT basis-independent access to these functionalities. A related tool is {\textsc DEFT} \cite{Gripaios:2018zrz}, a python code that can check if a set of operators forms a basis, generate a basis and change between bases.  Efforts are also underway to establish a common format for the Wilson coefficients \cite{Aebischer:2017ugx}, \p{WCxf}, which facilitates interfacing various programs computing the matching and running of the operators such as \p{DsixTools} \cite{Celis:2017hod}  and \p{Wilson} \cite{Aebischer:2018bkb}.
A public fitting framework that can be used to obtain constraints on the EFT is \p{HEPfit} \cite{deBlas:2019okz}, which is based on the Bayesian Analysis Toolkit, and includes Higgs and electroweak precision observables.

\p{JHUGen}\,\cite{jhugen,Gao:2010qx} is a coherent framework for modeling anomalous Higgs boson interactions 
with electroweak vector bosons, gluons, or fermions. The \p{JHUGenLexicon} interface allows for the parameterization 
of EFT effects either in the mass eigenstate or weak eigenstate basis of SMEFT, or directly as modifications 
of the anomalous interactions with either fermions or vector bosons.
The \p{MELA} package provides a library of matrix elements for MC sample re-weighting and optimal observable calculation. 
Details regarding EFT modeling of the off-shell processes with \p{JHUGen} are given in Section~\ref{sec:jhugen} and applications 
of this EFT framework to LHC data can be found in Refs.~\cite{CMS:2015chx,CMS:2019ekd,CMS:2022ley}.
In both on-shell $H$ and off-shell $H^*$ production with subsequent decay to two vector bosons, the most general Higgs boson couplings in gluon fusion, vector boson fusion, and associated production with a vector boson ($VH$) are implemented~\cite{Gritsan:2020pib}. In the off-shell case, additional heavy particles in the gluon fusion loop and a second resonance interfering with the SM processes are considered. Furthermore, interference with background processes is incorporated via the utilization of Standard Model matrix elements from \p{MCFM}~\cite{Campbell:2013una,Campbell:2015vwa}. In the $VH$ process, $ZH$ production via gluon fusion is included. The \p{JHUGen} framework also supports those processes with direct sensitivity to Higgs-fermion $Hff$ couplings, such as $t\bar{t}H$, $b\bar{b}H$, $tqH$, $tWH$, or $H\to\tau^+\tau^-$~\cite{Gritsan:2016hjl}.
The NLO QCD corrections for the EFT have been considered in the $VH$ and $t\bar{t}H$ processes~\cite{Gritsan:2016hjl,Gritsan:2020pib}. 

There is significant progress in computing NLO QCD corrections for the EFT, in both the top and Higgs sector 
\cite{Degrande:2016dqg,Mimasu:2015nqa,Alioli:2018ljm,Franzosi:2015osa,Zhang:2016omx,Bylund:2016phk,Maltoni:2016yxb,Degrande:2018fog,deBeurs:2018pvs}. 
This progress, on a process-by-process basis, will eventually lead to a full automation of QCD corrections for the SMEFT. As experimental 
measurements become increasingly systematics dominated, the importance of higher order calculations grows. The complete implementation of dimension-6 operators at NLO will enable the computation of NLO-QCD corrections to any tree-level process, bringing the Monte Carlo automation to the same level as the Standard Model.

\p{SMEFTatNLO} is a first complete implementation of the SMEFT which allows the computation of NLO QCD predictions. The implementation is based on the Warsaw basis of operators and includes all degrees of freedom consistent with the following symmetry assumptions: i) CP-conservation ii) $U(2)_Q \times U(2)_u \times U(3)_d \times U(3)_L \times U(3)_e$ flavour symmetry and iii) the CKM matrix is approximated as the unit matrix. The flavour symmetry imposes that only the top quark is massive. The model therefore implements the 5-flavour scheme for PDFs. The bosonic operators are implemented as in the Warsaw basis employing the $M_Z, M_W, G_F$ scheme of electroweak input parameters. The Standard Model input parameters that need to be specified are: $M_Z, M_W, G_F, M_H, M_t, \alpha_s(M_Z)$. The fermionic degrees of freedom (2 \& 4 fermion operators) are defined according to the common standards and prescriptions established by the LHC TOP WG \cite{AguilarSaavedra:2018nen} for the EFT interpretation of top-quark measurements at the LHC. The \p{SMEFTatNLO} model has been validated at LO with the \p{dim6top} implementation. A full list of the operators included in the implementation and the naming conventions can be found on the model website. 
The EFT modeling of the off-shell process with \p{SMEFTatNLO} is demonstrated in Section~\ref{sec:smeftatnlo}.

Over the coming years, progress is also expected for the computation of weak corrections in the SMEFT. A small sample of computations has been done, e.g. weak corrections to Higgs production and decay due to top quark loops \cite{Vryonidou:2018eyv} and due to modified trilinear Higgs coupling \cite{Degrassi:2016wml,Bizon:2016wgr,DiVita:2017eyz} as well as Higgs and Z-boson decays \cite{Hartmann:2015aia,Hartmann:2015oia,Hartmann:2016pil,Dawson:2018pyl,Dedes:2018seb,Dawson:2018liq}. Due to the behaviour of the Sudakov logarithms, weak corrections are typically important for high transverse momentum regions. Therefore at HE/HL-LHC their impact is expected to be enhanced. It can be expected that the recent progress on a process-by-process basis will eventually lead to the automation of the computation of weak loops in the EFT, as in the Standard Model. 

Finally, progress is expected in linking tools which compute the running and mixing of the operators with Monte Carlo tools. This will allow the automatic computation of cross-sections and differential distributions taking into account the mixing and running of the operator coefficients.


\chapter{Summary and conclusions \label{ch:conclusions}}

First, a general summary of the discussions in the subgroup was given.

To illustrate the potential impact of off-shell Higgs measurements on searches for BSM physics, the off-shell potential to resolve flat directions in parameter space for on-shell measurements was studied.  Furthermore, the sensitivity of off-shell measurements to SMEFT dimension-6 operators for the $gg\to ZZ$ process was discussed, and studies of explicit models that are testable in off-shell production were reviewed.

In a complementary contribution, the SMEFT effects in the off-shell gluon fusion and electroweak processes were discussed.  Subsequently, the computation of integrated and differential effects using \p{SMEFT@NLO} and \p{MG5\_aMC@NLO}, or \p{JHUGen} and \p{MCFM}, was demonstrated.  On that basis, a study of the prospects of obtaining additional SMEFT constraints -- beyond those from global fits -- by utilising the off-shell process was presented.

For clarification, a revised introduction, definition and discussion of the Higgs basis parametrisation of the SMEFT was given.

In short notes on the SMEFT, the Higgs basis with an additional constraint was discussed and relations between the Higgs and Warsaw bases were presented.

Lastly, an overview of EFT calculations and tools was given.

We note that theoretical work on explicit realisations of SM deviations continues to be important.  To ascertain the validity of the EFT, light new degrees of freedom need to be excluded.

In conclusion, it is worth reminding ourselves that determining bounds for selected EFT coefficients or BSM benchmark model parameters by making certain model assumptions and comparing to data in relevant search channels is an excellent start, but not the end.  Producing more/better limits is not the ultimate goal.  (Higgs) New Physics characterisation is our task -- or to rule it out.


\appendix

\chapter{Higgs basis parametrization of the SMEFT: Notation and conventions\label{app:notation}}

This appendix discusses notation and conventions used in Chapter~\ref{ch:higgsbasis}. 
Note that some of the  conventions are changed as compared  to Refs.~\cite{deFlorian:2016spz,Falkowski:hdr} in order to match those of WCxf~\cite{Aebischer:2017ugx}. 

The couplings of the $SU(3)_C \times SU(2)_L \times U(1)_Y$ gauge group are denoted by
 $g_s$, $g_L$, $g_Y$, 
 and the corresponding gauge fields by $G_\mu^a$, $W_\mu^i $, $ B_\mu $, 
 $a = 1\dots 8$, $i = 1 \dots 3$. 
 The covariant derivatives read~\footnote{%
 Note the sign difference with respect to~\cite{deFlorian:2016spz,Falkowski:hdr}.} 
\begin{equation} 
 D_\mu f =   \left (\partial_\mu  +  i  g_s G_\mu^a   T^a_f  +  i  g_L W_\mu^i T^i_f   +  i g_Y Y_f  B_\mu \right ) f .
\end{equation} 
Consequently, the covariant field strength tensors are expressed by the corresponding gauge fields as  
\begin{eqnarray}
B_{\mu \nu} &=& \partial_\mu B_\nu - \partial_\nu B_\mu,  
\nnl 
W_{\mu \nu}^i &=& \partial_\mu W_\nu^i - \partial_\nu W_\mu^i  -   g_L \eps^{ijk}  W_\mu^j W_\nu^k,  
\nnl 
G_{\mu \nu}^a &=& \partial_\mu G_\nu^a - \partial_\nu G_\mu^a -   g_s  f^{abc}G_\mu^b G_\nu^c,   
\end{eqnarray} 
where $\eps^{ijk}$ and $f^{abc}$  are the totally anti-symmetric structure tensors of $SU(2)$ and $SU(3)$. 

The fermions and Wilson coefficients of fermionic operators in the SMEFT Lagrangian are given in a particular flavor basis.  
Namely, the basis is chosen such that the fields $P_R u_J$, $P_R d_J$, $P_R e_J$ and $P_L L_J = (P_L \nu_J, P_L e_J)$  are mass eigenstates after electroweak symmetry breaking. 
Furthermore,  the left-handed quark doublets are given by $q_J= ([V_{\rm CKM}^\dagger]_{JK} P_L u_K, P_L d_J)$, 
where $P_L u_J$, $P_L d_J$ are mass eigenstates after electroweak symmetry breaking.\footnote{Note this feature is different than in~\cite{deFlorian:2016spz,Falkowski:hdr}, where the doublet was expressed as $q_J= (P_L u_J, [V_{\rm CKM}]_{JK}  P_L d_K)$  in terms of mass eigenstates.} 
Note that the definition of mass eigenstates may not be RG invariant in the presence of higher-dimensional operators. 
By convention, we choose to work with mass eigenstates defined by the Lagrangian at the scale $\mu = m_Z$.

Repeated Lorentz indices $\mu,\nu,\dots$ are implicitly contracted using the Lorentz tensor 
$\eta_{\mu\nu} = {\rm diag}(1,-1,-1,-1)$. 
Similarly, repeated generation indices $I,J,K$, as well as repeated group indices $i,j,k$, $a,b,c$  are implicitly summed over. 

The components of the Higgs double field $\varphi$ are parametrized as 
\begin{equation}
\varphi = {1 \over \sqrt 2} \bvec i \sqrt{2}  \hat G_+ \\ \hat v + \hat h + i \hat G_z \evec ,   
\end{equation} 
where $\hat v$ is the Higgs VEV,  $\hat h$ is the Higgs boson field, and $\hat G$ are the unphysical Goldstone boson fields. 

The SMEFT Lagrangian is given by 
\begin{equation}
  \mathcal L_\text{SMEFT} =
  \mathcal L_\text{SM} +
  \sum_i C_i \,Q_i ,  
\end{equation}
where $  \mathcal L_\text{SM}$ is the SM Lagrangian, and  $Q_i$ form a basis of dimension-6 operators. 
The Wilson coefficients $C_i$ have dimensions 
$[\rm mass]^{-2}$ and they count as $\cO(\Lambda^{-2})$ in the EFT expansion.
We ignore dimension-5 operators, as well as any effects  subleading to $\cO(\Lambda^{-2})$ (thus in particular, order $C_i^2$ effects are ignored).  
We work with the dimension-6 operators in the so-called Warsaw basis~\cite{Grzadkowski:2010es}. 
In the original reference the flavor structure of the operators was not specified. 
For that, we follow the notation and conventions established by the {\em Wilson coefficient exchange format} (WCxf)~\cite{Aebischer:2017ugx}. 
We also apply the  WCxf  convention that all components of  $\vec C_{\rm WB}$ have dimension  $1/{\rm mass}^2$ (thus, the SMEFT scale $\Lambda$, often displayed explicitly in the literature,  is absorbed into $C_i$ here).


\clearpage
\chapter*{Acknowledgements}
\label{acknowledgements}
\addcontentsline{toc}{chapter}{\nameref{acknowledgements}}

\indent A.A.\ was supported in part by the MIUR contract 2017L5W2PT.

\indent The work of J.B.\ has been supported by the FEDER/Junta de Andaluc\'ia project grant P18-FRJ-3735.

\indent A.F.\ is supported by the Agence Nationale de la Recherche (ANR) under
grant ANR-19-CE31-0012 (project MORA) and  by the European Union’s
Horizon 2020 research and innovation programme under the Marie
Sklodowska-Curie grant agreement No. 860881 (HIDDe$\nu$ network).

\indent C.G.\ benefited from the support of the Deutsche Forschungsgemeinschaft under Germany’s Excellence Strategy EXC 2121 ``Quantum Universe'' -- 390833306.

\indent N.K.\ acknowledges support by the Science and Technology Facilities Council under grant\\ ST/P000738/1 and thanks the Technical University of Munich for hospitality.

\indent E.V.\ and M.T.\ are supported by a Royal Society University Research Fellowship through grant URF/R1/201553.

\indent A.G.\  and L.K.\ are partially supported by the U.S.\ NSF under grant PHY-1707887. 

\indent U.S.\ is supported by the U.S.\ DOE under grant DE-SC0011702.

\indent We thank all participants for the informative and stimulating presentations and discussions during the task force meetings, and are grateful to the Off-Shell Subgroup conveners for the organisation.  We also thank the Steering Committee and WG1 conveners for support.  We are grateful to Ben Gripaios for clarifying comments.  We are obliged to CERN, in particular to the IT Department and to the Theory Unit, for technical support and hospitality.

\clearpage
\bibliography{report}
\bibliographystyle{atlasnote}


\end{document}